\documentclass[fleqn,usenatbib,useAMS]{mnras}

\usepackage{graphicx}	% Including figure files
\usepackage{amsmath}	% Advanced maths commands
\usepackage{amssymb}	% Extra maths symbols
\usepackage{multicol}        % Multi-column entries in tables
\usepackage{bm}		% Bold maths symbols, including upright Greek
\usepackage{pdflscape}	% Landscape pages
\usepackage{amsmath}
\usepackage{soul}
\usepackage[dvipsnames]{xcolor}
\usepackage{footnote}
\makesavenoteenv{tabular}
\usepackage{soul}

\defcitealias{Zibetti2009}{Z09}
\defcitealias{Taylor2011}{T11}
\defcitealias{Into2013}{IP13}
\defcitealias{Roediger2015}{RC15}
\defcitealias{Bruzual2003}{BC03}
\defcitealias{Maraston2005}{M05}
\defcitealias{Willick1997A}{W1997}
\defcitealias{Mould2000}{M2000}
\defcitealias{1997ApJ...484..145T}{TV97}
\defcitealias{de_Vaucouleurs1991rc3}{RC3}
\usepackage[bordercolor=white,backgroundcolor=gray!30,linecolor=black,colorinlistoftodos]{todonotes}

\usepackage[T1]{fontenc}
\usepackage{ae,aecompl}
\usepackage[flushleft]{threeparttable}

\usepackage{newtxtext,newtxmath}
\usepackage{tablefootnote}

\usepackage[normalem]{ulem}
\title[Disc cloaking]{Disc cloaking: Establishing a lower limit to the number density of local compact massive spheroids/bulges and the potential fate of some high-$z$ red nuggets}

\author[Hon et al.]{Dexter S.\ -H.\ Hon,$^{1}$\thanks{Contact e-mail: \href{mailto:shon@swin.edu.au}{shon@swin.edu.au},~\href{mailto:dex-hon-sci@outlook.com}{dex-hon-sci@outlook.com}}
    Alister W.\ Graham,$^{1}$
    Benjamin L.\ Davis,$^{1,2}$ and
    \newauthor
    Alessandro Marconi$^{3,4}$ 
\\
    $^{1}$Centre for Astrophysics and Supercomputing, Swinburne University of Technology, Hawthorn, Victoria 3122, Australia\\
    $^{2}$Center for Astro, Particle, and Planetary Physics (CAP$^3$), New York University Abu Dhabi\\
    $^{3}$Dipartimento di Fisica e Astronomia, Universitá degli Studi di Firenze, 50019 Firenze, Italy\\
    $^{4}$INAF Osservatorio Astrofisico di Arcetri, 50125 Firenze, Italy
}

\date{Last updated 21 Mar 2022}

\pubyear{\the\year}

\begin{document}
\label{firstpage}
\pagerange{\pageref{firstpage}--\pageref{lastpage}}
\maketitle

\begin{abstract}
The near-absence of compact massive quiescent 
{\em galaxies} in the local Universe implies a size evolution since $z\sim2.5$. 
It is often theorised that such `red nuggets' have evolved into today's elliptical (E) galaxies via an E-to-E transformation.
We examine an alternative scenario in which a red nugget develops a rotational disc through mergers and accretion, say, at $1\lesssim z\lesssim2$, thereby cloaking the nugget as the extant bulge/spheroid component of a larger, now old, galaxy.
We have performed detailed, physically-motivated, multi-component decompositions of a volume-limited sample of 103 massive ($M_*/\rm M_{\odot} \gtrsim 1\times 10^{11}$) galaxies within 110\,Mpc.
Many less massive nearby galaxies are known to be `fast-rotators' with discs.
Among our 28 galaxies with existing elliptical classifications, we found that 18 have large-scale discs, and two have intermediate-scale discs, and are reclassified here as lenticulars (S0) and elliculars (ES).
The local {\it spheroid} stellar mass function, size-mass diagram and bulge-to-total ($B/T$) flux ratio are presented.
We report lower-limits for the volume number density of compact massive spheroids, $n_\mathrm{c,Sph}\sim (0.17$--$1.2) \times 10^{-4}\,\rm  Mpc^{-3}$, based on different definitions of `red nuggets' in the literature.
Similar number densities of local compact massive bulges were reported by de la Rosa et al. using automated two-component decompositions and their existence is now abundantly clear with our multi-component decompositions.
We find disc-cloaking to be a salient alternative for galaxy evolution.
In particular, instead of an E-to-E process, disc growth is the dominant evolutionary pathway for at least low-mass ($1\times10^{10}<M_*/\rm M_{\odot} \lessapprox 4 \times 10^{10}$) red nuggets, while our current lower-limits are within an alluring factor of a few of the peak abundance of high-mass red nuggets at $1\lesssim z\lesssim2$.

\end{abstract}

\begin{keywords}
galaxies: evolution -- galaxies: bulges -- galaxies: elliptical and lenticular, cD -- galaxies: structure -- galaxies: abundances -- galaxies: disc
\end{keywords}

%%%%%%%%%%%%%%%%%%%%%%%%%%%%%%%%%%%%%%%%%%%%%%%%%%

%%%%%%%%%%%%%%%%% BODY OF PAPER %%%%%%%%%%%%%%%%%%

% The MNRAS class isn't designed to include a table of contents, but for this document one is useful.
% I therefore have to do some kludging to make it work without masses of blank space.
%\begingroup
%\let\clearpage\relax
%\tableofcontents
%\endgroup
%\newpage

\section{Introduction}

 Compact ($R_{\rm e} \lesssim 2$\,kpc) massive ($M_{\rm *,gal} \gtrsim 10^{10}$--$10^{11}$\,M$_{\odot}$), quiescent galaxies \citep[nicknamed `red nuggets' by][]{Damjanov2009} are common at high-redshifts \citep{Buitrago2008,Perez-Gonzalez2008}, but relatively rare in the local Universe according to many studies \citep[e.g.,][]{Trujillo2009,Taylor2010,Saulder2015}.
First detected in the local Universe by \citet{1968cgcg.bookR....Z} and \citet{1971cscg.book.....Z}, and later \citet{2016ASSP...42..177P}, examples include NGC~1271 \citep{2016ApJ...831..132G}, NGC~1277 \citep{2014ApJ...780L..20T, 2016ApJ...819...43G}, NGC~5252 \citep{Sahu2019A}, MRK 1216, PGC 032873 \citep{2018MNRAS.477.3886W}, and others \citep{Damjanov2013,Savorgnan2016disc,Tortora2018}.
It is frequently speculated that mergers converted the majority of the high-$z$ compact massive galaxies into large (pure) elliptical galaxies by today \citep[e.g.][]{Naab2009,Hopkins2009,Bezanson2009,Trujillo2011}.
However, \citet{2013pss6.book...91G}\footnote{First published online in 2011 (\url{https://arxiv.org/abs/1108.0997}).}, and later \citet{2013MNRAS.430.2622D}, \citet{Dullo2013}, and \citet{Dutton2013}, noted that this may not always be the case and that the spheroidal components of today's disc galaxies (including ES, S0, and S) are also compact and massive.

While `relic' galaxies, believed to be untouched red nuggets from the local Universe, have been found, they are certainly not common. In addition, relics show complex dynamical structures \citep{2014ApJ...780L..20T,Ferre-Mateu2017} that warrant being treated as more than a single component system.
Could the growth of two-dimensional discs, rather than three-dimensional envelopes, be a prelevant mechanism transforming these galaxies? 
Such a two-phase disc-building scenario, rather than an envelope-building process, was suggested by \citet{Graham2011}, and \citet{2012ApJ...747L..28N} subsequently reported on the rapid build up of discs at $z \sim 1$, which could yield S0 galaxies with old stellar populations by today.
At stake is the evolutionary pathway for some/many of the Universe's old and massive galaxies, and perhaps also for some of the less massive, high-$z$ 'red nuggets', which we term `red pebbles', which may now be the bulges of some spiral galaxies.

The distinct size difference between the passive galaxy populations at high and low redshifts sparked deliberations as to the dominant evolution mechanism of massive early-type galaxies (ETGs). 
While regular size ETG are not rare at low and high-$z$, one cannot say the same for the compact massive red nuggets.
The red nuggets are rather abundant at high redshifts. For instance, \citet{Saracco2010} reported a comoving number density of compact galaxies at \(0.9< z < 1.92\) of $2.3\times10^{-5}$\,Mpc$^{-3}$ for $M_*/\mathrm{M_{\odot}}> 3\times10^{10}$. 
\citet{Barro2013} plotted the evolution of the number density across different redshifts, finding that the number density, $n$, of red nuggets peaked at a lookback time of 8\,Gyr ($z=1$), with  $n_{\rm peak} \approx 1.5 \times 10^{-4}$ Mpc$^{-3}$.
However, regardless of the assorted size and mass criteria used to define red nuggets, it has repeatedly been claimed that these objects are virtually nonexistent at $z \sim 0$ \citep{Trujillo2009,Taylor2010}.
The sharp decrease in the compact massive galaxy population indicates a drastic change in galaxy structure over this period of time. 
However, disagreement among observations \citep[e.g.,][]{Trujillo2007,Valentinuzzi2010,Poggianti2013A} led to debates over whether there is indeed a drop in the number of compact massive quiescent systems across time.

The structure of an ETG is often overlooked and oversimplified.
Many are misclassified as purely elliptical (E) galaxies while in reality, most ETGs contain a disc \citep{Carter1987,1987AJ.....94.1519C,Nieto1988,1990AJ.....99.1813C,D'Onofrio_1995, Graham1998,Emsellem2011,Scott2015}; either a large-scale disc making it a lenticular (S0) galaxy, or an intermediate-scale disc making it an ellicular\footnote{The concatenated name (elliptical $+$ lenticular) `ellicular' was introduced for the ES galaxy type in \citet{Graham2017}.} (ES) galaxy  \citep{Liller1966,Savorgnan2016disc,Graham2019class}.
Given that different structures, for example, triaxial spheroids or relatively flat discs, form via different physical mechanisms, advances may be made by not simply treating galaxies as if they are single-component systems but rather examining their bulge/disc nature and the role of S0 and ES galaxies in the grand scheme of galaxy evolution.\footnote{Those unfamiliar with the historical discovery of, and continuity between, the different morphological sub-structures in large, high surface brightness galaxies may like to refer to the galaxy classification grid presented in \citet{Graham2019class}.
It builds upon past works by better recognizing the range of disc sizes in ETGs, which can also contain bars.}

From a random incomplete sample, \citet{Graham2015} reported a lower-limit to the volume number density for local compact massive {\it spheroids} in fair agreement with some of the number densities reported for compact massive {\it galaxies} at high-$z$ \citep[see also][]{de_la_Rosa2016,Costantin2020}.
Here, we investigate the issue more thoroughly, using a mass and a volume-limited sample of nearby galaxies combined with a careful, homogeneous, physically-motivated\footnote{Rather than randomly/automatically fitting two or three S\'ersic components, we fit physical structures such as bars, \textit{ansae} \citep[e.g.,][]{2007AJ....134.1863M}, discs, etc., as detected through imaging and kinematic data.}, multi-component decomposition of their surface brightness profile. 
We do this to detect and quantify the primary spheroidal component, a.k.a.\ bulge.

While we do not advocate any particular origin for the high-$z$ red nuggets, we do note that some
bulges may form in other ways.  For example, star-forming clumps may form in turbulent disks at high-redshift before migrating inwards to form the bulge \citep[e.g.,][]{2007ApJ...670..237B,2008ApJ...688...67E}. This process does not preclude subsequent disk accretion and growth but it does start with a disk rather than the monolithic collapse of a spheroid.  In addition, disks (with and without a stabilising dark matter halo) can experience instabilities leading to bars, which can in turn experience instabilities of their own, leading to the formation of buckled bars often with distinct (peanut shell)-shaped structures known as pseudobulges \citep[e.g.,][]{1975IAUS...69..297B, 1975IAUS...69..349H,1981A&A....96..164C, 1990A&A...233...82C}.  While both low-mass classical bulges and pseudobulges can rotate and have a S\'ersic index for their light profile around 1 \citep{2015HiA....16..360G}, the bar$+$pseudobulge has a tendency to result in elongated isophotes having a maximum $B_6$ Fourier harmonic term at around half the length of the bar \citep[][and references therein]{Ciambur2016}\footnote{In our decomposition, such pseudobulge structures are effectively folded into the bar component.}.
Classical bulges, bars and pseudobulges, and disks are known to coexist \citep[e.g.,][]{1996ApJ...462..114N, 2003ApJ...597..929E, 2005MNRAS.358.1477A}.  
In our multi-component decompositions, we focus on recovering what would be considered the "classical bulge". 
We will present the decompositions of the galaxy light in a forth-coming paper which will compare the sizes and masses obtained from fitting a single S\'ersic function, a S\'ersic plus an exponential function, and our multi-component analysis used here to obtain the spheroid size and mass. 

In Section~\ref{sec:data}, we describe the data selection of local galaxies that potentially contain a compact massive spheroid.
Section~\ref{sec:analysis} goes through the process of galaxy model fitting and some of the nuances of multi-component decomposition.
In Section~\ref{sec:results}, the spheroid mass function, the spheroid size-mass relation, and the number density of local compact massive spheroids are presented.  
A comparison with local spheroid and high-$z$ samples are shown in Section~\ref{sec:comparsion}.
Based on the newly-found local spheroids, 
we discuss the implications for galaxy evolution as a whole in Section~\ref{sec:Discussion}.
We summarise our findings in Section~\ref{sec:conclusion}.
Finally, in a set of appendices, we provide an accounting of distance corrections (Appendix~\ref{sec:dist_corr}), assumptions governing our adopted mass-to-light colour relations (Appendix~\ref{sec:MLCR_suppl}), host galaxy sample (Appendix~\ref{sec:Gal_can}), and structural parameters for the local spheroids (Appendix~\ref{sec:sph_para}).

Throughout this paper, we assume the following cosmographic parameters: $H_0=68$ km s$^{-1}$ Mpc$^{-1}$, $\Omega_{\rm m} = 0.3$, and $\Omega_{\Lambda} = 0.7$.
The absolute magnitude of the sun ($\mathfrak{M_{\odot}}$) in $i-$band is set to 4.53 (in AB mag).
Unless otherwise stated, all bulge parameters ($R_{\rm e}$, $\mu_{\rm e}$ and $n$) correspond to the geometric-mean axis ($R_{\rm eq}=\sqrt{R_{\rm maj}R_{\rm min}}$, {\it aka} the equivalent axis) which is equivalent to a circularised form of the isophotes.

\section{Data}
\label{sec:data}

In Section~\ref{sec:Selection}, we explain the parent data selection process. Section~\ref{sec:image} briefly describe the images sources.
Section~\ref{sec:stellar_mass} explores the various stellar mass-to-light ratio implemented in this study. 
Section~\ref{sec:distance} describes our process of finding the appropriate distance for each galaxy.
Section~\ref{sec:volume-limited} is a general summary of our volume-limited sample.

\subsection{Parent sample}
\label{sec:Selection}
 
To calculate the number of (compact, massive) spheroids per cubic Mpc, hereafter the number density, as well as study the statistical properties of these local spheroids, a well-defined volume-limited sample of galaxies is required.
The galaxy images should also be of sufficiently high spatial resolution for us to properly separate the substructures such as discs and bars from the spheroidal components.
This limits our sample's maximum co-moving distance to approximately 100~Mpc, because, beyond this point, the multi-component decomposition becomes challenging with $\sim$1$\arcsec$ seeing.

 We elected to use Sloan Digital Sky Survey \citep{York2000} photometric data because of its wide sky coverage of $\sim$14,555\,deg$^2$ \citep{Aihara2011_SDSSDR8}, and its completeness of galaxies at the bright end ($m_{i^{\prime}} <$ 23\,mag) of the galaxy magnitude distribution.
 Our sample selection is based on the NASA-Sloan ATLAS catalogue\footnote{\url{http://www.nsatlas.org/}}, a data set that consists of information from the SDSS DR8 \citep{Aihara2011_SDSSDR8}, NASA Extragalactic Database \citep[NED,][]{Helou1991}\footnote{\url{http://ned.ipac.caltech.edu}}, Six-degree Field Galaxy Redshift Survey \citep{Jones2004_6dF}, Two-degree Field Galaxy Redshift Survey \citep{Colless1999_2dF}, ZCAT \citep{Huchra1983_CfA}, and ALFALFA \citep{Giovanelli2005_ALFALFA}.
 The catalogue provides a list of unique objects as they remove all the redundant data and stars that were previously misclassified as galaxies.
 It contains 145,155 galaxies within a redshift \(z \sim 0.055\).

\begin{figure}
\centering
	\includegraphics[clip=true, trim= 13mm 2mm 12mm 0mm, width=\columnwidth]{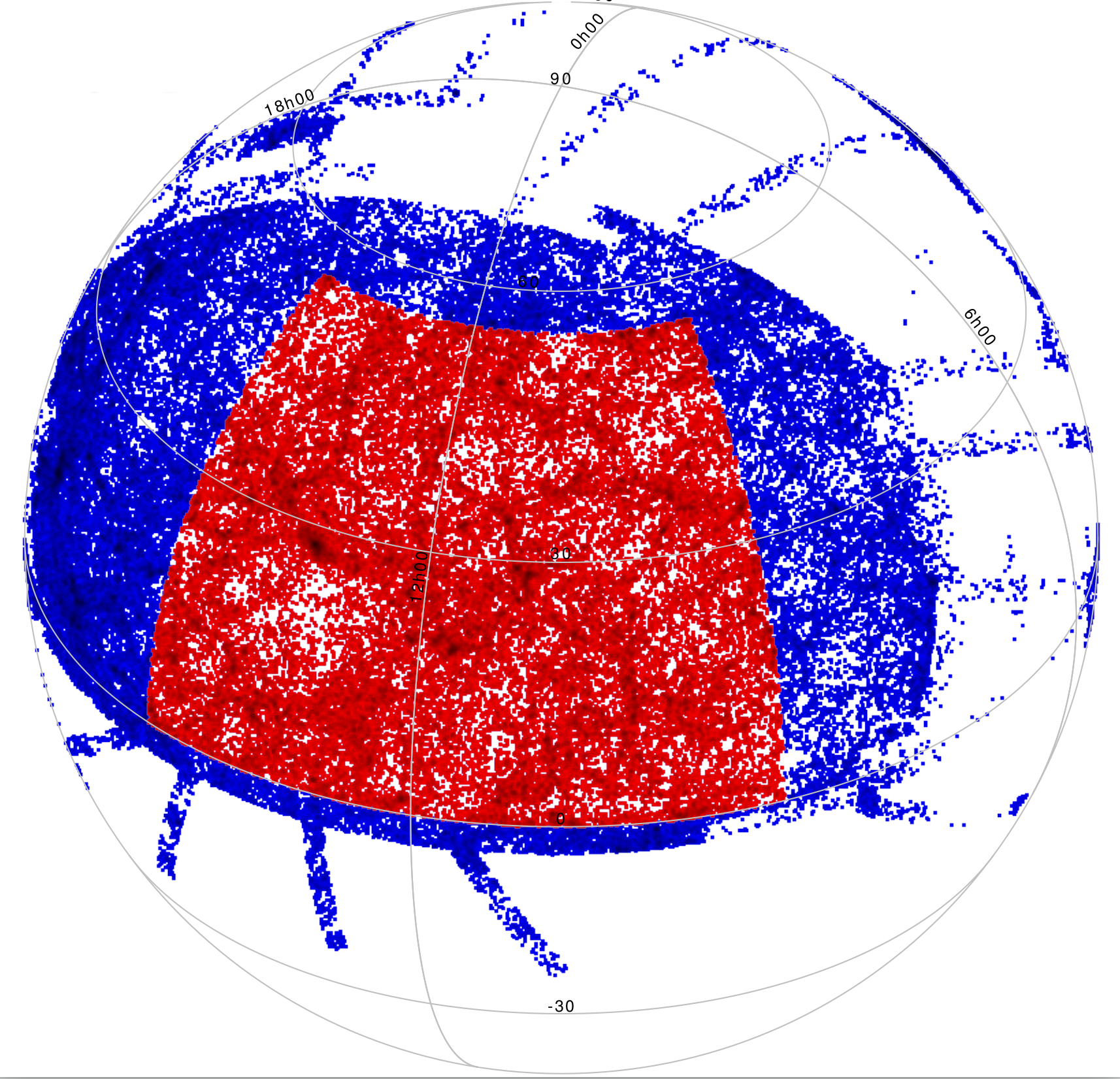}
    \caption{The angular boundaries of the volume-limited sample.
    The \textcolor{blue}{blue} points belong to the parent sample from the NASA-Sloan ATLAS catalogue, while the \textcolor{red}{red} points are the galaxies satisfying the criteria in  Equation~\ref{eq:RADEC}.} 
    \label{fig:RADEC}
\end{figure}

From the parent sample, we limit the angular boundaries to
\begin{subequations} 
\label{eq:RADEC}
\begin{align}
139\degr &< {\rm R.A.} < 214\degr, {\rm and} \\
0 &< {\rm Dec} < 55\degr. 
\end{align}
\end{subequations}
We define these arbitrary boundaries in the attempt to cover the majority of the northern sky in the SDSS field.
It contains 55,370 galaxies and captures 8.5\% of the sky, or a solid angle of 1.07\,sr.
Figure~\ref{fig:RADEC} shows the distribution of the angular selection.
The area contains notable supergalactic structures such as the Virgo and Coma clusters.

\subsection{The imaging data}
\label{sec:image}

Our imaging data come from SDSS $i-$band images.
We decided to use $i$-band photometry for its smaller galactic extinction.
The initial flux of the objects was previously measured with a single S\'{e}rsic component fit by the SDSS pipeline.
The NASA-Sloan ATLAS catalogue provided the flux in the $ugriz$ bands in units of \textit{nanomaggies}.
The SDSS magnitude of a galaxy in the AB magnitude system \citep{1974ApJS...27...21O} is:
\begin{equation} 
\label{eq:magSDSS}
m = [22.5\,{\rm mag}] - 2.5\log_{10}(f)
\end{equation}
where $f$ is the flux of the object in \textit{nanomaggies} and [$22.5\,{\rm mag}$] is the photometric zero-point magnitude for all bands in the SDSS. 

The seeing for each galaxy was estimated by using the Image Reduction and Analysis Facility \citep[\texttt{IRAF};][]{Tody1986IRAF, Tody1993IRAF} function \texttt{imexam}.
We typically measured 5 or 6 stars randomly distributed in each frame.
The \texttt{imexam} task was used to fit a Gaussian function to each star.
The median value of the full width at half maximum (FWHM) was used to represent the seeing of the frame.
The measurements are listed in Column~8 of Tables~\ref{tab:sample1}--\ref{tab:sample3}.

\subsection{Stellar masses}
\label{sec:stellar_mass}

Despite slight differences in definition between studies, compact massive galaxies generally have a stellar-mass larger than $M_*/{\rm M_\odot} \sim (1$--$7) \times 10^{10}$.
We will focus on local host galaxies with a total stellar mass of  $M_*/{\rm M_\odot} \gtrsim 1 \times 10^{11}$, in order to find compact massive spheroidal components.
The determination of the stellar mass depends heavily on different factors, including the initial mass function (IMF); stellar population synthesis (SPS) models; the treatment of dust attenuation; and the fine-tuning of the age, metallicity, and reddening parameters.
While it is a common practice to obtain the stellar mass by fitting a stellar model to the spectral energy distribution (SED) of the galaxy light, it is not quite appropriate for our purpose.
Having the {\em galaxy} SED does not provide us with the mass-to-light ratio ($M_*/L$) of the {\em spheroid}, which would require a structural decomposition in many bands to provide the SED of the spheroid.

While bulge$+$disc fits in many bands are increasingly produced on scale \citep[e.g.][]{2016MNRAS.460.3458K}, the proximity of our sample, in which bars and rings are resolved, mandates the implementation of multi-component decompositions.
To date, these are performed manually rather than automatically, which allows for greater care and recourse to additional information, such as kinematic maps in the literature.  Given time constraints, 
we elected to focus on the decomposition of $i-$band images (described in detail in Section~\ref{sec:decomposition}).
It has also be shown that the $i$-band mass-to-light ratio $M_*/L$ and $(g-i)$ relation has a low scatter intrinsically \citep{Taylor2011}.
We use the galaxy colour, rather than the full SED, as a proxy for the spheroid colour when estimating the spheroid $M_*/L$ ratio.

We utilise several mass-to-light versus ($g-i$)-colour relations (MLCRs), 
$\Upsilon_{i}(g,i$), 
available from the literature and applicable to the $i$-band magnitudes:
\citet[][hereafter Z09]{Zibetti2009};
\citet[][hereafter T11]{Taylor2011}; 
\citet[][hereafter IP13]{Into2013}; and 
\citet[][hereafter RC15]{Roediger2015}.
All of these works assume an exponentially declining star formation history (${\rm SFH} \propto e^{t/\tau}$) in their models.
The following equations can be used to calculate the host galaxy total stellar mass, and later the bulge/spheroid stellar mass:
\begin{subequations} 
\label{eq:MLCRs}
\begin{align}
 Z09:&\, \log_{10} \Upsilon_{i}(g,i)=1.032(g - i)-0.963\label{eq:MLCRs_a} \\
 T11:&\,  \log_{10} \Upsilon_{i}(g,i)=0.700(g - i)-0.680\\
 IP13:&\,  \log_{10} \Upsilon_{i}(g,i)=0.985(g - i)-0.669\\
 RC15:&\, \log_{10} \Upsilon_{i}(g,i)=0.979(g - i)-0.831
\end{align}
\end{subequations}
Given the conditions on their respective assumptions, each MLCR self-reported to have the following intrinsic dispersions, Z09: 0.10--0.15\,dex, T11: 0.1\,dex, IP13: 0.13\,dex, and RC15: 0.14\,dex.

\begin{figure}
\centering
	\includegraphics[clip=true, trim= 3mm 1mm 3mm 2mm, width=\columnwidth]{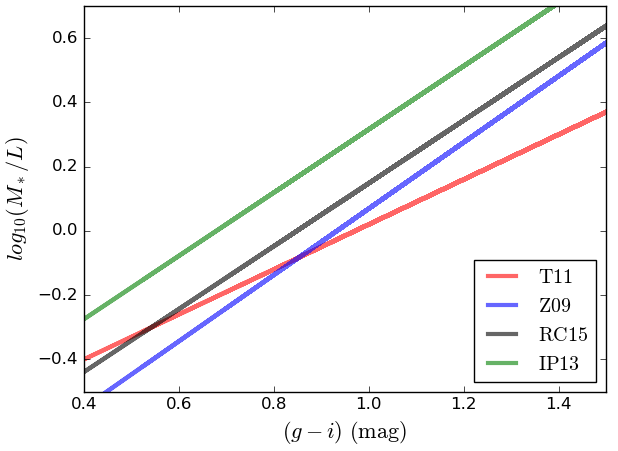}
    \caption{The ($M_*/L_i$)-to-$(g-i)$ relations.
    The MLCR equations (\ref{eq:MLCRs}) are shown as red, blue, black, and green lines for T11, Z09, RC15, and IP13, respectively.
    The massive galaxies ($M_*/\rm M_{\odot} > 1\times 10^{11}$) occupy the colour range $(g-i)\sim 1.20 \pm0.04~\rm mag$.
    Within this range, the $M_*/L$ ratio decreases monotonically in the following order:  IP13 > RC15 > Z09 > T11. 
    } 
    \label{fig:ML}
\end{figure}

For our sample, the galaxy $(g-i)$ colours exhibit a familiar bimodal distribution (\citealt[][]{1997ApJ...484..145T}; \citealt{2003AJ....126.1720S}; \citealt{2003ApJ...594..186B}; \citealt{Kauffmann2003A}; \citealt{2004ApJ...600..681B}; \citealt{2004MNRAS.351.1151B}; \citealt{2004ApJ...615L.101B}).
The most massive galaxies occupy the `red' distribution, centered at $(g-i) \sim 1.2$\,mag.
We demonstrate the $\mathrm{log_{10}}(M_*/L)$-to-$(g-i)$ relations for the four equations in Figure~\ref{fig:ML}. 
Within our sample's colour range ($\sim 1.2 \pm 0.04$\,mag), the MLCRs yield decreasing $M_*/L$ ratio in the following order: IP13 $>$ RC15 $>$ Z09 $>$ T11.  

Given the goal of this work is comparative in nature, it is important to know the stellar mass \textit{relative} to the high-$z$ galaxies.
While the underlying assumptions in calculating the stellar masses of high-$z$ galaxies have varied from paper to paper, we attempt to make the comparison using the MLCR best matches their assumptions\footnote{The theoretical background for each MLCR is discussed in detail in Appendix~\ref{sec:MLCR_suppl}.}. 
Throughout this work, we present the result mainly using RC15 stellar mass for the sake of consistency.
It represents an intermediate estimation between the lowest (T11) and the highest estimation (IP13\footnote{Considering there are many S0 and S galaxies in our sample, we apply the disc-model equation from IP13 (see their Table 5). Although, the two equations in IP13 (in their Table 3 and 5) for $(g-i)$ colour would return a similar result.
We provide several estimates of the stellar mass, with the MLCRs from IP13 yielding the highest masses here. It is noted that this is not an extreme mass estimate, with MLCRs from \citet{Bell2003} and \citet{Schombert2019} yielding yet higher masses.}).
Moreover, RC15 have the same priors as some prominent red nuggets studies \citep[][, etc.]{Barro2013,van_der_Wel2014,van_Dokkum2015}, namely the \citet{Chabrier2003} IMF and the \cite{Bruzual2003} SPS.
That being said, the effect of using a different $M_*/L$ ratio prescription should be considered thoroughly.  
Having several stellar mass estimations allow us to examine the potential bias inherent in the choice of stellar mass estimation measurement. 
We shall explore how the number density of the embedded compact massive spheroids may change for each MLCRs, later in Section~\ref{sec:z3}.

\subsection{Distances}
\label{sec:distance}
We try to account for the peculiar velocity influence on distance measurements.
There are several empirical methods to approximate the distance of nearby galaxies: the Tully-Fisher (TF) relation \citep{Tully1977}, Fundamental Plane \citep{Djorgovski1987}, and velocity field reconstruction \citep{Peebles1989,Peebles2001,Phelps2006,Shaya2013}.
To construct our volume-limited sample, the procedure involved several steps.

\begin{figure}
\centering
	\includegraphics[clip=true, trim= 1mm 1mm 1mm 1mm, width=\columnwidth]{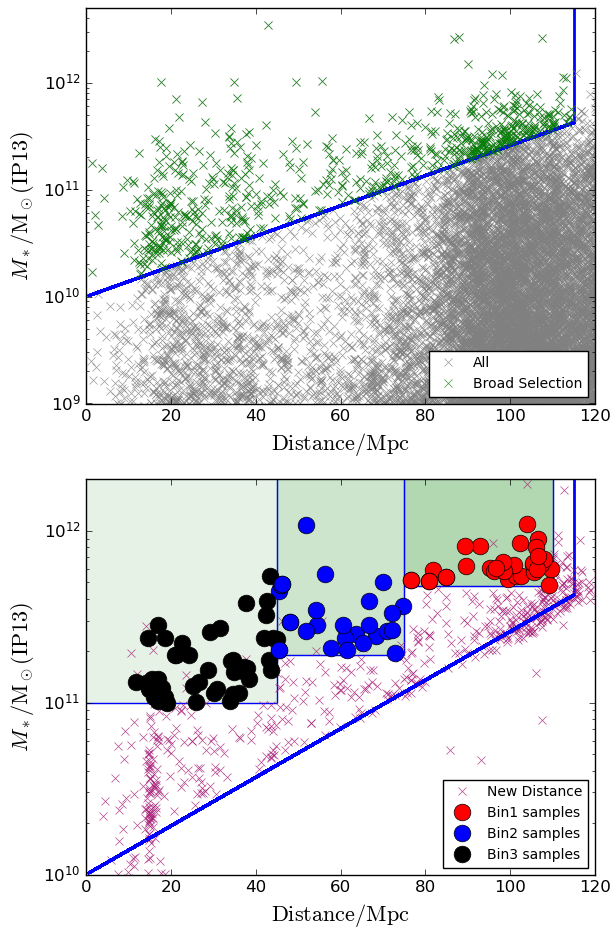}
    \caption{
    The data selection for potential host galaxies containing compact massive spheroids.
    \underline{Top}: The grey crosses (\textcolor{Gray}{$\times$}) are the galaxies satisfying the angular sample selection boundaries in Equation~\ref{eq:RADEC}.
    The green crosses (\textcolor{OliveGreen}{$\times$}) are the galaxies that satisfy Equation~\ref{eq:braod_cut}.
    All distances are from W1997.
    \underline{Bottom}: Bins~1, 2, and 3 are shaded in green for the right, middle, and left boxes, respectively.
    All distances have been updated (see Appendix~\ref{sec:dist_corr}).
    The \textcolor{red}{red}, \textcolor{blue}{blue}, and \textbf{black} points correspond to the samples listed in Tables~\ref{tab:sample1}--\ref{tab:sample3}, respectively.
    Violet crosses (\textcolor{RedViolet}{$\times$}) either do not satisfy the Equation~\ref{eq:Bin} conditions or do not consistently reside in the same bin if alternative distance estimates are adopted (see Figure~\ref{fig:dist_compare_bin1}).
    After careful examination, we excluded the false sample that are stars or galaxies with wrong velocities.
    The crosses not overlapping with the points in the lower panels are rejected from the analysis.
    }
    \label{fig:data_selection}
\end{figure}

(i) Redshifts.  We commenced this project by first calculating the distances using the corrected redshift (ZDIST) available in the NASA-Sloan ATLAS catalogue.
The redshift values are based on the Mark III Catalog of Galaxy Peculiar Velocities \citep[][hereafter W1997]{Willick1997A}, obtained using the TF or $\rm D_{n}$-$\sigma$ relation.
This catalog provides a Malmquist bias-corrected distance based on the velocity field provided by the IRAS $1.2$\,Jy redshift survey \citep{Fisher1995} density field.
The initial distances and stellar mass estimation are shown in the upper panel of Figure~\ref{fig:data_selection} as grey crosses.

(ii) A broad mass-distance sample selection boundary. 
We select our initial galaxy sample for further examination using the broad mass-distance selection criteria (blue line) in the upper left-hand corner of Figure~\ref{fig:data_selection}, satisfying the arbitrary condition
\begin{align}
\label{eq:braod_cut}
    \log_{10}(M_*/{\rm M_\odot}) &> 10 + (0.014 \times \rm Dist.), 
\end{align}
and $ \rm Dist. < 115 \, \rm M p c$.
This is intended to produce a manageable sample size of galaxies for us to further check if redshift-independent distances are available on NED.
There are 708 data points (green crosses) remaining after imposing the mass-distance sample selection boundaries.

(iii) Updated redshift-independent distances.
Each of these green points was checked in NED to see if there were redshift-independent measurements. 
We primarily focussed on obtaining distances related to stellar phenomena, such as surface brightness fluctuation (SBF), Cepheids, Supernovae Ia (SNIa), the tip of the red giant branch (TRGB), etc.  
When such measurements were not available for a galaxy, we resorted to using the velocity correction from NED, based on the linear infall velocity model in \citet[][hereafter M2000]{Mould2000}.
This model corrects for the influence imposed by major gravitational bodies, namely the Virgo Cluster, the Great Attractor, and the Shapley supercluster. 

The revised distances are shown as the coloured dots and purple crosses (labelled `New Distance', see Column (4) in Table~\ref{tab:sample1}-\ref{tab:sample3}) in the lower panel of Figure~\ref{fig:data_selection}\footnote{Note that, because we rely on redshift independent distances measured using stellar information, our distance is luminosity distance. Although, in this redshift range ($0<z<0.025$), luminosity distance and comoving distance are similar.}.
One can see changes from the upper panel.  For example, 
one can see how the Virgo Cluster at $\sim$17\,Mpc materializes from the data, 
and quite a number of galaxies escape the blue-line selection boundary.
One can similarly expect that some galaxies initially outside of the blue-line selection boundary (and ignored by us) would be shifted inside. 
We mitigated against this by limiting the final selection to galaxies in three regions, which contained the galaxies previously outside these bins based on the older distances.  
Within the broad mass-distance selection boundary sample of 708 galaxies, we defined three volumes, \emph{aka} bins, to acquire both a manageable\footnote{By the term `manageable', we are referring to the workload required to perform careful multi-component decomposition.} and useful number of galaxies.
\begin{subequations} 
\label{eq:Bin}
\begin{align}
 \mathrm{Bin~1:}\, &  M_*/ \rm  M_\odot (IP13)\gtrsim 5.0 \times 10^{11},\nonumber\\
 & 75\,\mathrm{M p c}< \rm Dist. <  110\, \rm M p c \\
 \mathrm{Bin~2:}\, &  M_*/ \rm  M_\odot (IP13) \gtrsim 2.0 \times 10^{11} ,\nonumber\\
 & 45\,\mathrm{M p c}< \rm Dist. < 75\, \rm M p c \\
 \mathrm{Bin~3:}\, &  M_*/ \rm  M_\odot (IP13) \gtrsim 1.0 \times 10^{11} ,\nonumber\\
 & \rm Dist. < 45\,\rm M p c.
\end{align}
\end{subequations}

We chose these arbitrary bins to include the massive galaxies that potentially host a massive bulge ($M_*/\rm M_{\odot} > 10^{10}$).
The data points of each bin are highlighted in different colours in the lower panel of Figure~\ref{fig:data_selection}.

(iv) Further refinements.  
In this step, we further checked the appropriate distance of each galaxy in the three bins.
In addition to the redshift-independent measurements, as well as the W1997 and M2000 velocities, we collected another independent set of distances using the newly available \textit{Cosmicflow-3} distance calculators \citep{Kourkchi2020}.
The \textit{Cosmicflow} project uses a collection of diverse distance measurements \citep{Tully2008,Tully2013} to derive the velocity and density field fluctuations in the local Universe.
Their two online calculators provide a distinct approach for galaxies with $ \rm Dist. < 38 \rm \, M p c$ and distances greater than this and up to 200 Mpc. 
In the former case, the velocity field is based on the non-linear numerical action orbit reconstruction in \citet{Shaya2017} and the velocity model of \citet{Tully2014}.
For the latter case, the distance is inferred from the three-dimensional linear velocity field model in \citet{Graziani2019}, with an emphasis on mitigating inherent biases such as Malmquist bias 
and the lognormal distribution of peculiar velocities.
The main advantage of \textit{Cosmicflow} is the large library (17,647 individual galaxies) of distances from which the velocity field is constructed.

Armed with 3 to 4 distinct distance estimations for each galaxy in our set of three bins, we found that no matter which distance is chosen, all of our sample lies within the three bins selection volume (within 110\,Mpc). 
The sample provides a good representation of galaxies within these boundaries.

The distance we used for the calculation is chosen on a case-by-case basis.
The $z$-independent distance is prioritized if they are available.
We have 20 and 1 $z$-independent distances in Bin 3 and Bin 1, respectively. 
Unless the measurement has a significant error or disagrees significantly with the other measurements, $z$-independent distance is used (in total 18).
When $z$-independent distances are not available, we resort to using the \textit{Cosmicflow-3} distances (in total 81). 
There are a few of \textit{Cosmicflow-3} distances that fall within the "triple-value" region, as the Virgo Cluster bends the distance-velocity relation into a cubic function and returns three distinct distance solutions.
In such cases (in total, only 4 galaxies), we choose the W1997 distances instead.

The final distance we used in the calculation is shown in Column~(2) in Tables~\ref{tab:result1}--\ref{tab:result3}, and the details of specific cases are discussed in the Appendix~\ref{sec:dist_corr}.
After eliminating all the false samples (stars or galaxies with obviously wrong velocities), this leaves us with 115 galaxies spread over three bins. 

\begin{table}
   \caption{The volume of each bin.}
   \begin{threeparttable}[b]
   \label{tab:volume}
   \begin{tabular}{cll}
   \hline
   & Distance ($\rm Mpc$) & Volume ($\rm Mpc^3$)\\
   \hline
   \hline
   Bin~1 & $\rm 75-110$ & $3.25 \times 10^{5}$\\  
   Bin~2 & $\rm 45-75$ & $1.18 \times 10^{5}$\\
   Bin~3 & $\rm 0-45$ & $3.26 \times10^{4}$\\
   \hline
   Total & $\rm 0-110$ & $4.76\times10^{5}$\\    
   \hline
   \end{tabular}
    \end{threeparttable}
\end{table}

\subsection{Volume-limited galaxy sample}
\label{sec:volume-limited}
We defined the bins in a manner that would produce a manageable sample size for careful multi-component decomposition work. 
The volume of each bin and the total survey volume are shown in Table~\ref{tab:volume}.
It is labour intensive to create a detailed decomposition.
As we will discuss in Section \ref{sec:decomposition}, the process involves manual attention and cross-referencing evidences from the literature.
While it would be best to conduct large-scale studies on all the local massive host galaxies, the current limitation in automatic decomposition programmes prevent us from doing so.
We mitigate the challenge by limiting a local volume with respect to the relevant mass range.

In the parent sample selection (Equation~\ref{eq:Bin}), we have proceeded by intentionally using the highest mass estimations, which come from IP13, in order not to miss potential host galaxies of compact massive spheroids. 
This ensures the inclusion of even the least likely (the least massive) candidates within our stellar mass selection criteria.
For easy comparison across different MLCRs, we present a table of conversion, the lower mass limit for each bin across the four MLCRs in Table~\ref{tab:mass_limit}.
As one applied another MLCRs instead of IP13, the galaxies stellar mass within the three bins decreased by roughly a fixed amount ($\delta M_*/ \rm M_{\odot}$)\footnote{
While $\delta M_*/ \rm M_{\odot}$ is different for each galaxy, the deviation is insignificant.
The deviation between the maximum and minimum $\delta M_*/ \rm M_{\odot}$ is equal to 0.087\,dex for Bin~1, 0.01\,dex for Bin~2, and 0.002\,dex for Bin~3.
}. 
The lower mass limits for other MLCRs are obtained by reducing the IP13 lower mass limits by $\delta M_*/\rm M_{\odot}$.
Ultimately, the choice of MLCR would not affect the analysis.
As we move forward to discuss in RC15 stellar mass terms, the lower mass boundaries of the host galaxies are now $3.4\times10^{11}\,\mathrm{M}_{\sun}$, $1.3\times10^{11}\,\mathrm{M}_{\sun}$, and $6.7\times10^{10}\,\mathrm{M}_{\sun}$ for Bin 1, 2, and 3, respectively.

Given each bin's differences in the lower galaxy mass that was included, throughout this paper, the result of each bin is presented separately.
Tables~\ref{tab:sample1}--\ref{tab:sample3} show the galaxies' basic information from the three bins.

\begin{table}
   \caption{The lower mass limit of each bin using different MLCRs.}
   \begin{threeparttable}[b]
   \label{tab:mass_limit}
   \begin{tabular}{cllll}
   \hline
         & T11 & Z09 & RC15 & IP13\\
         & ($M_*/\rm M_{\odot}$) & ($M_*/\rm M_{\odot}$) & ($M_*/\rm M_{\odot}$) & ($M_*/\rm M_{\odot}$)\\
   \hline
   \hline
   Bin~1 & $1.8\times10^{11}$ & $2.9\times 10^{11}$ & $3.4\times10^{11}$ &   $5.0\times 10^{11}$ \\  
   Bin~2 & $7.2\times10^{10}$ & $1.1\times 10^{11}$ & $1.3\times10^{11}$ &  $2.0\times 10^{11}$\\
   Bin~3 & $3.0\times10^{10}$ & $5.7 \times 10^{10}$ & $6.7\times10^{10}$ &  $1.0 \times 10^{11}$\\
   \hline
   \end{tabular}
    \end{threeparttable}
\end{table}

\section{Methodology}
\label{sec:analysis}

This section showcase our data processing methods, including the reduction process (Section~\ref{sec:reduction}), modelling of the galaxy image (Section~\ref{sec:ISO}), and the multi-component decompositions (Section~\ref{sec:decomposition}). 
Based on the result of the decomposition, we explore several important aspect about the embedded spheroids such as: the spheroid colour (Section~\ref{sec:sph_colour}), mass (Section~\ref{sec:sph_mass}) and size (Section~\ref{sec:sph_size}).
Additionally, we reclassify the morphology for each galaxy based on our multi-component decompositions (Section~\ref{sec:morph_reclass}).

\subsection{Data reduction}
\label{sec:reduction}
We obtained cutout images of the galaxies from the SDSS DR12 Science Archive Server (SAS)\footnote{\url{https://dr12.sdss.org/bulkFields},\\ \url{https://dr12.sdss.org/sas/dr12/boss/photoObj/frames/301/}}. 
Each frame spans 2048 pixels $\times$ 1489 pixels ($819\farcs2\times595\farcs6$).
The flux originates from multiple sources, including the Poissonian background noise plus the signal coming from foreground stars and our target galaxy.
Naturally, the distribution of the pixel intensity values in frames with a sufficient field-of-view to capture the sky background will be a combination of a symmetrical Poissonian bell curve (random noise) centred on a low photon count, plus pixels with target-galaxy light stretching to brighter intensities.  
We define our sky value as the median of the Poissonian distribution and subsequently subtract it from the frame \citep{Almoznino1993,Davis2019,Sahu2019A}.

The use of wide-frame cutouts resolves the reported SDSS sky subtraction problem in the SDSS DR8 \citep{West2005PhDT, Blanton2005, Bernardi2007, Lauer2007, Hyde2009, West2010, Blanton2011}. 
The SDSS standard photometric pipeline \citep{Lupton2001} estimates the background from a $ 100\arcsec \times 100\arcsec$ frame and subtracts it from the image.
In the case of large extended local galaxies, such an approach will result in over-subtraction because the frame is not wide enough to reach beyond the galaxy.
Since our frame is significantly bigger than the target galaxy, the sky subtraction issue is under control.

After the sky subtraction, the images proceed to a masking procedure.
It is important to mask out the excess light from foreground stars and nearby galaxies because they can prevent the fitting algorithm (Section~\ref{sec:ISO}) from creating an accurate model for the target galaxy.
We divided the masking process into two stages.
\begin{enumerate}
    \item 
First, we exclude the majority of the bright sources using the \texttt{SExtractor} \citep{Bertin1996} automatic process.
The programme identified the contamination by threshold sigma detection.
While it is good at 
tracking isolated sources, \texttt{SExtractor} is not as effective in deblenidng overlapping objects, in particular, foreground stars in front of a galaxy along the line-of-sight.
We handle this issue through the second stage: manual masking.
    \item 
Additional foreground objects are identified by human eyes and concealed using the polygon tool from \texttt{DS9} \citep{2003ASPC..295..489J}.
We masked the bright sources, as well as the dust lanes that influence the stellar light.
\end{enumerate}
We construct the mask using the \texttt{IRAF} function \texttt{mskregion}.
Combining both the automatic and manual masking, all the irrelevant objects are effectively removed from the frame.

\subsection{Isophotal fitting}
\label{sec:ISO}
The galaxy images are modelled with the isophotal fitting programme \texttt{ISOFIT}  \citep{Ciambur2015}.
The method was first described by \citet{1978MNRAS.182..797C, Carter1987} and implemented by \citet{Jedrzejewski1987}.  It has since been a staple technique in modelling early-type galaxies, most notably through the \texttt{IRAF} package \texttt{ellipse}.
The algorithm fits a series of nested isophotes centred around the peak brightness region.
It performs a least-square fit on each isophote with intensity, $I(\theta)$, expressed by the Fourier series:
\begin{equation} 
\label{eq:Iso1}
I(\theta)=\left\langle I_{e l l}\right\rangle+\sum_{n}\left[A_{n} \sin (n \theta)+B_{n} \cos (n \theta)\right]
\end{equation}
where \(\left\langle I_{e l l}\right\rangle\) is the average intensity, $A_n$ and $B_n$ are the harmonic coefficients of $n^\mathrm{th}$ order, and $\theta$ is the azimuthal angle.
\texttt{ISOFIT} made two significant changes to the \citet{Jedrzejewski1987} algorithm.
It replaces the azimuthal angle ($\theta$) with the eccentric anomaly ($\psi$), with the following relation:
\begin{equation} 
\label{eq:Iso2}
\psi=-\arctan \left(\frac{\tan \theta}{1-\epsilon}\right), 
\end{equation}
where $\epsilon$ is the varying ellipticity \(\epsilon \equiv 1-\left(b/a\right)\) of each nested isophote, and $a$ and $b$ are the lengths of the semi-major and semi-minor axes, respectively.

The incorporation of the ellipticity information in the eccentric anomaly gives us a better description of the shape of each isophote, in particular those that have high $\epsilon$.
It substantially improves the modelling of edge-on disc galaxies and those with elongated bars. Given the dramatic improvement, \texttt{ISOFIT} additionally allows more higher harmonic coefficients, with amplitudes $A_n$ and $B_n$ typically asymptoting to zero by $n\lesssim12$.
The high-order Fourier coefficients capture the radial perturbations of each isophote.
Each integer $n$ represents a specific type of deformation from a perfect ellipse.
For example, $A_4$ and $B_4$ describe the amplitude of a four-cornered deformation of an ellipse, in turn showing the `boxyness' and `discyness' of the isophote.
As illustrated in \citet[][Figure~4]{Ciambur2015}, the high-order ($n\geq 6$) coefficient reveals additional information about a galaxy that is rarely studied, i.e., the $B_6$ profile captures (peanut shell)-shaped structures nestled around bars \citep{2016MNRAS.459.1276C, 2017MNRAS.471.3988C, 2018ApJ...852..133S}.

\begin{figure}
\centering
	\includegraphics[clip=true, trim= 3mm 3mm 2mm 2mm, width=\columnwidth]{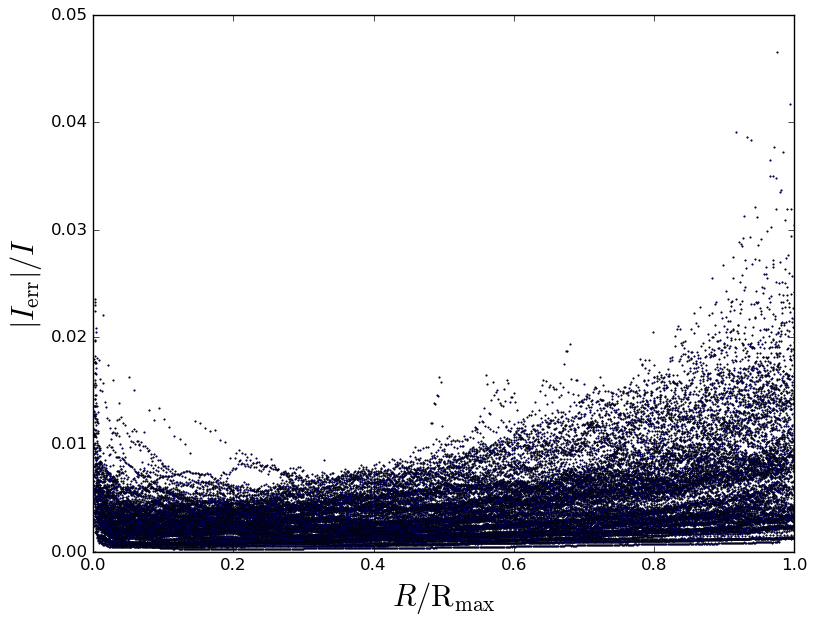}
    \caption{(Model error)-to-intensity ratio along the normalised radius of 103 galaxies.  
    The ($|I_\mathrm{err}/I|$) ratio is defined by $I_\mathrm{err} = \sigma/\sqrt{N}$, where $\sigma$ is the standard deviation and $N$ is the number of pixels comprising the isophotal ring at a given radius.
    The normalized radius ($R / R_\mathrm{max}$) is defined by the maximum radius $R_\mathrm{max}$ to which we modelled each galaxy. 
    We display the average ratio with black points. 
    }
    \label{fig:rad_err}
\end{figure}

The 2D intensity map described by the \texttt{ISOFIT} task is comprised of a series of 1-D radial distributions, capturing the surface brightness, ellipticity, position angle, and Fourier harmonic terms. 
The ratio of the model error \(I_{\rm err} = \sigma / \sqrt{N}\) to intensity ($I$) at each isophote is shown in Figure~\ref{fig:rad_err}, where $\sigma$ is the standard deviation of the pixel values (data) about the two-dimensional reconstruction of the image created with the IRAF task \texttt{Cmodel} \citep{Ciambur2015} and $N$ is the number of pixels in each isophote. 
$R_\mathrm{max}$ is the cutoff radius to which we modelled each galaxy\footnote{For further information on the galaxy size, see Appendix~\ref{sec:Gal_can}.}.
While the model error-to-intensity ratio increases with radius, they do not deviate much from $I_{\rm err}/I = 0$, with most points having $|I_{\rm err}|/I <0.02$.
The error from the isophotal fitting is small and the surface brightness information is well captured by the fitting process.

It is common in the community to use two-dimensional (2-D) fitting algorithms such as \texttt{GIM2D} \citep{Simard1998}, \texttt{GALFIT} \citep{2002AJ....124..266P, Peng2010}, \texttt{BUDDA} \citep{2004ApJS..153..411D} and \texttt{IMFIT} \citep{Erwin2015}, to fit  2-dimensional analytic functions directly onto the image.
While there are pros and cons concerning whether 1-D profile or 2-D image analysis is more adequate \citep[e.g.][]{Ciambur2016}, we favour the use of the above-mentioned 1-D radial brightness profiles. 
As shown in \citet{1990MNRAS.245..582V} and \citet{Lasker2014}, some triaxial spheroids exhibit twisting in their position angle ($PA$) and ellipticity $\epsilon$ profile.
The aforementioned 2-D codes can not account for the shape changes because they assume a fixed $PA$ and $\epsilon$ for each galaxy component. 
Furthermore, \citet{Savorgnan2016A} conducted an empirical comparison between 1-D and 2-D methods on the same galaxies and remarked on some key advantages of the 1-D approach.
With the 1-D radial profiles, weak features such as embedded discs and small bars were more visible.
The changes in the $PA$, $\epsilon$, and harmonic coefficient profiles can signify the presence of these components.
They also found that the 2-D routine has a substantially harder time converging to a meaningful solution.  We, therefore, adopted the more manually-intensive, but also more reliable, 1-D approach over the automated 2-D approach.

\begin{figure*}
\centering
	\includegraphics[clip=true, trim= 12mm 6mm 5mm 6mm, width=\textwidth]{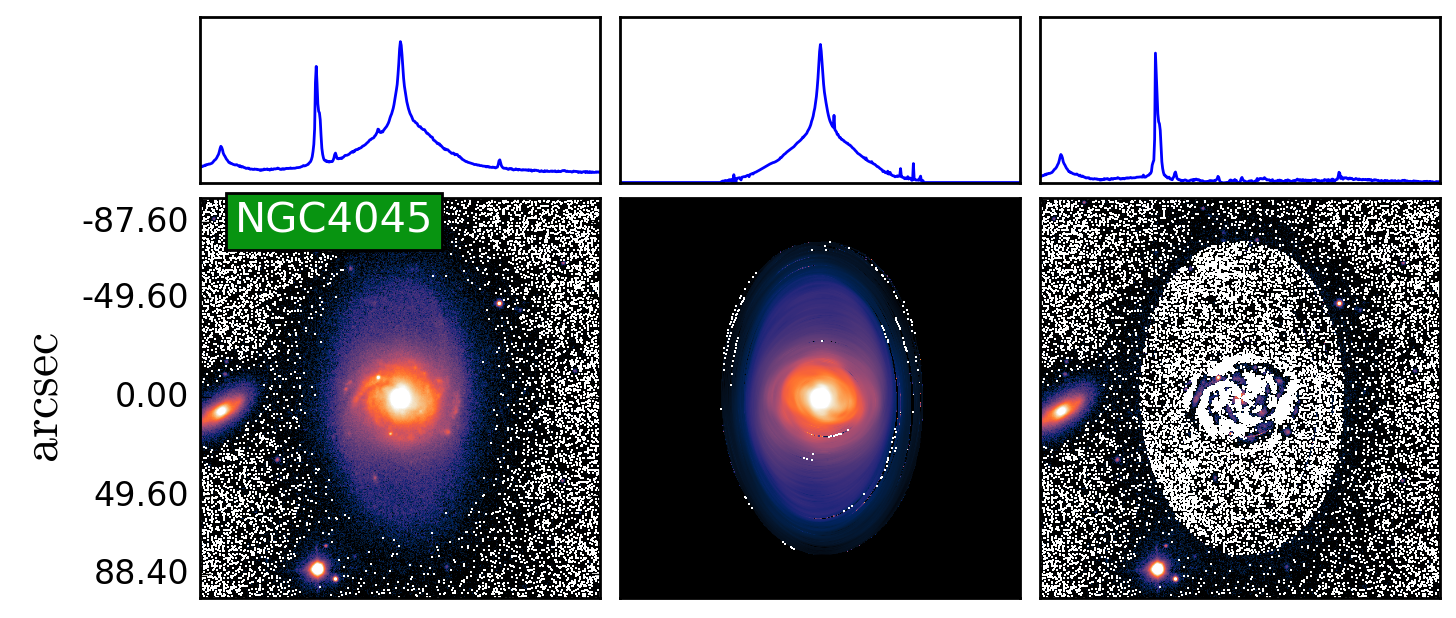}
    \caption{NGC~4045, an example of an early-type spiral galaxy.
    \underline{Bottom row}: The galaxy images, the pixel value are shown in the same contrast setting across the three panels.
    \underline{Upper row}: The average pixel value along the y-axis of the image.
    \underline{Left column}: The original image.
    \underline{Middle column}: The 2-D isophote model created by {\sc Cmodel}.
    \underline{Right column}: the residual image.
    The images are orientated with their right side pointing north and up side pointing east.}
    \label{fig:ISOFIT1}
\end{figure*}

A few of the galaxies from our three bins had to be excluded from the analysis due to various reasons:
\begin{enumerate}
    \item 
edge-on galaxies with substantial dust covering the centre of the galaxy (see Section \ref{sec:ISO});
    \item 
galaxy pairs where the gravitational interaction significantly distorts the structure of both galaxies; or
    \item 
unique cases for which the fitting programme failed to produce a convincing 2-D model.
\end{enumerate}
Ultimately, we have reliable data for 27, 24, and 52 galaxies in Bins 1, 2, and 3, highlighted as red, blue, and black in Figure~\ref{fig:data_selection}, respectively.
12 galaxies are rejected in our analysis.

We have listed the perceived Hubble types by their \citep[][hereafter RC3]{de_Vaucouleurs1991rc3} classifications in Column~(8) of Tables~\ref{tab:sample1}--\ref{tab:sample3}.
For each bin, they are Bin 1 (E: 13, S0: 6, and S: 8), Bin 2 (E: 9, S0: 7, and S: 8), and Bin 3 (E: 6, S0: 18, and S: 28).

\subsection{Multi-component decompositions}
\label{sec:decomposition}

In this section, we present the justification and describe the decomposition process.
The projected radial intensity profile $I(R)$ of a spheroid or bulge can be described by the 
\citet{Sersic1968} function, expressed in \citet{Graham2005PASA} as: 
\begin{equation} 
\label{eq:prof1}
I(R)=I_{\rm e} \exp \left\{-b_{n}\left[\left(\frac{R}{R_{\rm e}}\right)^{\frac{1}{n}}-1\right]\right\}, 
\end{equation}
where $R_{\rm e}$ is the effective `half-light' radius, $n$ is the S\'{e}rsic index,  $I_\mathrm{e}$ is the surface brightness at $R_\mathrm{e}$, and $b_n$ can be obtained by solving  \(\Gamma(2 n)=2 \gamma\left(2 n, b_{n}\right)\), involving the complete and incomplete Gamma functions.
The surface brightness profile is calculated by substituting $I(R)$ into the following equation:
\begin{equation} 
\label{eq:mag}
\mu(R)=zp-2.5 \log _{10}\left[\frac{I(R)}{ps^{2}}\right]
\end{equation}
where $zp$ is the zero-point magnitude and $ps$ is the pixel angular size.

Many conventional studies perform Bulge$+$Disc  decompositions involving a S\'{e}rsic $R^{(1/n)}$-bulge plus an exponential-disc \citep[e.g.,][etc.]{Andredakis1995,Seigar1998,Iodice1999,Khosroshahi2000,D'Onofrio2001,Graham2001A,Mollenhoff2011,Simard2002,Allen2006,Simard2011}.
While a gravitationally bound system can be broken into rotational (disc) and triaxial, pressure-supported components (bulge), in the case of a galaxy, there is an extra layer of complication.

Additional substructures, such as \textit{ansae} and rings, are common. 
They have distinct features and stellar orbits from a bulge or a disc.
Furthermore, spheroids and discs can have a more complex structure than the S\'{e}rsic and exponential function permit. 
For instance, massive galaxies often have a depleted core \citep{King1978, Trujillo2004}, where the light in the centre is fainter than what a S\'{e}rsic function predicts.

Bulges/spheroids are also known to rotate.
Some rotate along their short axes \citep{Binney1978, Miller1978}.
There are evidences where spheroid rotate around both major \citep{Davies1986,Davies1988,Franx1989a,Jedrzejewski1989} and minor \citep{Schechter1979, Efstathiou1980, Davies1983, Bender1988, Bender1990, Bender1994} axes.

Discs also often exhibit bending in their outer region \citep{van_der_Kruit1987, Pohlen2004, Erwin2005}, classified by the different behaviour as either Type~I (no bend), II (a downward-bend), and III (a temporary upward-bend).
Moreover, substructures such as nuclear discs, intermediate-scale discs, and nuclear stellar clusters also exist.
They can contribute an indelible amount of stellar light to the galaxy.

For years, these extra features have motivated the use of multi-component decompositions for well-resolved galaxies  \citep[e.g.,][]{Martin1995,Prieto1997,Aguerri1998,Graham2003,Laurikainen2005,Laurikainen2010,Vika2012,Lasker2014,Savorgnan2016A,Davis2019,Sahu2019A}.
Indeed, \citet{Stone2021} concluded `that two-component fits (e.g., S\'{e}rsic plus exponential) are insufficient to describe late-type galaxies with high fidelity'.

In addition, one need to be cautious while modelling with S\'{e}rsic function.
A part of its usefulness comes from the function's flexibility.
It is able to fit a range of light profile shapes. 
The flexible nature, however, also presents a problem.
If one ignores a prominent substructure, such as a bar or an outer ring, the fitting code will attempt to compensate by bending the S\'{e}rsic function, intended to describe the bulge, upward at the tail end, leading to a higher S\'{e}rsic index $n$ and radius $R_{\rm e}$ than is correct for the bulge. 
Therefore, in a counter-intuitive sense, the most accurate way to model a spheroid is to not solely focus on the S\'{e}rsic component but also account for potentially biasing substructures.
With that in mind, we proceed to model galaxies' structures.

\begin{table}
   \caption{\texttt{Profiler} functions and their corresponding galactic structures.}
   \begin{threeparttable}[b]
   \label{tab:structure}
   \begin{tabular}{ll}
   
   \hline
   Analytic Functions & Galactic Structures\\
   \hline
   S{\'e}rsic$^\mathrm{a}$ & Spheroid/Bulge or Lens\\  
   Core-S{\'e}rsic$^\mathrm{b}$ & Core-depleted Spheroid\\
   Exponential$^\mathrm{c}$ & Type I Disc\\
   Broken Exponential$^\mathrm{d}$ & Type II/III Disc\\
   Edge-on Disc model$^\mathrm{e}$ & Inclined Disc\\
   Ferrers$^\mathrm{f}$  & Bars\\
   Gaussian  & Excess light, spiral arms, rings, and \textit{ansae}\\
   \hline
   \end{tabular}
    (a) \citet{Sersic1968}; (b) \citet{Graham2003coreSersic}; (c) \citet{1940BHarO.914....9P, 1959ApJ...130..728D}; (d) \citet{2001MNRAS.324.1074D, 2002A&A...392..807P}; (e) \citet{van_der_Kruit1981}; (f) \citet{Ferrers1877,Sellwood1993Bar}
   \begin{tablenotes}
      \small
      \item 
      \begin{flushleft}
      \end{flushleft}
    \end{tablenotes}
    \end{threeparttable}
\end{table}

We use the programme \texttt{Profiler} \citep{Ciambur2016} to perform the decomposition.
The analytic functions in \texttt{Profiler} that were used to depict the different galactic structures are listed in Table~\ref{tab:structure}.
We start our fitting using \texttt{ISOFIT}'s major axis surface brightness profile, 
where the galaxy components often appear more prominent than in comparison with the geometric-mean axis\footnote{
The geometric mean of each isophotes' major- and minor-axis, equivalent to a circularised version of the galaxy.}.
The key to the decomposition is to identify deviations in the various 1-D profiles for each galaxy, such as their surface brightness, position angle, and ellipticity profiles. 
Each component that we include in the fitting process has a corresponding physical substructure that is visible and identified in either the 2-D image or (more readily) the 1-D profiles.
We additionally spent some time scouring the literature for additional evidence to justify the inclusion of components.  This typically included confirmation of bars already detected by others or rotating discs seen in kinematic data.

\begin{figure}
\centering
\includegraphics[clip=true, trim= 6mm 0mm 5mm 6mm, width=\columnwidth]{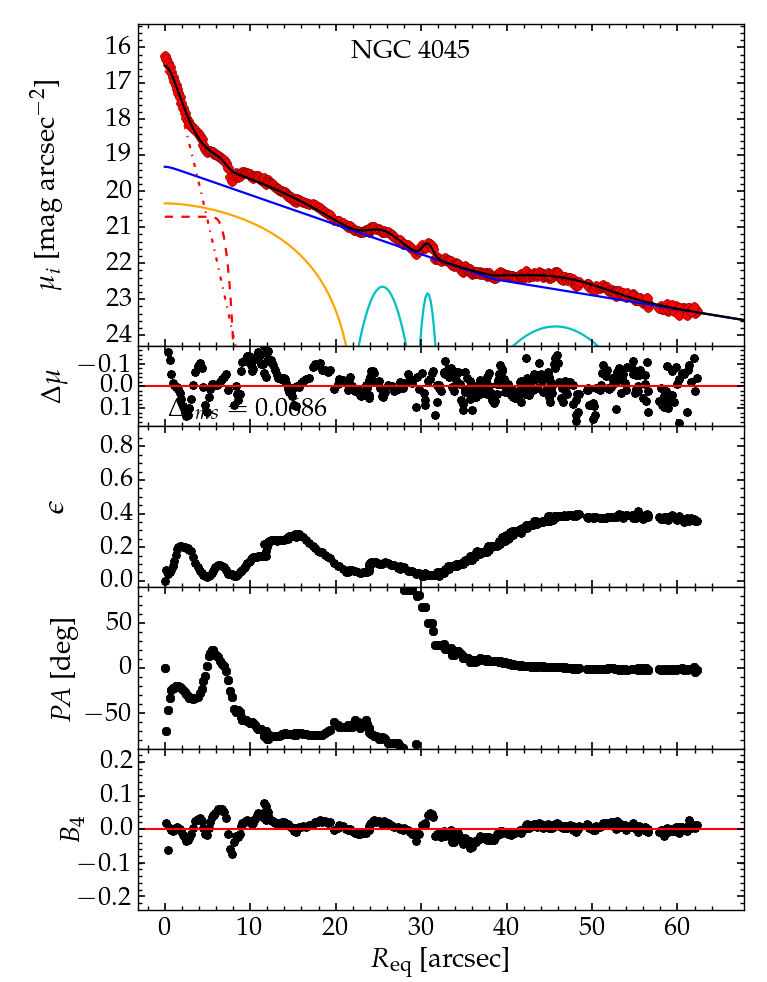}
    \caption{
    NGC~4045 `equivalent axis' surface brightness profile, with $R_{\rm eq}$ equivalent to the geometric mean of the major (a) and minor (b) axis: $R_{\rm eq} = \sqrt{ab}$.
    From top to bottom panel: $i$-band surface brightness ($\mu$); (model $-$ data) residual ($\Delta \mu$); ellpticity ($\epsilon$); Position Angle ($PA$); and 4$^\mathrm{th}$-order Fourier harmonic coefficient ($B_4$).
    The galaxy components are the Bulge (dashed-dotted \textcolor{red}{red} line), 
    nuclear lens (dashed \textcolor{red}{red} line) , Bar (\textcolor{orange}{orange}), Type III truncated disc (\textcolor{blue}{blue}), plus a ansae, a faint inner, and outer spiral arm (\textcolor{cyan}{cyan}).
    }
    \label{fig:profiler1}
\end{figure}

To illustrate the process, here we present a case study using the late-type galaxy NGC~4045.
In Figure~\ref{fig:ISOFIT1}, we present the \texttt{ISOFIT} modelling result.
In the bottom row, from left to right, they are the image of NGC~4045, the model produced by the  \texttt{ISOFIT} model, and the residual by subtracting the image by the model.
The top row illustrates a 1-D representation of the image, model and residual, by averaging the pixel value along the vertical direction. 
One can see the \texttt{ISOFIT} model has done a marvellous job capturing the surface brightness profile of the galaxy.

\begin{figure}
	\includegraphics[clip=true, trim=3mm 2mm 2mm 3mm, width=\columnwidth]{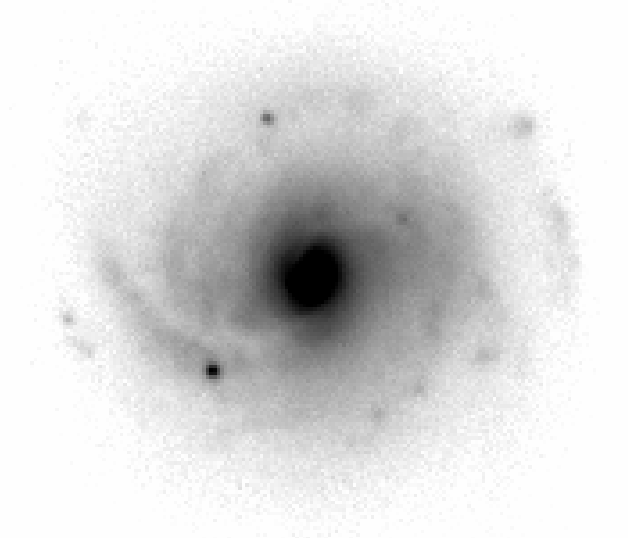}
    \caption{A zoomed-in image for the inner part ($\sim20~\arcsec$) of NGC 4045 with the same orientation as Figure~\ref{fig:ISOFIT1}. One can see a four armed `pinwheel' shaped pattern, aka the nuclear lens shown as dashed red line in Figure~\ref{fig:profiler1}. The two bright fringes of dust lanes spiraling towards the centre obscure some of the light and decreases the ellipticity at $\sim 7~\arcsec$, making the isophote appears more circular than it should be.}
    \label{fig:zoomin}
\end{figure}

Figure~\ref{fig:profiler1} shows the decomposition for NGC~4045 along its geometric-mean axis, a.k.a.\ equivalent axis, $R_{\rm eq}$.
The galaxy has a small bulge (dashed-dotted \textcolor{red}{red} curve), a large extended type III truncated disc (\textcolor{blue}{blue} curve), and a bar \citep[\textcolor{orange}{orange} curve, see][]{Erwin2005bar, Erwin2013, Diaz-Garcia2016, Font2017, Font2019}.  
Along the $\epsilon$ and $PA$ profiles, there are two `bumps' at $R_{\rm eq} \sim 3\arcsec$ and $7\arcsec$, corresponding to the inner spiral arm-like structures.
\citet[][]{Comeron2014} classified NGC 4045 as a $\rm (R_{1}'L)SAB(rs,nl)ab$ galaxy containing two outer rings and a `nuclear lens'\footnote{
The `nuclear lens' appears to be the head of the spiral arms that run perpendicular to the so-called `bar' in this galaxy.}.
Upon closer inspection (see Figure~\ref{fig:zoomin}), there are two dust lanes obscuring the light at $\sim 7$-$8~\arcsec$ \citep[see also][]{Erwin2002}, resulting in a dip in $\epsilon$, making the isophote appear more circular. We modelled the nuclear lens with a S\'ersic function (dashed \textcolor{red}{red} line) with a very low S\'ersic index ($n\sim0.01$).
The strong position angle ($PA$) twist at $R_{\rm eq} \sim 27\arcsec$ coincides with the small bump in $\epsilon$.
This feature associated with the bar and elevation in $\mu$ is an `\textit{ansae}', where the bar transitions into the disc rotation. 
Finally, the outermost plateau of $\epsilon \sim 0.4$ beyond $R_{\rm eq} \sim 40\arcsec$ is the projection of the inclined, rotationally-supported disc \citep[see][]{Erwin2005, Erwin2008}.
We use an type III truncated exponential model for this disc, plus a Gaussian function to account for the brightening in $\mu$ at $R_{\rm eq} \sim 43\arcsec$, which thereby captures the weak spiral arm within this disc.
Failing to account for the elongated bar structure in NGC~4045
would result in the fitted S\'ersic function substantially over-estimating the mass and size of the spheroidal component of this galaxy.
As shown in Column~(9) of Table~\ref{tab:sample1}, our sample has a range of morphologies.

Our decomposition experience also reveals that the outer components can significantly affect the estimation of a bulge.  That is, 
if a prominent structure in the outer region of a galaxy, such as a strong spiral arm or disc truncation, is not considered, then the parameters of the disc will be distorted.
Like a domino effect, an incorrect disc model impacts upon the S\'{e}rsic model at the centre, changing its parameters to compensate for the ill-modelled disc. 

We analysed 103 galaxies with the same individual level of attention to detail, following that done by \citet{Davis2019} and \citet{Sahu2019A} in their decomposition of late- and early-type galaxies with directly measured supermassive black hole masses. 
Tables~\ref{tab:result1}--\ref{tab:result3} provide the best-fitting, equivalent-axis, S\'ersic parameters for the 103 spheroids.
This axis provides circularised component sizes which enable \texttt{Profiler} to integrate each component's model surface brightness profile to obtain the associated luminosity.
The resulting spheroid apparent magnitudes are given in Column~(7) of the tables.
From there, we calculated the stellar mass based on the four MLCRs from equation array~\ref{eq:MLCRs}.  These are listed in Columns~(9)--(12).

\subsection{Morphological reclassifications}
\label{sec:morph_reclass}

We are able to reclassify the morphology of the host galaxies based on our decomposition of their light.
The previous morphologies were determined by visual inspection of photographic images \citep{Dressler1980,Binggeli1987,Sandage1985,de_Vaucouleurs1991rc3}, which can drown out features and limit the accuracy of the past classifications.
For instance, a faint disc in an early-type galaxy can be hard or impossible to see in images at certain fixed contrasts.
In the case of a face-on disc, from a photograph, it is very difficult to distinguish a barless lenticular galaxy from an elliptical galaxy.  For this reason, in our search for compact massive spheroids, we have not restricted the initial galaxy sample to galaxies classified as either lenticular or spiral.

Many, and no doubt most, galaxies classified as elliptical are either lenticular or ellicular galaxies with discs and bulges. 
Analysing the surface brightness profiles is much more reliable for identifying host galaxy substructures due to the bumps and bends in the profiles.
Our decompositions, therefore, provide a more complete picture of the host galaxies than past visual inspections.
The new information will be particularly valuable to understand the morphological evolution of galaxies.

Among the 28 initially elliptical galaxies (E, according to the RC3), 64\% (18/28) of them are relabelled as lenticular galaxies (S0).
 Two of the original S0s are now classified as ellicular galaxies (ES): NGC~3805 and NGC~5382.

\begin{figure*}
\centering
	\includegraphics[clip=true, trim= 4mm 2mm 8mm 2mm, width=\textwidth]{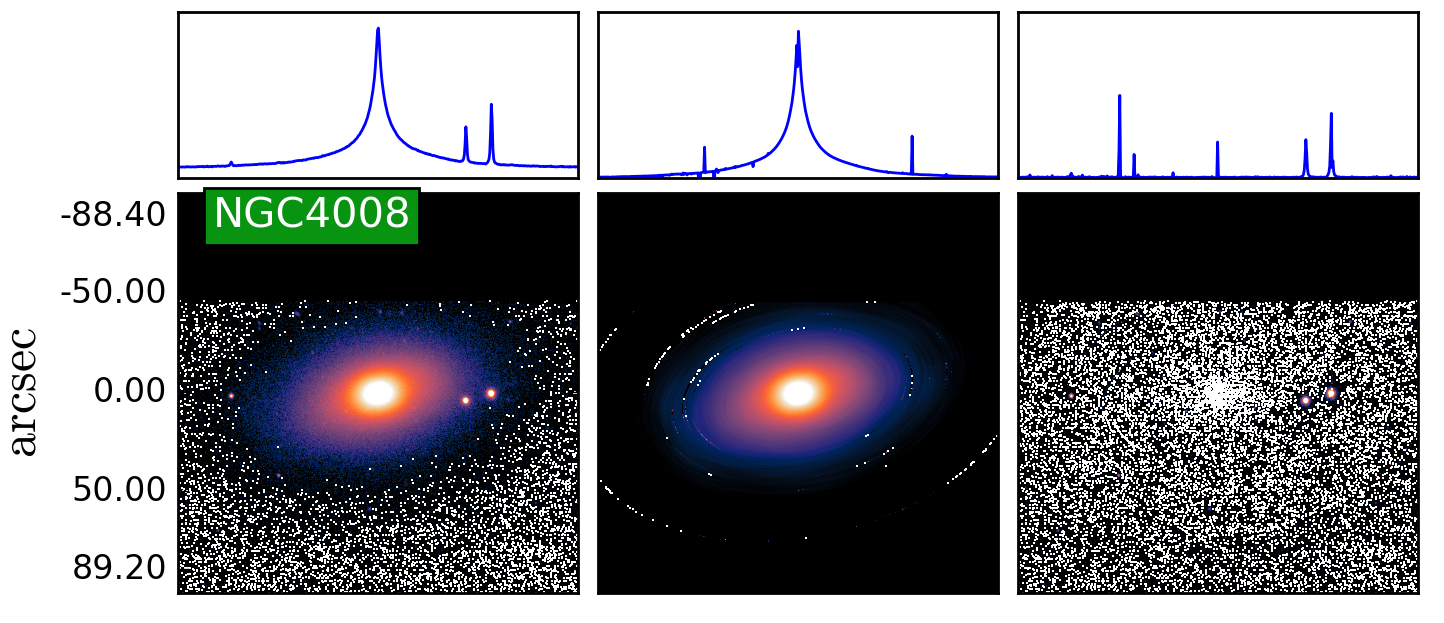}
    \caption{
    NGC~4008, an example of an ellipitcal (E5) galaxy according to RC3. 
    See also Figure~\ref{fig:ISOFIT1}.
    We reclassified it as a lenticular (SA0) galaxy based on the decomposition in Figure~\ref{fig:NGC4008profile}.}
    \label{fig:NGC4008img}
\end{figure*}

\begin{figure}
\centering
\includegraphics[clip=true, trim= 6mm 0mm 5mm 6mm, width=\columnwidth]{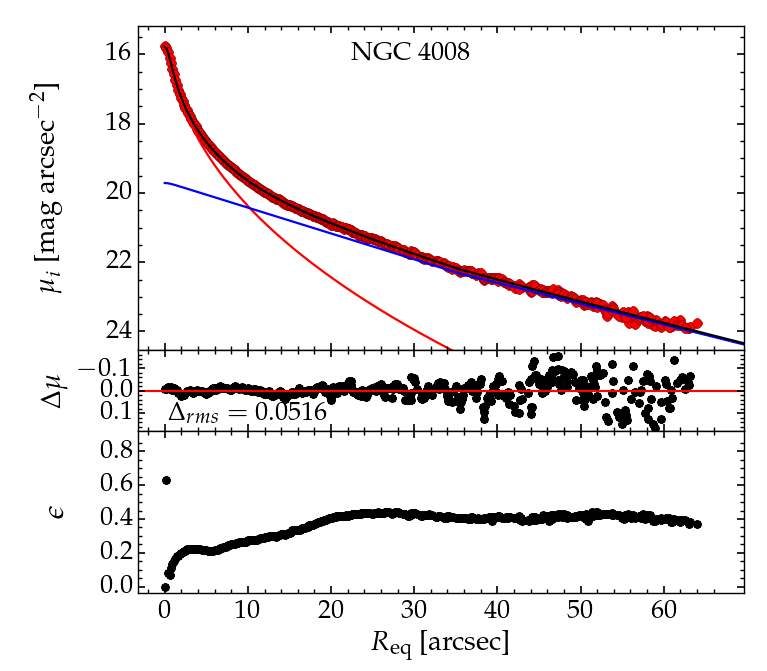}
    \caption{
    NGC~4008's surface brightness profile in `equivalent axis' (circularised radius). See also the general description in Figure~\ref{fig:profiler1} for reference. The profile is fitted with a bulge (red line in the top panel) and a disc (blue line in the top panel).}
    \label{fig:NGC4008profile}
\end{figure}

We present one such `elliptical galaxy', NGC~4008, to
illustrate why we reclassified it as an S0 galaxy instead of an E galaxy.
NGC~4008 was classified as an `E5' in RC3.
Figure~\ref{fig:NGC4008img} shows the galaxy, our model, and our residual image of NGC~4008.
The galaxy has a central stellar velocity dispersion of $\sigma_0\sim 208$--$240\rm\,km\,s^{-1}$ \citep{Faber1989,Simien2002}.
In its ellipticity profile ($\epsilon$, middle panel in Figure~\ref{fig:NGC4008profile}), one can see that the ellipticity reaches $\epsilon\sim0.4$, and the surface brightness profile ($\mu$, top panel in Figure~\ref{fig:NGC4008profile}) has a smooth exponential decline from $R_\mathrm{eq} \sim 20\arcsec$.
Although the $PA$ and $B_4$ seem to be featureless, the galaxy has clear signs of an extended disc.
\citet{Simien2002} found the stellar velocity to be $v = 123~\rm\,km\,s^{-1}$ at $R = 10\arcsec$ (in major axis) and appear to keep rising beyond this point.
We first tried to use a S\'{e}rsic bulge plus an exponential disc to model the galaxy, but we found that it can be better described with a bulge (top panel in Figure~\ref{fig:NGC4008profile}, red line) and a Type~III exponential disc with a slight slope difference separated at  $R_\mathrm{eq} = 35\arcsec$ (top panel in Figure~\ref{fig:NGC4008profile}, blue line).
The final fit is an excellent agreement with the isophotal model and has a root mean square residual $\Delta_{\rm rms} = 0.0516$\,mag\,arcsec$^{-2}$ (see the middle panel in Figure~\ref{fig:NGC4008img}). 
This decomposition also aligns with \citet{Simien2002}, who show that at $R_\mathrm{eq} > 10\arcsec$, NGC~4008 transitions from bulge-dominant to disc-dominant.
The presence of an extended disc means NGC~4008 is not an `E5' but an `SA0' galaxy.
There are 17 other galaxies similar to NGC~4008 where we reclassified them as S0.

Two other galaxies are worthy of note.
NGC~4459 was the only galaxy originally labelled as an S0 prior to our changing it to an E2 designation.
It contains a nuclear disc in the centre, as reported by \citet{Peterson1978} and \citet{Krajnovic2011}.
\citet{Gutierrez2011} has suggested that it contains a Type~III anti-truncated disc.
\citet{Savorgnan2016A} broke down the galaxy into a S\'{e}rsic bulge, a Gaussian nucleus (for the nuclear disc), and an exponential disc.
However, the neighbouring disc galaxy is likely the main contribution to the light in the `extended disc' at large radii.
As we limit the fitting range to within $77\arcsec$ along the major axis, a nuclear exponential disc and a S\'{e}rsic bulge are found to suffice.
Additionally, we sustain the E classification for NGC~2872, but it is noteworthy that it contains a dusty nuclear disc \citep[as shown in][]{Tran2001}.

In summary, only eight elliptical galaxies (NGC~2832, NGC~3615, NGC~3812, NGC~4073, NGC~4261, NGC~4555, NGC~4889, and NGC~5444) in our sample are devoid of (detectable) discs, while three of them have nuclear discs (NGC~2872, NGC~4459, and NGC~4494) and two have intermediate-scale discs (NGC~3805 and NGC~5382).
The abundance of pure elliptical galaxies is much lower than the morphologies previously indicated.
In total, it leaves five E galaxies in Bin~1, three in Bin~2, and three in Bin~3.
True elliptical (E+ES) galaxies occupy only $\sim$13\% (13/103) of our sample\footnote{Among the true ellipticals, there are six cD galaxies: NGC~2832, NGC~4073, NGC~4874, NGC~4889, NGC~4914, and NGC~5444. NGC~4914 is relabelled as a SAB0 galaxy, for its weak bar \citep{Buta2010} and an extended disc with $\epsilon\sim0.4$. Informed by the three components model by \citet{Dullo2019}, NGC~4874 is now labelled as SA0 (see later in Section~\ref{sec:B/T_ratio}).}, and they are only $\sim$46\% (13/28) old E class objects. 
This equates to an E+ES galaxy number density in Bins~1, 2, and 3 of $1.84\times 10^{-5} \rm \, M p c^{-3}$, $3.39 \times 10^{-5} \rm \, M p c^{-3}$, and $9.20 \times 10^{-5} \rm \, M p c^{-3}$, respectively.

This result is comparable to the $\rm ATLAS^{3D}$ kinematic analysis of 260 nearby early-type galaxies \citep{Cappellari2011a, Krajnovic2011, Emsellem2011, Cappellari2011b, Krajnovic2013a}.
Studying a volume-limited sample within $<$40\,Mpc, \citet{Emsellem2011} reveal only $\sim$14\% (36/260) of their early type galaxies are `slow-rotators', defined by a luminosity-weighted specific angular momentum $\lambda_\mathrm{R} \leq 0.1$ \citep[see][]{Emsellem2007}.
They found that one-third ($\sim$34\%) of the previously-classified elliptical galaxies are slow rotators, and $\sim$47\% (17/36) of them contain a kinematically-distinct core (KDC).  
We share a mutual sentiment that reclassification is needed for many early-type galaxies. 
Indeed, \citet{Cappellari2011b} advocated the kinematic classification scheme from \citet{1988A&A...193L...7B} and \citet{1992ASSL..178...99C} to correct for the misunderstanding caused by the old visual-based classification.  Although, as illustrated in \citet{2017MNRAS.470.1321B}, this kinematic scheme fails to accommodate the ES galaxies which are either fast or slow rotators depending on the radial extent of one's observations, it does help identify which early-type galaxies have discs and therefore require a decomposition to measure the size and mass of the spheroidal component.

\subsection{Spheroid colours}
\label{sec:sph_colour}

Local early-type galaxies typically, although not universally, exhibit a negative colour gradient such that they are redder in their centre than their outskirts \citep[e.g.][]{Franx1989b, Peletier1990, Gargiulo2012, Peletier2012, Marian2018}. 
In the absence of substantial dust and gas, 
a negative gradient suggests younger (more blue) stellar population in the outer regions, which can be a sign of inside-out disc growth. 
However, from galaxy colour profiles, 
it can be problematic to readily establish the colours of the individual overlapping structural components. 
For instance, a young nuclear disc can make a galaxy core colour blue and result in a more positive galaxy gradient than compared to the surrounding red quiescent spheroid \citep[e.g.][]{Graham2017}.

While we would like individual $g-i$ colours for the bulges, or better yet, the spectral energy distributions of the bulges for use in software like MAGPHYS \citep{da_Cunha2012} to obtain the stellar mass, we make use of the galaxy colours, recognising that there is not a big difference between bulge and disc colours of early-type galaxies and early-type disk galaxies \citep[e.g.,][]{Peletier1996}.  Our galaxies have a mean $g-i$ colour of $\sim$1.2, in accord with the value reported in \citet{Fukugita1996} for what they thought were elliptical galaxies, but undoubtedly included many S0 galaxies. 

In general, the bulges of galaxies are redder than their discs \citep{1990ApJ...350...73B, 1999MNRAS.310..703P}. 
For bright red galaxies, \citet[][see their Figures~10--13]{2016MNRAS.460.3458K} find $u-r$ bulge colours $\sim$0.5\,mag redder than the disc colours, although they remind the reader that they performed a bulge-disc decomposition of every galaxy regardless of whether there was a physical need for two components. 
As such, their colour difference will also capture metallicty gradients in disk-less elliptical galaxies. 
However, for some of our galaxies with bulges, their galaxy $g-i$ colour might be bluer (a smaller value) than the bulge colour.

Our use of the galaxy colour in the MLCRs to derive the stellar mass of the bulge might, therefore, underestimate the stellar mass of the bulges.
Although, \citet[][see their Figure~10 and Table~2]{2014MNRAS.444.3603V} report that E-S0-Sa galaxies tend to have bulges and discs with a similar $(g-i)$ colour \citep[see also][]{1996AJ....111.2238P}.
Using morphologically-classified galaxies, they find it is the Sb and later spiral galaxies that have notably bluer discs than their bulges, with average $(g-i)$ disc colours that are 0.3\,mag bluer than their bulges. 
Throughout this work, we use the \textit{galaxy} colour as a proxy for the \textit{bulge/spheroid} colour.
The effect of underestimating how red any bulge is will be to underestimate the $M_*/L$ ratio assigned to that bulge and thus underestimate the stellar mass of that bulge.
That is, the number density of local, massive, compact spheroids will not be erroneously overestimated due to this issue.

\begin{figure}
	\includegraphics[clip=true, trim= 1mm 1mm 2mm 2mm, width=\columnwidth]{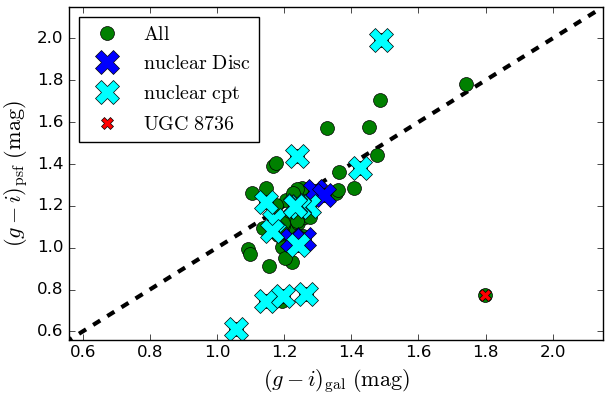}
    \caption{
    The comparison between galaxy colour difference $(g-i)_\mathrm{gal}$ and SDSS PSF colour difference$ (g-i)_\mathrm{psf}$.
    The dashed \textbf{black} line is the centre line where $y=x$. 
    The galaxies that contain a nuclear disc are marked with blue crosses.
    Those containing other prominent nuclear components, such as a nuclear ring, secondary bar, or AGN, are marked collectively with cyan crosses, labelled as `nuclear cpt'.
    The particular outlier UGC~8737 is marked with a red cross.}
    \label{fig:psf_compare}
\end{figure}

In an effort to check on this matter, 
in Figure~\ref{fig:psf_compare}, we compare the galaxy $(g-i)_\mathrm{gal}$ colour with the SDSS point-spread function (PSF)  $(g-i)_\mathrm{PSF}$ colour. 
Figure~\ref{fig:psf_compare} reveals how some galaxies with a global $(g-i)_\mathrm{gal} \sim 1.2\pm0.1$\,mag have relatively blue cores.
The PSF magnitude is obtained by fitting a point-spread function to the light source, stars, and galaxies alike \citep[see in][]{Stoughton2002}. 
A PSF is a good description for point sources like stars, but not so for extended sources like galaxies.
The PSF fit on the galaxies should mainly sample the spheroidal part of the galaxies.
PSF size variation is biased to the red.
We highlighted those with unique central components in Figure~\ref{fig:psf_compare}. 
The sample with a nuclear disc is marked with blue crosses. 
Other prominent central components, such as nuclear rings, secondary bar, and AGN are marked with cyan crosses.

All of this simply provides a rough guide for what the error in the spheroid colour may be.
Of the 103 galaxies in our sample, 69 have PSF magnitude available in the NED database. 
The median difference between $(g-i)_\mathrm{gal}$ and 
$(g-i)_\mathrm{PSF}$ is 0.09\,mag and the standard deviation $\pm0.26$\,mag.
Most galaxies reside at $1.0<(g-i)_\mathrm{gal} <1.5$ along the dashed centre line. 
Some of the major outliers can be explained by the influence of prominent central components, as evidenced by the four galaxies at $(g-i)_\mathrm{PSF} <0.9$\,mag and the one with $(g-i)_\mathrm{PSF} >1.8$\,mag.
One outlier deviates from the centre line significantly at $(g-i)_\mathrm{gal} \sim 1.8$\,mag and $(g-i)_\mathrm{PSF} \sim 0.8$\,mag.
This galaxy is UGC~8736 (marked with a red cross in Figure~\ref{fig:psf_compare}).
Its image has severe foreground contamination. 
The galaxy is situated next to a bright star. 
We suspect its PSF colour might not be accurate due to such an influence. 
The galaxy $(g-i)_\mathrm{gal}$ colour provides a reasonable estimation of the spheroid $(g-i)$ for our purpose. 

Given that galaxies do not have $g-i$ colours redder than $\sim$1.5 (e.g., T11), we recognise the five galaxies\footnote{The five outliers are: UGC~8736, NGC~4527 NGC~3718, NGC~2968, and NGC~2894. All of which are spiral galaxies from Bin~3. We marked their spheroids in Figure~\ref{fig:sizemass} with black star signs ($\star$). Three of them contains a low mass spheroids ($M_*/\mathrm{M_{\odot}}< 1\times10^{10}$). Only two lie within a 'compact massive' region. The outliers does not alter the number of compact massive spheroids, nor the shape of the spheroids distribution in any significant way.} in our sample with a $g-i$ colour of $\sim$1.8 as erroneous, perhaps due to dust or a poor single S\'{e}rsic fit to obtain the $g-$ and $i-0$band magnitudes. We reset these galaxies $g-i$ colour to 1.5 so as to not overestimate the stellar mass-to-light ratio used and thus not overestimate the galaxies' stellar mass.
We move on to calculate the stellar masses of the spheroids accordingly.

\subsection{Spheroid masses}
\label{sec:sph_mass}

Multiple high-$z$ red nugget studies \citep{Barro2013,van_der_Wel2014,van_Dokkum2015} use the SED fitting code \texttt{FAST} \citep{Kriek2009} to determine the stellar masses of their high-$z$ galaxies.
These works assume the \citet{Chabrier2003} IMF and \citet[][hereafter BC03]{Bruzual2003} SPS, with a variety of dust extinction treatments.
Among the available MLCRs, only the T11 and RC15 relations are constructed with these priors.
For the sake of simplicity, the result will be presented in  RC15 stellar mass. 
However, to ensure the result is not biased by the choice of stellar mass measurements, the effect of using the different MLCRs will be later discussed in Section~\ref{sec:diff_assumption}.

Our final uncertainty on the stellar masses have been calculated using the following error propagation equation:
\begin{equation} 
\label{eq:mass_error}
\delta \log_{10} M_*=\sqrt{\left(\frac{\delta \mathfrak{m}}{2.5}\right)^{2}+\left(2 \frac{\delta D}{D \ln (10)}\right)^{2}+\left(\frac{\delta \Upsilon_{*}}{\Upsilon_{*} \ln (10)}\right)^{2}},
\end{equation}
where $\delta D$ is our assumed uncertainty in the distance, $\delta \mathfrak{m}$ is our uncertainty in the bulge apparent magnitude, and $\delta \Upsilon_{*}$ is our uncertainty in the $M_*/L$ ratio.  
The uncertainty in the magnitude, $\delta \mathfrak{m}$, is determined by the component modelling process.
The unknown degeneracy amongst components is the primary source of error. 
As explored in \citet{Savorgnan2016A}, adding or subtracting a galaxy component can change the spheroid parameters.
Based on our extensive practical experience with the fitting routine, and informed by the results of previous studies \citep{Savorgnan2016A,Davis2019,Sahu2019A}, we have assigned a probable magnitude error of $\delta \mathfrak{m} = 0.3 \, \rm mag$ for all of our spheroids.
This level of uncertainty encapsulates the average error commonly encountered throughout the combined efforts of this work and its predecessors.
The uncertainty in the mass-to-light ratio is taken to be the intrinsic scatter of each MLCR, as recorded in Section \ref{sec:stellar_mass}. 
The error in distance is taken directly from their respective measurement sources (see Appendix~\ref{sec:dist_corr}).

\subsection{Bulge sizes}
\label{sec:sph_size}

\begin{figure}
	\includegraphics[clip=true, trim= 1mm 1mm 2mm 2mm, width=\columnwidth]{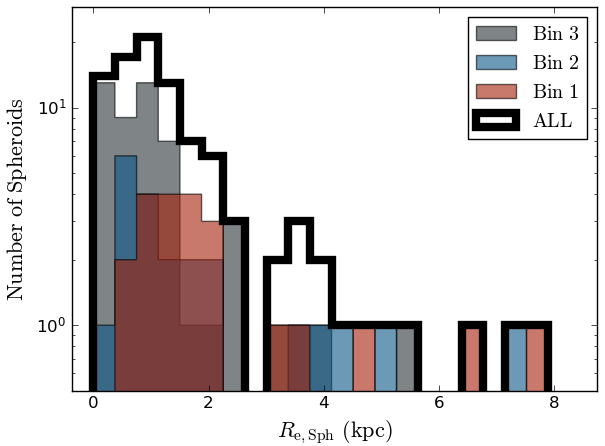}
    \caption{The size ($R_\mathrm{e,Sph}$, in equivalent axis) distribution of our bulges/spheroids.
    The spheroids from Bins~1, 2, and 3 are depicted by the red, blue, and black histograms, respectively.
    The thick black line depicts the size distribution of the entire sample set. 
    The figure only shows the 96 spheroids with $R_\mathrm{e,Sph} < 10~\rm kpc$ of the complete sample, see the spheroids' size-mass distribution later in Section~\ref{sec:size-mass} for further reference.}
    \label{fig:size_dist}
\end{figure}

We present the bulge/spheroid size (in equivalent axis, $R_{\rm e,Sph}$) distribution in Figure~\ref{fig:size_dist}.
The spheroids from Bin~1, 2, and 3 are depicted with red, blue, and black filled histograms, respectively.
The median size $\langle R_\mathrm{e,Sph} \rangle$ for Bin 1, 2, and 3 are 1.7 kpc, 1.2 kpc, and 0.8 kpc, respectively.
The majority of our spheroids (96/103, 93\%) are smaller than $R_\mathrm{e,Sph} < 7~\rm kpc$.

Over time, as the stars in galaxies evolve and eject their stellar winds, the amount of dust builds up, at least in late-type galaxies where dust sputtering does not operate efficiently. 
While the presence of dust is known to reduce the apparent bulge and disc magnitudes, and increasingly so with a more edge-on disc inclination ($i$), the impact on the effective half-light radius of the bulge is not well known.
An investigation of how $R_{\rm e}$ changes with $i$, at different wavelengths, would be welcome.
Ideally, this information would be known for different galaxy morphological types.  

The dust is known to be more centrally concentrated in galaxies than in their outskirts.
Qualitatively, this acts to reduce the inner flux relative to the outer flux, which has the effect of increasing the half-light radius of the galaxy, and the disc component of the galaxy \citep[e.g.,][see their Figure~2]{2008MNRAS.388.1708G}.
It is not known how much the apparent half-light radius of the bulge increases, however, the effect is such that we may be underestimating the true number of local, compact spheroids.  

Combined with potentially underestimating the masses of the bulges, we move forward in the knowledge that we are presenting a conservative estimate of the number density of compact massive spheroids.
That is, these potential biases do not act to overestimate the number density.

\section{Results}
\label{sec:results}
We present our result in the following subsections: Section~\ref{sec:B/T_ratio} showcase the bulge-to-total flux ratio ($B/T$) newly obtained from multi-component decomposition; Section~\ref{sec:mass-function} shows the bulge/spheroid mass function; and Section~\ref{sec:size-mass} present the size-mass relation of the local spheroids and the number of compact massive spheroids under different definitions.

\subsection{Bulge-to-total (\texorpdfstring{$B/T$}~)~flux ratios}
\label{sec:B/T_ratio}

\begin{figure}
	\includegraphics[clip=true, trim=2mm 2mm 1mm 1mm, width=0.95\columnwidth]{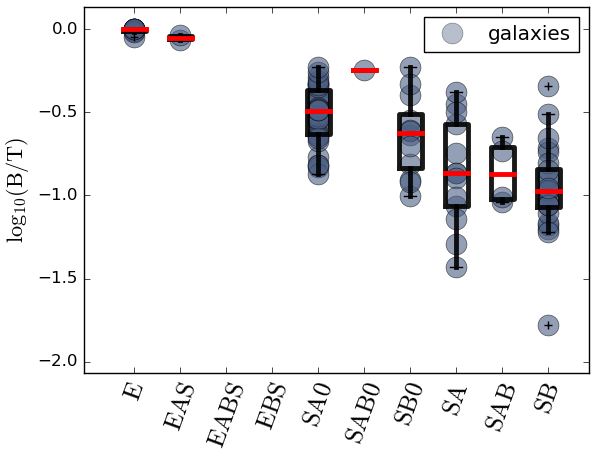}
    \caption{SDSS $i$-band Bulge-to-Total ($B/T$) flux ratios.
    The spiral (SA--SB) galaxies presented here are dominated by the so-called `early-type' spirals (Sa--Sb).
    The boxes denote the lower (Q1) to upper (Q3) quartiles of the data, and the whiskers extend from the minimum to the maximum data points (excluding any outliers).
    The red lines represent the median $\langle B/T \rangle$ ratios for each morphology.
    The `+' marker denotes an outlier (NGC~5350) that lies beyond the $\rm Q1-1.5(Q3-Q1)$ limit \citep{Tukey1977}.}
    \label{fig:BT_ratio}
\end{figure}

We present the $i$-band bulge-to-galaxy flux ratios in Figure~\ref{fig:BT_ratio}, plotted against by their morphologies.
Our flux ratios are the observed (no internal dust correction) ratios taken from the equivalent-axis decompositions.
For simplicity, we have applied the same $M_*/L$ ratio to all galaxy components; this poses a slight limitation and is an area where future improvements can be made. However, we note that the (dust-free) colours of bulges and discs are, in general, not too dissimilar from each other in massive disc galaxies \citep[][]{Peletier1996}. As such, the $M_*/L$ ratio will typically, although not necessarily, be similar.
The grey points indicate the bulge-to-total ($B/T$) flux ratios for individual galaxies while the red line is the median ratio in each subgroup.
We obtained a disc-dominated median $\langle B/T \rangle \pm 1 \sigma \approx 0.33\pm 0.15$ for the S0 galaxies (49 in sample). 
Our spiral galaxies exhibited a notably lower median ratio of $\approx 0.13\pm 0.12 $, (41 in sample). 
In both cases, the barred galaxies consistently have a lower $B/T$ ratio than the unbarred galaxies.

\begin{table*}
   \caption{The Bulge-to-Total ($B/T$) flux ratio of our galaxy sample.
   This table presents the median flux ratio $\langle B/T \rangle $ and the $1\sigma$ range for each morphological type in each of our three volume bins.
   The number in parentheses is the sample size for each subdivision, i.e., median $\langle B/T \rangle \pm 1\sigma~(\rm sample~size)$.}
   \begin{threeparttable}[b]
   \label{tab:B/T_table}
   \begin{tabular}{lllllllll}
   \hline
     & E & EAS & SA0 & SAB0  & SB0 & SA & SAB  & SB \\
   \hline
   \hline 
    Bin~1 & $1.0\substack{+0.0\\-0.01}$ (5) & $0.86$ (1) & $0.33\substack{+0.15\\-0.17}$ (13) & $--$ (0)  & $0.25\substack{+0.10\\-0.10}$ (2)& $0.19$ (1) & $0.19\substack{+0.03\\-0.06}$ (3)  & $0.10\substack{+0.01\\-0.01}$ (2) \\
    Bin~2 & $1.0\substack{+0.0\\-0.04}$ (3) & $0.92$ (1) & $0.30\substack{+0.14\\-0.06}$ (9) & $0.56$ (1)  & $0.53\substack{+0.04\\-0.04}$ (2) & $0.2\substack{+0.08\\-0.08}$ (2) & $0.10$ (1) & $0.09 \substack{+0.05\\-0.01}$ (5)\\
    Bin~3 & $1.0\substack{+0.0\\-0.02}$ (3)& $--$ (0) & $0.33\substack{+0.12\\-0.10}$ (14) & $--$ (0) & $0.21\substack{+0.03\\-0.09}$ (8) & $0.13\substack{+0.18\\-0.10}$ (10) & $--$ (0) & $0.11\substack{+0.10\\-0.04}$ (17) \\
   \hline
    Total & $1.0 \substack{+0.0\\-0.07} $ (11) & $0.89 \substack{+0.02\\-0.02} $ (2) & $0.32\substack{+0.14\\-0.10}$ (36)& $0.57$  (1)& $0.24\substack{+0.18\\-0.11}$ (12) & $0.14\substack{+0.18\\-0.07}$ (13) & $0.14\substack{+0.06\\-0.05}$ (4) & $0.11\substack{+0.08\\-0.04}$ (24) \\
    \hline
   \end{tabular}
    \end{threeparttable}
\end{table*}

In Table~\ref{tab:B/T_table}, we present the median $\langle B/T \rangle \pm 1 \sigma$ value for each morphological type from each bin.
The two unbarred ES galaxies have an average $\langle B/T \rangle \pm 1\sigma \approx 0.86 \pm 0.02$.
While this simply reflects the range of masses and galaxy types, it is important for understanding the transition from galaxy to spheroid mass at the high-mass end of the galaxy stellar-mass function. 

Our inclusion of minor components, such as nuclear discs, spiral arms, lenses, and secondary bars can change the $B/T$ ratio.  This is not so much because of the flux held in these components but because of how their presence can skew the galaxy decomposition if they are not properly accounted. 
Although one may expect this to decrease the $B/T$ ratio, it is more complicated than that.
For example, the more sophisticated models, which can involve truncated-exponential discs rather than a single exponential disc model, can result in a fainter central disc surface brightness and thus a greater assignment of flux to the bulge if the  single-disc model was biased by the outer disc profile.
We have 32 truncated discs in our sample: 14 Type~II (downward-bending) and 18 Type~III (upward-bending) discs.
Echoing the same point made by \citet{Kim2014}, which claimed that ignoring Type~II disc breaks will result in an $\approx$10\% decrease in the $B/T$ ratio, we agree that broken, and inclined, exponential disc models play an important role in correctly modelling a galaxy.
Contrarily, the application of these more sophisticated disc models to Type~III discs can be expected to decrease the $B/T$ ratio, compared to the result using a single exponential model, if the single-disc model is biased by the outer disc slope.

\begin{figure}
	\includegraphics[clip=true, trim=3mm 3mm 2mm 2mm, width=\columnwidth]{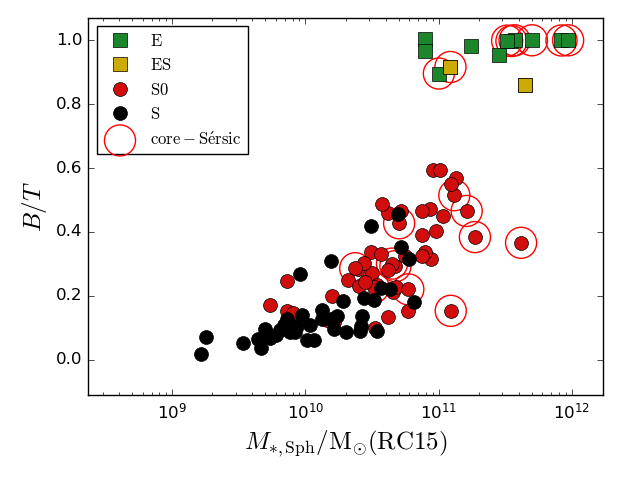}
	\includegraphics[clip=true, trim=3mm 6mm 2mm 2mm, width=\columnwidth]{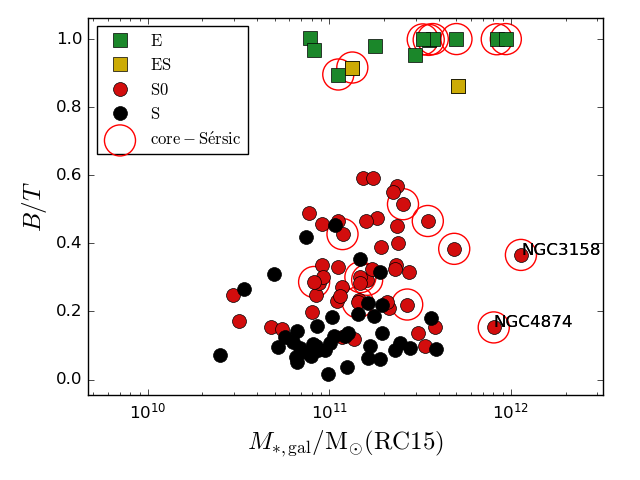}	
    \caption{The bulge-to-total ($B/T$) flux ratio-versus-spheroid stellar mass ($M_\mathrm{*,Sph} / \rm M_{\odot}$, upper panel) and galaxy stellar mass ($M_\mathrm{*,gal} / \rm M_{\odot}$, lower panel). We labelled the data points from E, ES, S0, and S galaxies as green squares, yellow squares, red points, and black points, respectively. Bulges modelled by core-S\'{e}rsic function are marked by red circles while the rest are modelled by S\'{e}rsic function. The outliers: NGC~3158 (S0) and NGC~4874 (S0) are marked by their name in the plots.}
    \label{fig:BT_stellar_mass}
\end{figure}

In Figure~\ref{fig:BT_stellar_mass}, we show the distribution of $B/T$ ratio and the stellar mass of the spheroids ($M_\mathrm{*,Sph} / \rm M_{\odot}$, upper panel) and the galaxies ($M_\mathrm{*,gal} / \rm M_{\odot}$, lower panel), separated by their morphology. The outliers: NGC~3158 (S0), and NGC~4874 (S0)\footnote{
The two S0 galaxy outliers are particularly massive ($M_\mathrm{*,gal}/\rm M_{\odot} > 7\times10^{11}$). The MASSIVE survey \citep{Veale2017} confirms our stellar mass calculation, where in their Table 1, NGC~3158 and NGC~4874 reports a galaxy stellar mass of $M_\mathrm{*,gal}/\rm M_{\odot}\sim1.05\times10^{12}$ and $M_\mathrm{*,gal}/\rm M_{\odot}\sim9.54\times10^{11}$, respectively. Similar to our classification, \citet{Veale2017} labelled NGC~3158 as a 'fast rotator'. 
For NGC~4874, according to the multi-component decomposition in \citet{Dullo2019}, it contains three components, an inner core-S\'ersic bulge, an intermediate component, and an outer exponential halo. The intermediate component is dominant between $\sim10\arcsec$ and $\sim 100 \arcsec$. In their $F606W - F814W$ colour map (their figure~2), the colour becomes gradually bluer towards larger radii.
We found that an exponential component can describe the intermediate component well. The outer envelope comes into play beyond our cut-off radius of $90\arcsec$ in the major axis and was therefore not required in our decomposition}
 are labelled individually in the lower panel.
Similar to Figure~\ref{fig:BT_ratio}, E+ES galaxies occupy the high $B/T$ region, between $0.8 \lesssim B/T < 1.0$, while S0 and S galaxy have a $B/T\lesssim0.6$.
Note that E galaxies do not have $B/T = 1$ exactly due to the presence of nuclear discs and other components. 
Regardless of their $B/T$ ratio and morphological type, most\footnote{Some (13/103) galaxies are below our lower mass selection limit in ($M_*/\rm M_{\odot}($RC15$) = 6.7\times10^{10}$). This is somewhat expected because SDSS $i-$band total galaxy magnitude is measured using a single S\'{e}rsic function, while ours is measured by a multi-components model (Bulge+Disc+others). For galaxies with prominent discs (S0 \& S), a single S\'{e}rsic model can overestimate the size and luminosity of the galaxy (see later in Figure~\ref{fig:ReRmax}). The implication of which, however, shall be better explored in future works.}  of our galaxies span the galaxy stellar mass range of $6 \times 10^{10} \lesssim M_\mathrm{*,gal} / \rm M_{\odot}(RC15)< 1.5 \times10^{12}$. It appears that many S0 galaxies, as well as a smaller amount of S galaxies, can be as massive as $M_\mathrm{*,gal}/\rm M_{\odot}($RC15$)>2\times10^{11}$.
Spheroids follow a positive correlation between B/T ratio and spheroid mass by morphology where S galaxies tend to have less massive bulges than S0 galaxies (see the upper panel of Figure~\ref{fig:BT_stellar_mass}).

The lack of a strong correlation between $B/T$ ratio and galaxy stellar mass is reminiscent to the result from \citet{Mendez-Abreu2017}, where they performed detailed 2D multi-component decomposition (including nuclear, bar, and broken disc components) to 404 galaxies from the Calar Alto Legacy Integral Field Area (CALIFA, \citet{Sanchez2016}) data.
They presented the mean $\langle B/T \rangle \pm 1\sigma$ value for S0, Sa, and Sb to be $\sim 0.32\pm0.17$, $0.28\pm0.17$, $0.12\pm0.11$ respectively.
The mass of their S0, Sa, and Sb galaxies occupy a similar range of $\langle \mathrm{log_{10}}(M_*/\mathrm{M_{\odot}})\rangle\pm1\sigma\sim 10.79\pm0.6$, $10.80\pm0.4$, and $10.48\pm0.4$, respectively.
%We show the galaxy stellar mass ranges and the B/T ratio $1\sigma$ range about from }\citep[][]{Mendez-Abreu2017}\hl{ with coloured crosses, divided by morphology, in the lower panel of Figure~}\ref{fig:BT_stellar_mass}\hl{. One can see that they overlap nicely with our sample}\citep[see also][]{Mendez-Abreu2021}\hl{. 
It has been shown that, if the bar component is not considered, the bulge's S\'{e}rsic index $n$ and the B/T ratio will be overestimated \citep{Gadotti2009,Salo2015}.
The rather low median $B/T$ ratio ($B/T\sim0.1$--$0.3$) in both massive S0 and S galaxies reflect a reality that stellar discs and their induced structures take up the majority of the mass budget \citep[see also,][]{Laurikainen2005}.

\subsection{Spheroid stellar mass function}
\label{sec:mass-function}

\begin{figure*}
\centering
	\includegraphics[clip=true, trim=3mm 2mm 2mm 3mm, width=1.0\textwidth]{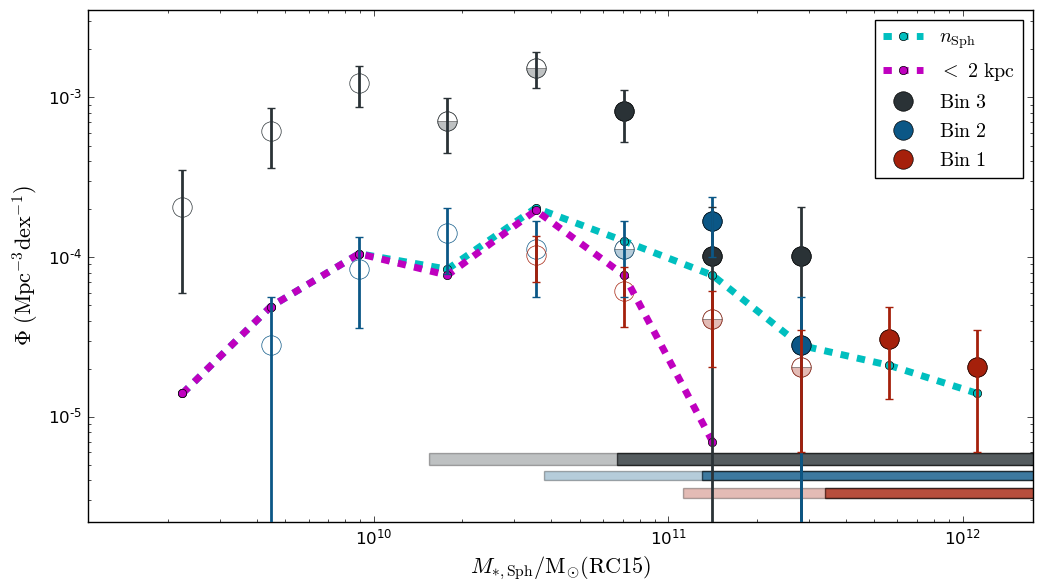}
    \caption{
    The local bulge/spheroid `stellar mass' function, with the stellar mass defined using the RC15 MLCR.
    Red, blue, and black points represent the spheroid sample from the three `galaxy mass bins' 1, 2, and 3, respectively (see Figure~\ref{fig:data_selection}).
    The spheroid stellar masses are divided into 0.3\,dex mass bins.
    The error bars represent the Poisson error.
    The cyan dashed line, labelled `$n_\mathrm{Sph}$', is the lower limit of the number density for each mass interval, obtained by summing up the total number of spheroids in each interval and dividing it by the total volume of our selection space.
    The purple dashed line, labelled `$<2~\rm kpc$', is the same as the cyan line but including only the spheroids with $R_\mathrm{e}<2\, \rm k p c$ (in equivalent axis).
    The three colour-coded bars in the bottom right represent the mass range of confidence, in which the solid bars depict the mass `selection limit' from the host galaxies, and the transparent bars depict a `partially-complete' region.
    The points in the `selection limit' are plotted in solid circles, the ones in the `partially-complete' region are in half-filled circles, and the ones below these regions are in open circles.
    The spheroid mass function is complete down to each host galaxies' `selection limit' where all the hidden spheroids within this range are accounted.
    In the `partially-complete' region, our sample includes all spheroids coming from  high-mass galaxies ($M_*/\rm M_{\odot} \gtrsim 0.7$--$3.0\times10^{11}$) with low $B/T$ ratios, but not the spheroids coming from low-mass galaxies ($M_*/\rm M_{\odot} \sim 0.18$--$1.5\times10^{11}$) with high $B/T$ ratios. Since Bin 1 has a volume of $3.25\times10^{5}~\rm Mpc^{3}$, 1 object per 0.3 dex range of spheroid mass will give a number density of $\Phi\approx1\times10^{-5}~\rm Mpc^{-3}dex^{-1}$. See Table~\ref{tab:volume} for the volume of each bin.
    }
    \label{fig:massfunc}
\end{figure*}

We present the high-mass end of the local bulge/spheroid stellar mass function in Figure~\ref{fig:massfunc}. 
It includes bulges in the mass range $2\times10^{9} \lesssim M_*/{\rm M_\odot} ($RC15$) \lesssim 1.5 \times 10^{12}$.
To our knowledge, these results are the first of their kind derived from highly-detailed, multi-component decompositions on a volume- and mass-limited sample.
For each Bin (1 to 3), the mass function is constructed by counting the number of spheroids per 0.3\,dex range in mass, then dividing that number by 0.3 and by the volume of the associated Bin (1 to 3).

Based on the $B/T$ flux ratios from Section~\ref{sec:B/T_ratio}, we use the individual galaxy $B/T$ ratios in each bin to estimate the median $\langle B/T \rangle \pm 1\sigma$ range for each bin. They are:
\begin{enumerate}
    \item 
Bin~1 \boldmath$(\langle B/T \rangle\sim 0.33 \pm 0.32)$: E (5), EAS (1), SA0 (13), SB0 (2), SA (1), SAB (3), and SB (2); 
    \item 
Bin~2 \boldmath$(\langle B/T \rangle\sim 0.29 \pm0.30)$: E (3), EAS (1), SA0 (9), SAB0 (1), SB0 (2), SA (2), SAB (1), and SB (5); and
    \item 
Bin~3 \boldmath$(\langle B/T \rangle\sim 0.21 \pm0.22)$: E (3), SA0 (14), SB0 (8), SA (10), and SB (17).
\end{enumerate}

While we have a well-defined stellar mass `selection limit' for the \textit{galaxies}, it is unclear to what completeness the \textit{spheroid} mass function is covered below this limit. 
Spheroids/bulges more massive than the `selection limit' cutoff are of course accounted for, but since we did not sample galaxies below the limit, the spheroid mass function is only `partially complete' at lower masses. 
For instance, a massive galaxy of $M_\mathrm{*,gal}\sim 2\times 10^{11}\,\rm M_{\odot}$ with $B/T =0.1$ will have a bulge of $M_\mathrm{*,Sph}\sim 2\times 10^{10}\,\rm M_{\odot}$, below our selection limit.
Our sample provides a good representation for embedded spheroids coming from galaxies more massive than $6.7 \times10^{10}\, \rm M_{\odot}$ (RC15), but increasingly less so for less massive spheroids.
For example, a less massive galaxy ($M_\mathrm{*,gal}\sim 4 \times 10^{10}\,\rm M_{\odot}$, outside of the limit) with a bulge-to-total ratio $B/T\sim0.5$ will also have a spheroid of the same stellar mass as above ($M_\mathrm{*,Sph}\sim 2\times 10^{10}\,\rm M_{\odot}$), but it will not be in our sample. 
We refer to the region below the `selection limit' as the `partially-complete region'. 
The number of spheroids found in this mass range represents a lower limit to the true number. 

We estimate a rough lower bound to this `partially-complete region' by multiplying the galaxy mass limit (see Table~\ref{tab:mass_limit}) by the average $\langle B/T \rangle$ ratio for the galaxies with masses greater than this limit.
This provides the following rough lower bound to the spheroid stellar masses (when using the $M/L$ prescription from RC15) for each bin:
$1.12\times10^{11}$\,M$_\odot$ (Bin 1), 
$3.77\times10^{10}$\,M$_\odot$ (Bin 2), and 
$1.54\times 10^{10}$\,M$_\odot$ (Bin~3).  
Between the galaxy mass limit and these lower values, our spheroid mass function is only partially complete. Above the galaxy mass limit, our numbers are complete.  
The galaxy mass `selection limit' has been depicted by the colour-coded solid bars, and the `partially-complete region' by the semi-transparent colour-coded bars, in the lower right of Figure~\ref{fig:massfunc}. 
The data points above the `selection limit' are shown with solid circles, while those in the `partially-complete region' are shown using transparent half-filled circles, while those less massive than that are depicted using open circles.

Given that Bin-1 has a volume of $3.25\times10^5$\,Mpc$^3$, a count of 1, 2, 3, or 4 objects per 0.3\,dex range in spheroid mass will result in a number density of $\approx$1, 2, 3, or 4 $\times10^{-5}$\,Mpc$^{-3}$\,dex$^{-1}$. 
The number densities coming from the smaller volumes of Bin-2 and Bin-3 will be roughly 3 and 10 times greater for the same number of objects per 0.3\,dex in spheroid mass.  As can be seen in Figure~\ref{fig:massfunc}, there is a rising number density from $\sim$$2\times10^{-5}$\,Mpc$^{-3}$\,dex$^{-1}$ at $10^{12}\,{\rm M}_\odot$, 
to $\sim$10$^{-4}$ Mpc$^{-3}$\,dex$^{-1}$ at $10^{11}\,{\rm M}_\odot$, 
and reaching $\sim$10$^{-3}$ Mpc$^{-3}$\,dex$^{-1}$ by $0.7\times10^{11}\,{\rm M}_\odot$. 
Bin-3, which is more complete at lower masses, reveals a lower-limit to the number density of $\sim$$10^{-3}$\,Mpc$^{-3}$\,dex$^{-1}$ at $10^{10}\, {\rm M}_\odot$.

We present the overall mass function across the three bins using a  dashed cyan line in Figure~\ref{fig:massfunc}, obtained by summing up the number of spheroids in each mass interval from all three bins and dividing them by the total survey volume ($4.76\times10^{5}\rm\,Mpc^{3}$) and then by 0.3\,dex.
The purple dashed line depicts the overall mass function of small spheroids with $R_\mathrm{e} < 2\,\rm k p c$ (equivalent axis).
The overall mass function serves as a lower limit of spheroids in each mass interval.
At stellar masses above $M_{\rm *,Sph}/\rm M_\odot$ (RC15) $ \sim 1.5\times10^{11}$, there are no spheroids smaller than $2\,\rm k p c$ in our sample.
However, there is no strict definition in the literature for `compact massive' galaxies.
To better explore the situation requires us to next look at the size-mass diagram. 

\subsection{Size-mass relation of the local spheroids}
\label{sec:size-mass}

\begin{figure*}
\centering
	\includegraphics[clip=true, trim=2mm 4mm 4mm 4.5mm, width=0.74\textwidth]{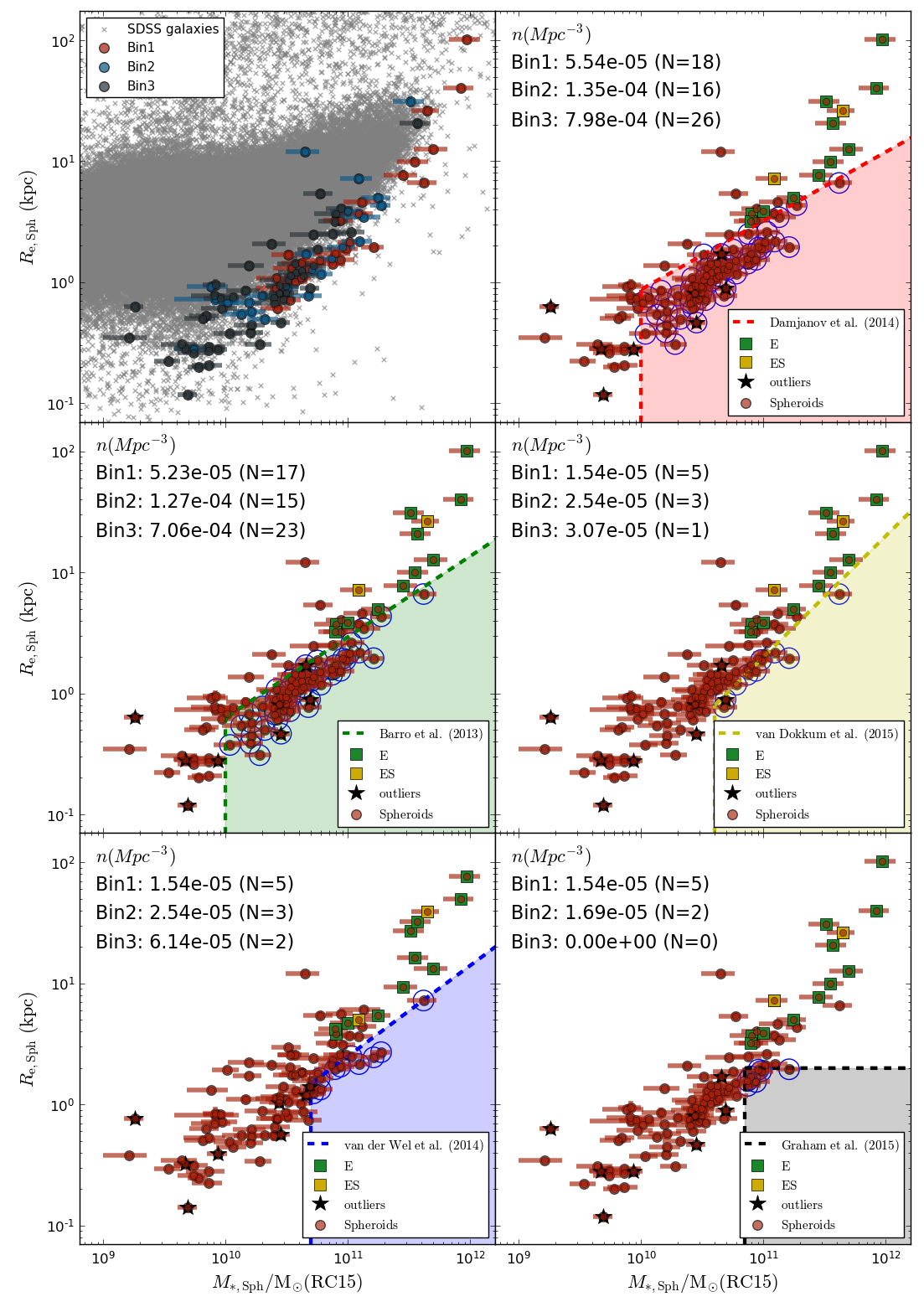}
    \caption{
    Size-mass distribution for our local spheroids (marked as $\bullet$).
    The stellar masses of the local spheroids are calculated via the RC15 MLCR.
    The effective half-light radii for the local spheroids come from their geometric mean axes, with the exception that the major-axis half-light radii have been used in the bottom left panel to provide a better comparison with \citet{van_der_Wel2014}.
    In the top left panel, we have the size-mass distribution separated by our three-volume bins, shown against the SDSS galaxy sizes and stellar masses (\textcolor{Gray}{$\times$}).
    The SDSS galaxy radius and flux data are taken from the NASA-Sloan ATLAS catalogue (see description in Section~\ref{sec:data}).
    We calculate their stellar mass using the same MLCR as our spheroids, in this case, with the RC15 equation.
    The other plots illustrate some of the different `compactness' criteria used in the literature.
    The number of galaxies ($N$) that satisfied the conditions in each bin and the corresponding number densities ($n$) are shown in the panels.  These are, of course, lower limits to the true number density of local compact massive spheroids due to the lower galaxy-mass limits used to construct each bin.
    The compactness criteria are from: 
     \citet{Damjanov2014} (top right), 
    \citet{Barro2013} (middle left), 
     \citet{van_Dokkum2015} (middle right)
    \citet{van_der_Wel2014} (bottom left), and 
    \citet{Graham2015} (bottom right).
    Additionally, the 13 E+ES class galaxies are highlighted in squares ($\blacksquare$), where E galaxies are in green and ES galaxies are in yellow. The outliers with galaxy colour $(g-i)>1.5$ are marked with black star signs ($\star$, see Section~\ref{sec:sph_colour}). 
    In our sample, there are 55 (53\%), 10 (9.7\%), 60 (58.3\%), 9 (8.7\%), and 7 (6.7\%) compact massive spheroids according to \citet{Barro2013}'s, \citet{van_der_Wel2014}'s, \citet[][]{Damjanov2015A}'s, \citet[][]{van_Dokkum2015}'s, and \citet{Graham2015}'s compactness criteria, respectively.
    }
    \label{fig:sizemass}
\end{figure*}

In the top left panel of Figure~\ref{fig:sizemass}, we show the distribution of local spheroids in terms of their (equivalent-axis) effective half-light radius $R_{\rm e,eq}$  and their stellar masses derived from the RC15 MLCR, assuming a BC03 SPS.
The spheroids are colour-coded according to the volume-bins (1, 2, and 3) shown in 
red, blue, and black, respectively.
We plot our {\em spheroids} against the opaque grey cloud of the general population of bright SDSS {\em galaxies}, consisting of both early- and late-type galaxies.
It is clear that, for a given mass, the local spheroids are considerably more compact than most galaxies. 
As expected, the spheroids extracted from Bin~3, with their lower galaxy mass-boundary, occupy the lower to the medium mass range, and the giant spheroidal elliptical galaxies reside in the (high mass)--(large radius) region of the diagram.

In the other panels, we report the number density ($n~(\rm Mpc^{-3})$) based on different selection criteria. 
The arbitrary selection boundary often contains a diagonal line separating the sample into `compact' and `non-compact' systems, with a lower-mass limit.

The compactness criteria are as follows:
\begin{itemize}
    \item 
In the middle left-hand panel of Figure \ref{fig:sizemass}, we applied the \citet{Barro2013} selection boundary (green) with mass and circularised size limits:
\begin{equation}
\label{eq:Barro_cut}
\begin{split}
M_*/{\rm M_\odot} &> 10^{10}\\
\log_{10} \left(R_{\mathrm{e}} / \mathrm{kpc}\right)&<\left[\log_{10} \left(M_{*} / \mathrm{M}_{\odot}\right)-10.3\right] / 1.5.
\end{split}
\end{equation}
The compact massive quiescent galaxies in the CANDELs survey \citep{Grogin2011,Koekemoer2011} within this boundary roughly follow the \citet{Newman2012} size-mass relation.
The slope of the selection boundary is motivated by the apparent trend of the quiescent galaxies (with specific star-formation rate $ \log_{10}(sSFR_\mathrm{UV+IR}) > - 0.5\, \rm [Gyr^{-1}]$) at $ z > 1.5$.
    \item
In the bottom left-hand panel, the \citet{van_der_Wel2014} boundary (blue) was applied:
\begin{equation} 
\label{eq:vdWel_cut}
\begin{split}
    M_*/{\rm M_\odot} &> 5 \times 10^{10}\\
    R_{\mathrm{e,major}}/\mathrm{kpc}&<\left(M_{*} / 10^{11}\,\mathrm{M}_{\odot}\right)^{0.75}\times2.5.
\end{split}
\end{equation}
Note, that unlike most studies, \citet{van_der_Wel2014} selected their compact sample by the major axis effective radius $R_{\mathrm{e,major}}$ instead of in the equivalent axis  $R_{\mathrm{e,eq}}$, where the two quantities follow the relation $R_{\mathrm{e,eq}} \equiv R_{\mathrm{e,major}} \sqrt{b/a}$, in which $b/a$ is the semi-minor to semi-major axes ratio of a galaxy's apparent elliptical shape. Therefore, the size-mass plot in the lower left of Figure~\ref{sec:size-mass} is plotted in major axis instead. 
    \item
In the top right-hand panel, the sample is divided by the \citet{Damjanov2014} compactness criteria (red):
\begin{equation} 
\label{eq:Dam_cut}
\begin{split}
    M_*/{\rm M_\odot} &> 10^{10}\\
    \log_{10} \left(R_\mathrm{e}/\mathrm{kpc}\right) &<0.568\log_{10} \left(M_{*}/{\rm M_\odot}\right) -5.74.
\end{split}
\end{equation}
Unlike the high-$z$ studies, \citet{Damjanov2014} applied this selection criteria on a sample of intermediate redshift galaxies from the COSMOS field at $0.2 < z < 0.8$.
While having discussed the different choice in the mass limit, we set it to the lowest limit for easier comparison with \citet{Barro2013}.
    \item
In the middle right-hand panel, we select for the \citet{van_Dokkum2015} boundary (yellow): 
\begin{equation} 
\label{eq:vDokkum_cut}
\begin{split}
    M_*/{\rm M_\odot} &> 4 \times 10^{10}\\
    \log_{10} \left(R_{\mathrm{e}} / \mathrm{kpc}\right)&<\log_{10} \left(M_{*} / \mathrm{M}_{\odot}\right)-10.7.
\end{split}
\end{equation}
This selection criteria are constructed based on the argument that the \citet{Barro2013} boundary is not restrictive enough because it will also include $60 \%$ of the star-forming galaxies, which have a factor of two bigger effective radii than the quiescent galaxies.
It could enhance the `progenitor bias', where the quenching of star-forming compact galaxies may inflate the number density of the passively evolving quiescent galaxies.
It does not, however, concern our sample because the stellar population of local bulges has exist for a long time \citep[][]{MacArthur2009}. 
It is unlikely the quenching process between $0 < z < 1.0$ to produce such system.
    \item
In the bottom right, we select the compact spheroids via the \citet{Graham2015} selection conditions (grey):
\begin{equation} 
\label{eq:Graham_cut}
\begin{split}
    M_{*}/{\rm M_\odot} &> 7 \times 10^{10}\\
    R_{\mathrm{e,\rm major}}/\rm kpc &< 2.0.
\end{split}
\end{equation}
Building on \citet{2013pss6.book...91G}, 
\citet{Graham2015} furthered the proof of concept for this study as it supplied generic selection criteria on small, but massive bulges in the local Universe from different studies \citep{Seigar1998,Graham2001A,Mollenhoff2011,Balcells2007,Laurikainen2010,van_den_Bosch2012,Dullo2013,Savorgnan2016A}.
They found 21 small, but massive spheroids within 90\,Mpc and gave an initial (lower) estimate of the volume number density $n \sim  6.9 \times 10^{-6}\, \rm Mpc^{-3}$ (or per unit dex stellar mass, it is $ 3.5\times 10^{-5}\, \rm M p c^{-3}\,dex^{-1}$) as a lower limit for such systems.
\end{itemize}

While we have focused on galaxies more massive than $10^{11}$\,M$_\odot$, given that the presence of a disc results in a bulge-to-total ratio less than 1, we have uncovered some spheroids less massive than $10^{11}$\,M$_\odot$.  Although our spheroid sample is incomplete below this mass, we are able to obtain lower-limits to the actual number density of compact spheroids defined by a range of selection criteria in the literature which have encompassed masses down to 1, 4, 5, and 7$\times10^{10}$\,M$_\odot$. 
It must, therefore, be kept in mind that the number densities, from our three bins, shown in Figure~\ref{fig:sizemass} are lower-limits.

The majority of the spheroids (66/103) lie within the size-mass range $ 0.4\,\mathrm{kpc} < R_\mathrm{e} < 6.0\, \mathrm{kpc}$ and $ 1.0 \times 10^{10} < M_*/\mathrm{M_{\odot}}($RC15$) < 2.5 \times 10^{11}$. 
The lower bound simply reflects the mass boundary of our host galaxy sample and reveals that some galaxies have $B/T$ ratios less than 0.1. 
The scatter along either the size or mass axis is rather small in this region. 
With (geometric mean)-axis half light radii less than $\approx$1\,kpc and stellar masses $M_{\rm *,sph} < 2\times10^{10}$\,M$_\odot$ (RC15), the scatter is much more prominent. 
This variation of scatter is less apparent when using the major-axis size, as shown in the lower-left panel of Figure~\ref{fig:sizemass}.  We include this because \cite{van_der_Wel2014} used the major-axis galaxy sizes for their investigation.

Considerable effort has been made searching for the ultra-compact massive galaxies (UCMGs) in the local Universe, defined by as those with $M_\mathrm{*,gal}/\rm M_{\odot} > 8\times10^{10}$ and $R_\mathrm{e} < 1.5~\rm kpc$ \citep{Trujillo2009, Ferre-Mateu2017,Tortora2018,Tortora2020,Scognamiglio2020}. These objects are extremely rare compared to the regular red nuggets, with the number density of $n_\mathrm{UCMG}\sim 1$--$9\times 10^{-6}~\rm Mpc^{-3}$ at $z < 0.5$ \citep[][]{Tortora2018,Tortora2020}. 
In our sample, there are two spheroids that satisfy the UGMG definition, giving a lower limit to the number density of ultra-compact massive spheroids of $n_\mathrm{c,Sph}\sim 4.2\times10^{-6}\rm~Mpc^{-3}$. 

We provide an assortment of information in Tables~\ref{tab:result1}--\ref{tab:result3} regarding our spheroid sample.
The parameters of the S{\'e}rsic functions are listed in Columns~(4)--(6).
Column~(7) gives the spheroid apparent magnitudes as measured by \texttt{Profiler}.
Column~(8) lists the absolute magnitudes of the spheroids, based on the distances shown in Appendix~\ref{sec:dist_corr}. 
The magnitudes are corrected for galactic extinction from \citet{Schlegel1997}.
Columns~(9)--(12) tabulate the spheroid stellar masses, obtained using each of the four MLCRs (Equation Array~\ref{eq:MLCRs}).

\section{Cosmic evolution of compact massive spheroids}
\label{sec:comparsion}

In this section, we compare our local spheroids with the red nuggets in low-to-intermediate-$z$ (Section~\ref{sec:z1}) and high-$z$ (Section~\ref{sec:z2}) from the literature, as well as the embedded local bulges obtained by other multi-decomposition works (Section~\ref{sec:z0}).
Subsequently, we are able to plot out the potential evolution of compact massive spheroids across cosmic time.
The effect of adopting a different $M_*/L$ ratio is explored in Section~\ref{sec:z3}.

\subsection{Low redshifts (\texorpdfstring{$z < 0.03$}{low-z})}
\label{sec:z0}

We have compared our local spheroids with the bulges from other studies of $z\approx0$ galaxies. 
We have a preference for decompositions with a similar degree of detail. 
Relevant studies of the galaxy size evolution \citep[e.g.,][]{Trujillo2006,Trujillo2007,Taylor2010,Newman2012,McLure2013,Damjanov2014,Fang2015} have measured the effective radii and stellar masses of local early-type galaxies, but not their bulges nor the bulges of any massive early-type spiral galaxies. 
This may have prevented them from establishing the evolutionary pathway taken by the high-$z$ compact massive galaxies.
The underlying assumption of those works was the preservation of the morphological types throughout their evolutionary history.
Due to our different approach, a comparison with the {\em galaxy} sizes and masses from those studies would be of limited value to test our theory.  We can, however, explore if our bulge sizes and masses agree with those from 
multi-component decompositions of local galaxies.  

\begin{figure}
\centering
	\includegraphics[clip=true, trim= 3mm 2mm 3mm 2mm, width=1.0\columnwidth]{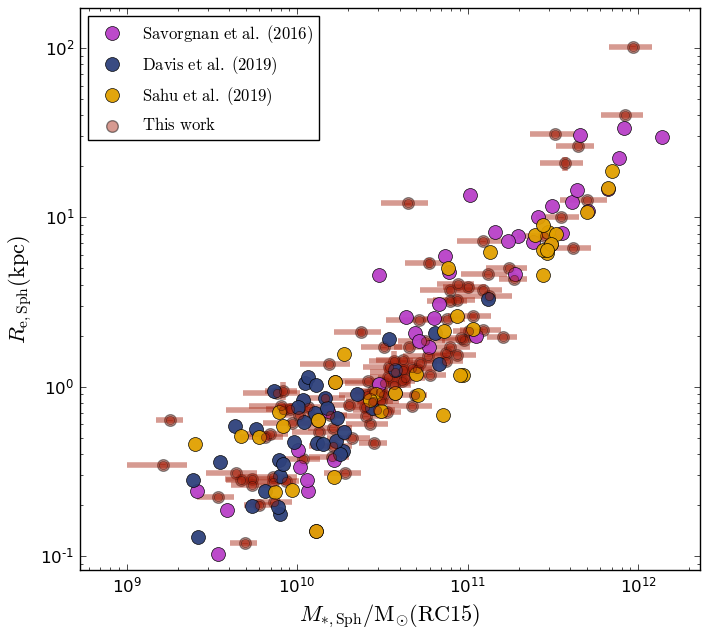}
    \caption{A comparison of our spheroids' stellar masses ($M_\mathrm{*,Sph}/{\rm M_{\odot}}$(RC15)) and circularised effective radius ($R_\mathrm{e,Sph}$, transparent red points) with those from \citet{Savorgnan2016A} (purple points), \citet{Davis2019} (blue points), and \citet{Sahu2019A} (orange points).
    The spheroids stellar masses from \citet{Savorgnan2016A}, \citet[][]{Davis2019}, and \citet[][]{Sahu2019A} are based on $3.6\,\micron$.
    Both bulges and elliptical galaxies are included.}
    \label{fig:comp_z0}
\end{figure}

Several studies of supermassive black hole scaling relations \citep[e.g.,][]{Savorgnan2016A, Davis2019, Sahu2019A} have performed a similar style of careful galaxy analysis using IRAC $3.6\,\micron$ images of galaxies in the local Universe.
Figure~\ref{fig:comp_z0} shows the size-mass distribution of our spheroids (red points) and the local bulges and giant elliptical galaxies from these studies.

\citet{Savorgnan2016A} obtained their surface brightness profiles via \texttt{IRAF} task \texttt{ellipse} from 66 early-type host galaxies and measured the sizes and masses by using a bespoke fitting code.
\citet{Davis2019} and \citet{Sahu2019A} extracted their surface brightness profiles with \texttt{ISOFIT} and performed decompositions via \texttt{Profiler}.
The \citet{Davis2019} spheroids are from 44 spiral host galaxies, while the \citet{Sahu2019A} spheroids are from 40 early-type galaxies.
These studies examine each galaxy on a case-by-case basis and assign physical components (e.g., bars and ansae rather than random S\'ersic components) when modelling each surface brightness profile (see the appendices of the respective papers for detailed discussions on each galaxy).

Despite the decompositions being conducted by different personnel, the resulting distribution of bulge sizes and masses appear remarkably consistent.
The high-mass end of the near-linear distribution in Figure~\ref{fig:comp_z0} resembles the bright arm of the spheroid distribution seen in \citet[][see Figure~1]{2013pss6.book...91G}.
%As the population increases in size and mass, it tends to converge into a tighter relation.
Here, we mainly want to illustrate the consistency across the different studies; that is, there is no unusual offset in our sample, nor sign of dichotomy.
The implications of this bulge/spheroid size-mass diagram shall be explored in future work.

In passing, we briefly note that we may be seeing evidence for a steepening of the relation at high masses. This is expected for merger-built elliptical galaxies for which the mass scales with $\sigma^2 R_{\rm eq}/G$ yet $\sigma$ barely increases over the value in the progenitor galaxies \citep[e.g., ][]{2013ApJ...768...29V}. Also at play is the influence of the intracluster light (ICL) surrounding BCGs, and we may have over-estimated the galaxy sizes at the top-end.
\citet{2007ApJ...657L..85D} produced a $B$-band bulge luminosity function from two-component S\'ersic$+$exponential fits \citep{2006MNRAS.371....2A} at $0.013< z <0.18$. 
While that work contained no dust corrections, Figure~2 from that work reveals, for their brightest bulges, a declining number density at $\mathfrak{M}_B=-$20 to $-$20.75\,mag (AB magnitude system) --- where S0 galaxies likely dominate and there is no substantial dust --- of around $3\times10^{-4}$ to $0.5\times10^{-4}$\,Mpc$^{-3}$\,dex$^{-1}$. 
For $\mathfrak{M}_{B,\odot}=+$5.44\,mag, and $M_*/L=8$, this roughly corresponds to 1.2--2.4$\times10^{11}\,{\rm M}_\odot$.
Their number density, therefore, overlaps well with our result at a similar mass in Figure~\ref{fig:massfunc}.
We do, however, find the same number density at 5$\times10^{11}\,{\rm M}_\odot$ in our data, while \citet{2007ApJ...657L..85D} has no bulges at these higher masses.

Our number density drops to $\sim$0.2$\times10^{-4}$\,Mpc$^{-3}$\,dex$^{-1}$ at $M_{\rm *,Sph} \sim 10^{12}\,{\rm M}_\odot$. 
The absence of the highest mass bulges in \citep{2007ApJ...657L..85D} may be due to the `logic filter' employed by \citet{2006MNRAS.371....2A} for identifying the more secure bulge+disc decompositions.
It meant that the ES galaxies, and importantly perhaps some misfit S0 galaxies, remained in the single-component E galaxy bin at these highest masses.
Resolving this is, however, beyond the scope of the current investigation.
However, it is important to note that in this mass range, there are only two spheroids in our sample. It reflects the reality that spheroids with mass $M_\mathrm{*,gal}/\rm M_{\odot} \sim 10^{12}$ are rare indeed.

\subsection{High redshifts (\texorpdfstring{$1 < z < 3$}{high-z})} 
\label{sec:z2}

The different selection criteria for compact massive systems have been used to illustrate the migration pattern of galaxy distributions, over cosmic time, in the size-mass diagram. 
As shown by \citet{Barro2013}, \citet{van_der_Wel2014}, and \citet{van_Dokkum2015}, with decreasing redshift, less and less galaxies occupy the `compactness' selection boundaries.
\citet{Damjanov2015A} demonstrated that while the actual number densities are different --- depending on the selection criteria for compact massive systems --- the overall trend/reduction with time remains.

\begin{figure} 
\centering
	\includegraphics[clip=true, trim= 1mm 1mm 1mm 2mm, width=0.85\columnwidth]{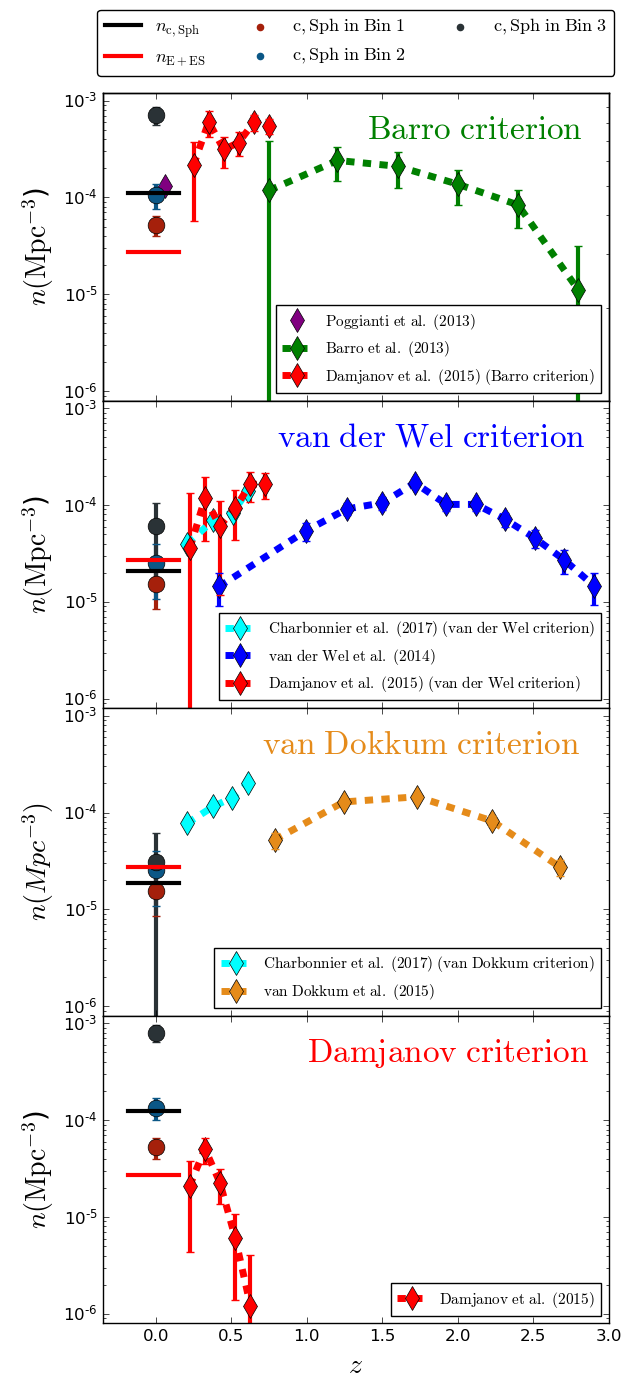}
    \caption{
    The number density of compact massive systems across cosmic time, based on four different size-mass selection criteria or defining boundaries.
    It shows the evolution of compact objects from the literature, marked in diamonds ($\blacklozenge$) under the \citet{Barro2013} boundary (top panel, in \textcolor{OliveGreen}{green}, from their figure~5), the \citet{van_der_Wel2014} boundary (upper-middle panel, in \textcolor{blue}{blue}, from their figure~13), the \citet{van_Dokkum2015} boundary (lower-middle panel, in \textcolor{orange}{orange}, from their figure~19), and the \citet{Damjanov2014} boundary (bottom panel, in \textcolor{red}{red}, from \citealt{Damjanov2015A}, figure~11).
    The \textcolor{red}{red}, \textcolor{blue}{blue}, and \textbf{black} circles ($\bullet$), at $z\sim0$, represent the number densities from Bins~1, 2, and 3, correspondingly.
    The error bars of our compact massive spheroids are assumed to be Poissonian.
    The horizontal \textbf{black} line, labelled `$n_\mathrm{c,Sph}$', is the lower limit of the number densities across the three bins, and the \textcolor{red}{red} line, labelled `$n_\mathrm{E+ES}$', is the lower limit for the true elliptical galaxies and ellicular galaxies in the three bins (given by Equation~\ref{eq:n_sum}). 
    In addition, the \citet{Damjanov2015A} compact sample (lower panel)
    is additionally shown by the  \textcolor{red}{red} diamonds  (\textcolor{red}{$\blacklozenge$}) in the top and upper-middle panels, 
    while the \citet{Poggianti2013B} compact sample is marked with a \textcolor{Fuchsia}{purple} diamond (\textcolor{Fuchsia}{$\blacklozenge$}) in the top panel.
    The number density of the compact massive galaxies from \citet{Charbonnier2017}, defined using the \citet{van_der_Wel2014} and \citet{van_Dokkum2015} criteria are depicted by the \textcolor{cyan}{cyan} diamonds (\textcolor{cyan}{$\blacklozenge$}) in the upper-middle and lower-middle panels, respectively.}
    \label{fig:nd_z}
\end{figure}

We have compared the number density of our local, compact massive spheroids with that of compact massive galaxies in the intermediate and high redshift Universe.
Figure~\ref{fig:nd_z} shows the evolutionary trend of number density for compact systems across time, based on four different compactness criteria, from top to bottom: Equation~\ref{eq:Barro_cut} (top), \ref{eq:vdWel_cut} (upper-middle), \ref{eq:vDokkum_cut} (lower-middle), and \ref{eq:Dam_cut} (bottom).
To represent the high-$z$ ($ 1.0 < z < 3.3 $) red nuggets, we have included the number density trends from three works: \cite{Barro2013} (\textcolor{OliveGreen}{green} diamonds \textcolor{OliveGreen}{$\blacklozenge$} in the top panel), \cite{van_der_Wel2014} (\textcolor{blue}{blue} diamonds \textcolor{blue}{$\blacklozenge$} in the upper-middle panel), and \cite{van_Dokkum2015} (\textcolor{orange}{orange} diamonds \textcolor{orange}{$\blacklozenge$} in the lower-middle panel).

For intermediate redshifts ($0.2 < z < 1.0$), we included the data from \cite{Damjanov2015A}.
They applied several different size-mass sample selection criteria to their data.
Their evolutionary trend based on the \citet{Barro2013} size-mass selection criteria is plotted in the top panel (see figure~9 in \citealt{Damjanov2015A}), and that based on the \citet{van_der_Wel2014} selection criteria is shown in the upper-middle panel (see figure~12 in \citealt{Damjanov2015A}), while that based on the \citet{Damjanov2014} selection criteria is shown in the bottom panel (from figure~11 in \citealt{Damjanov2015A}).
These trends are marked with \textcolor{red}{red} diamonds (\textcolor{red}{$\blacklozenge$}) in  Figure~\ref{fig:nd_z}. 
In addition, the number density provided by \cite{Poggianti2013B} is included, marked by the \textcolor{Fuchsia}{purple} diamond (\textcolor{Fuchsia}{$\blacklozenge$}) in the top panel.
The sample selection criteria from \citet{Barro2013} was used by  \cite{Poggianti2013B}.
Because the other three criteria were not implemented by \cite{Poggianti2013B}, we do not include their data in the other panels. \citet{Charbonnier2017} selects for compact massive galaxies in the SDSS Stripe 82 \citep[CS82, ][]{2014RMxAC..44..202M} survey at $0.2<z<0.6$. Their number density trends, based on \citet{van_der_Wel2014} and \citet{van_Dokkum2015} criteria, are presented as \textcolor{cyan}{cyan} diamonds (\textcolor{cyan}{$\blacklozenge$}) in the upper-middle and lower-middle panels of Figure~\ref{fig:nd_z}, respectively.

In Figure~\ref{fig:nd_z}, the three colour-coded circles ($\bullet$) at $z=0$ are our lower limits on the number density of compact massive spheroids, as obtained using the different selection criteria applied to our three bins.  
In all the panels, Bin~3 tends to have the highest number density, and Bin~1 the lowest. This is a result of the bin design.
As Bin~3 is complete down to the lowest host galaxy stellar mass ($M_*/\rm M_\odot($IP13$)\sim 1\times10^{11}$) compared to Bin~2  ($M_*/\rm M_\odot(IP13)\sim 2\times10^{11}$) and Bin~1 ($M_*/\rm M_\odot(IP13)\sim 5\times10^{11}$), we are able to capture more compact massive spheroids in Bin~3.
Note that in the lower-middle panel (\citet{van_Dokkum2015} criterion) of Figure~\ref{fig:nd_z}, the number densities of all three bins are closer than the other panels.
This is because \citet{van_Dokkum2015}'s criterion selects only for high-mass spheroids and red nuggets with $M_\mathrm{*}/\rm M_{\odot}(RC15) \gtrsim 4 \times 10^{10}$. 
If relatively low-mass ($1\times10^{10}\lesssim M_\mathrm{*}/\rm M_{\odot}(RC15) \lesssim 4 \times 10^{10}$) systems are also included, such as when using the selection criteria from \citet[][]{Barro2013} and \citet{Damjanov2015A}, an abundance of low-mass compact massive spheroids can be found in Bin~3.
These `red pebbles' are more likely to live inside of galaxies within the mass range of $6.7\times10^{10} < M_\mathrm{*,gal}/\rm M_{\odot} (RC15)< 1.3\times10^{11}$. 
As a result, there is roughly a factor of 10 more compact massive spheroids in Bin~3 in the upper- and lower-middle panels than the top and bottom panels in Figure~\ref{fig:nd_z}.

For the sake of discussion, we categorise the number density of our spheroids into $n_\mathrm{c,Sph}$ and $n_\mathrm{E+ES}$.
In Figure~\ref{fig:nd_z}, across all four panels, the black horizontal line (`$n_\mathrm{c,Sph}$') depicts the total number of compact spheroids from all three bins divided by the volume $V=4.76\times \rm 10^5\,M p c^{3}$ (our sample volume encompassed within $ 110\,\rm  Mpc$), 
while the red horizontal line (labelled `$n_\mathrm{E+ES}$') is the total number of the `true ellipticals' (E) and elliculars (ES) divided by this volume $V$:
\begin{subequations} 
\label{eq:n_sum}
\begin{align}
 n_\mathrm{c, Sph} &= (N_\mathrm{c,Sph,Bin1} + N_\mathrm{c, Sph,Bin2} + N_\mathrm{c,Sph,Bin3}) / V \\
  n_\mathrm{E+ES} &= (N_\mathrm{E+ES,Bin1} + N_\mathrm{E+ES,Bin2} + N_\mathrm{E+ES,Bin3}) / V.
\end{align}
\end{subequations}

These two values depict lower limits to the number densities. 
For $n_\mathrm{c, Sph}$, it is the lower limit of compact massive spheroids based on a volume-limited sample of galaxies with $M_* > 6.7\times10^{10}$ M$_{\odot}$ (RC15) and within a distance of 110\,Mpc.  
The value for $n_\mathrm{E+ES}$  represents the lower limit to the true number density of ellipitcal (E) and ellicular (ES) galaxies that can be found above $M_*/{\rm M_{\odot}} (RC15)\sim 6.7\times10^{10}$ . 
%within $110\,\rm Mpc$. 
Note that $n_\mathrm{E+ES, Bin3}$ is likely to be close to the true value because true elliptical galaxies tend to be massive; that is, ETGs less massive than $6.7\times10^{10}$\,M$_{\odot}$ tend to be S0 galaxies\footnote{Our E+ES galaxy with the lowest stellar mass is roughly $\sim 0.7\times 10^{11} \rm M_{\odot}$}   \citep[][]{Emsellem2011,Krajnovic2013a,Cappellari2013ATLAS3D,Cappellari2016review}. 
For Bins 1 and 2, we may have missed some E galaxies due to the higher stellar-mass selection criteria required to keep the number of galaxies requiring careful multi-component decomposition at a manageable level.

From Figure~\ref{fig:nd_z}, it is apparent that the definition of `compact and massive' makes a difference to the reported number density \citep[see also,][]{Charbonnier2017}. 
For the less restrictive size-mass sample selection criteria \citep{Barro2013,Damjanov2014}, the number density of compact massive spheroids is consistently higher than the number density of ellipitcal galaxies ($n_\mathrm{c,Sph} > n_\mathrm{E+ES}$), while applying the more  restrictive criteria \citep{van_der_Wel2014, van_Dokkum2015}, the two numbers are comparable ($n_\mathrm{c,sph}\simeq n_\mathrm{E+ES}$).
This is, in part, because $n_\mathrm{E+ES}$ does not depend on any size or mass selection criteria, unlike $n_\mathrm{c,sph}$ (see Figure~\ref{fig:sizemass}). 

The discrepancy between the restrictive and the less-restrictive size-mass sample selection criteria in use in the literature is informative.
It means that among the local compact systems, the elliptical galaxies and high-mass spheroids ($>$(4--7)$\times10^{10}$\,M$_\odot$), 
formed via an E-to-E process or disc-cloaking, respectively, may have occurred with similar frequency.  
The disc-cloaking process is likely the dominant mechanism in shaping galaxy evolution within the stellar mass range $1 \times 10^{10} < M_* /\rm M_{\odot}(\rm RC15) < 4\times10^{10}$.

\subsection{Low-to-intermediate redshifts (\texorpdfstring{$0.03 < z < 1$}{intermediate-z})}
\label{sec:z1}
As a bridge between the local spheroids and the high-$z$ galaxies, the information from intermediate redshifts provides insight into the transitioning process.
In Figure~\ref{fig:nd_z}, we compare our results with \citet{Damjanov2015A}.
They studied the number density of quiescent, compact massive galaxies at $0.2<z<0.8$ from the COSMOS field \citep{Leauthaud2007}.
While they acknowledged that most of their sample exhibited a Bulge+Disc structure, they concluded that the single S{\'e}rsic profile captures the size of the spheroids well in this redshift range, with perhaps just a slight overestimation in size. 

This sample's number density shows consistency with the \citet{Barro2013} and \citet{van_der_Wel2014} high-$z$ sample's peak abundance (see the top and upper middle panels in Figure~\ref{fig:nd_z}). 
In the upper panel of Figure~\ref{fig:nd_z}, the data from \citet{Damjanov2015A}  straddles the region between our Bin~2 and Bin~3 number densities.
It is also more abundant in comparison to the red nuggets' peaks number densities at $z=1.2$, and $z=1.7$. 
The lower panel shows the trend in number density, with redshift, using the size-mass selection criteria from \citet{Damjanov2014}.
Unfortunately with this selection, the complete set of data is not available, and \citet{Damjanov2015A} only show a partial result that they selected from the COSMOS quiescent galaxies classified as point sources in the photometric SDSS database.
Their Figure~11  depicts only a lower limit to the number density at intermediate-redshift. 
Nonetheless, our number of compact spheroids is vastly more numerous than the trend shown in the bottom panel.
The matching number densities with the high-$z$ data led them to conclude that the number of compact objects does not depend strongly on redshift.

Both \citet{Valentinuzzi2010} and \citet{Poggianti2013B} have argued for a constant compact quiescent galaxy number density \citep[see also][]{Saracco2010}.
\citet{Valentinuzzi2010} found a substantial number of compact quiescent galaxies in the WIde-field Nearby Galaxy-cluster Survey \citep[WINGS;][]{Fasano2006} at $0.04 < z < 0.07$. 
They reported a minimum number density of $1.3\times10^{-5}\,\mathrm{Mpc^{-3}}$ 
and possibly up to $1.3\times10^{-2}\,\mathrm{Mpc^{-3}}$ in cluster environments. 
Crucially, the majority (67\%) of their compact galaxies are S0.
In the Padova Millennium Galaxy and Group Catalogue \citep[PM2GC;][]{Calvi2011}, \citet{Poggianti2013B} found $n\sim4.23\times10^{-4}\,h^{3}\,\mathrm{Mpc^{-3}}$ at $0.03 < z < 0.11$. 
We have plotted their number density in the top panel of Figure~\ref{fig:nd_z} (marked as \textcolor{Fuchsia}{$\blacklozenge$}) for comparison.
Both studies acknowledge that more compact quiescent galaxies were found in dense cluster environments \citep[see also, ][]{Tortora2020}.
Indeed, \citet{Damjanov2015B} highlighted the environmental dependency of compact galaxies in the COSMOS field.
At intermediate-$z$, the compact quiescent galaxies align with the high-density regions (see their Figure~8). 
Although, it is important to note that this might simply be a result of more massive quiescent galaxies (of all sizes) living in cluster environments compared to the field \citep{Tortora2020}.

The depth and field-of-view in our parent sample selection (see Section~\ref{sec:data}, and Equation~\ref{eq:RADEC}) is wide enough to cover both clusters and fields. It captures the small Virgo Cluster overdensity at a distance of 17\,Mpc and a slight void at 40--60\,Mpc.
We explore the influence of the Virgo Cluster has on our analysis, later in Section~\ref{sec:VCC}.

\subsection{Variation arising from different MLCRs}
\label{sec:z3}

\begin{figure}
	\includegraphics[width=1.0\columnwidth]{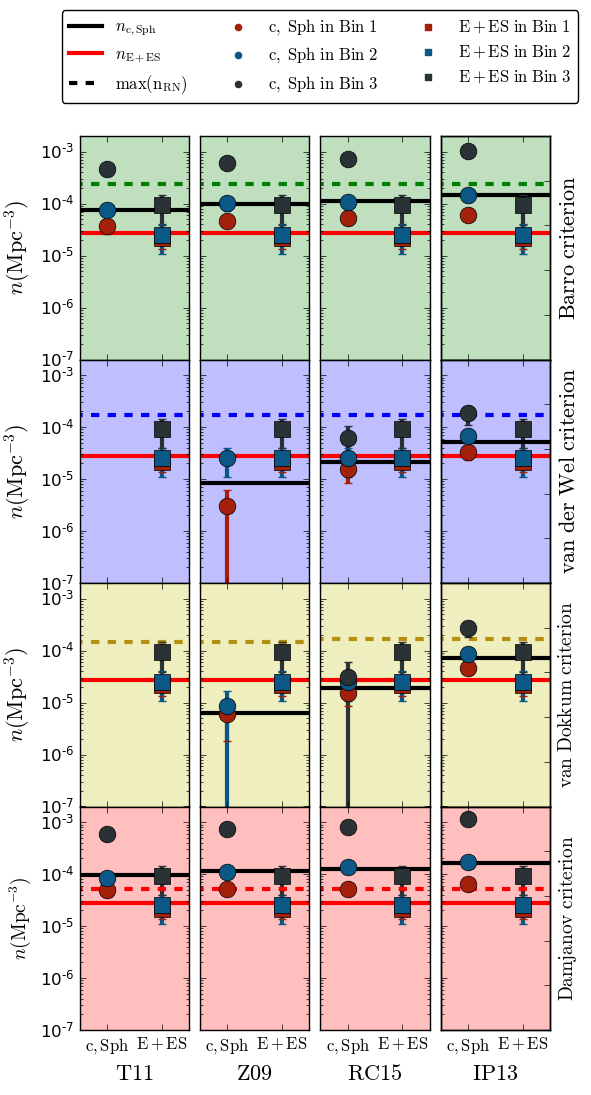}
    \caption{
    The number densities of local compact spheroids ($\bullet$) and ellipticals+elliculars ($\blacksquare$), using four different MLCRs, based on four size-mass selection criteria.
    The selection criteria applied from top to bottom are given by Equations \ref{eq:Barro_cut}, \ref{eq:vdWel_cut}, \ref{eq:vDokkum_cut}, and \ref{eq:Dam_cut}, respectively. 
    From left to right, the MLCRs are from T11, Z09, RC15, and IP13, respectively.
    The \textcolor{red}{red}, \textcolor{blue}{blue}, and \textbf{black} colours represent the samples from Bins~1, 2, and 3, respectively.
    The \textbf{black} and \textcolor{red}{red} horizontal lines are $n_\mathrm{c,Sph}$ and $n_\mathrm{E+ES}$, as depicted in Figure~\ref{fig:nd_z}.
    Here, the number density of `E+ES' includes 11 true E galaxies and 2 ES galaxies. 
    None of these E and ES galaxies are not `compact massive', that is, these 13 galaxies did not satisfy any of the four `compact massive' selection criteria. 
    The colour-coded dashed lines are the maximum number density of the red nuggets (labelled `$\rm max(n_\mathrm{RN})$') from the respective high-$z$ publications which defined the size-mass selection criteria.
    }
    \label{fig:all_mass}
\end{figure}

One of the uncertainties on our, and everyone's, spheroid number density comes from the adopted MLCRs and thus the stellar mass measurement.
For Figures~\ref{fig:massfunc} to \ref{fig:nd_z}, we presented the result using the RC15 MLCR.
Here, we will explore the consequences of applying different MLCRs. 
Figure~\ref{fig:all_mass} shows the number density of the local compact spheroids (marked with a $\bullet$) and the ellicular and elliptical galaxies (marked with a $\blacksquare$) based on the four different MLCRs presented in Section~\ref{sec:stellar_mass} and the four different selection criteria presented in Section~\ref{sec:size-mass}.
We have also highlighted the maximum number density of red nuggets from various works with colour-coded dashed lines.

Using the size-mass selection criteria from \citet{Barro2013} and \citet{Damjanov2015A}, the number densities of compact massive spheroids are higher than elliptical$+$ellicular (E+ES) galaxies, for all four MLCRs (see the upper green panel and the lower red panel in Figure~\ref{fig:all_mass}). 
Applying the more restrictive size-mass selection criteria from 
\citet{van_der_Wel2014} and \citet{van_Dokkum2015} reduces the number of compact massive spheroids included in the sample selection boundaries (see the upper-middle panel shaded blue and the lower-middle panel shaded yellow), lowering it to roughly less than or equal to the fixed number (density) of E+ES galaxies among our sample of 103 galaxies. 
When using the T11 stellar masses, no compact massive spheroid was found with the restrictive size-mass selection criteria. 
However, when using the higher IP13 stellar masses, the observed number densities of compact spheroids increased to greater than the number of E+ES galaxies, in all criteria. 

The comparison showcases how the number density varies under the different stellar mass measurement and compactness selection criteria.
Importantly, the MLCRs from T11 and RC15 assume the same IMF and SPS as used by those studying the high-$z$ red nuggets. 
The result from the less restrictive boundaries in the size-mass plane is consistent when using these two MLCRs.
With the more restrictive boundaries in the size-mass plane, and since no compact system was found using the MLCR from T11,  
we can conclude that the number density of local `compact massive' spheroids is subject to the particular $M_*/L$ ratio employed.  Resolving this, beyond having reported its impact, is clearly outside the scope of the current investigation. 

Our results are, however, robust when using the more inclusive size-mass selection criteria from \citet{Barro2013} and \citet{Damjanov2014}. 
For all stellar mass estimates, we found a consistent trend of $n_\mathrm{c,Sph} > n_\mathrm{E+ES}$. 
Considering that $n_\mathrm{c,Sph}$ is just a lower limit for the local compact massive spheroids, while 
$n_\mathrm{E+ES}$ is close to the true value, our result calls for a reevaluation of the nature of the size-evolution experienced by red nuggets.
At a minimum, disc growth has to play an essential role in shaping the high-$z$ galaxies in the lower mass range ($1\times 10^{10} < M_*/\rm M_{\odot}(\rm RC15) < 4\times 10^{10}$).

\section{Discussion}
\label{sec:Discussion}

We have established that only $\sim$11\% of the galaxies in our sample of massive galaxies are true elliptical galaxies; the majority are multicomponent systems with bulges, discs, bars, and other lesser components. 
Our trend between the bulge, {\it aka} spheroid size, and mass agrees well with data from comparable, multi-component decomposition analyses of nearby galaxies.  That is, we do not detect any systematic bias in our measurements. 
We derive a lower limit to the number density per unit volume of (hidden) compact massive spheroids 
using four different size/mass selection criteria, giving $n_\mathrm{c,Sph}\sim$ (0.17--1.20)$\times10^{-4}~\rm Mpc^{-3}$. 
We additionally examined how $n_\mathrm{c,Sph}$ changes when adopting the four different MLCRs.  
We found that if only high-mass spheroids ($M_*/\rm M_{\odot}(RC15) > 4\times 10^{10}$) are considered, $n_\mathrm{c,Sph}$ can be either lower, comparable or higher than $n_\mathrm{E+ES}$.
The uncertainty in stellar mass makes estimating the correct number density challenging.
In the case of low-mass spheroids ($1\times 10^{10} < M_*/\rm M_{\odot}(\rm RC15) < 4\times 10^{10}$), such an issue does not exist.

In what follows, we further compare local compact massive spheroids with distant red nuggets, and we address proposed size-evolution hypotheses based on the results from our findings. 

\subsection{Differing evolutionary views, definitions and assumptions}
\label{sec:difference}

\subsubsection{Alternative evolutionary processes}
\label{sec:brief_diff_theory}

Here, we provide a brief overview of different conjectures regarding the size-evolution pathway taken by high-$z$ compact massive quiescent galaxies. 

\vspace{2mm}
{\it (i) Compact massive quiescent galaxies experience rapid size evolution, an E-to-E scenario} 

\citet{Trujillo2007} plotted the half-light galaxy radii of local, luminous galaxies 
with S\'{e}rsic indices $n>2.5$. These galaxies are a factor of four larger than the compact spheroidal-like systems at $z\sim1.5$. From this, they concluded that the $z\sim1.5$ compact massive galaxies evolved through dry mergers with themselves, i.e., equal-mass major mergers without star formation. 
\citet{van_Dokkum2008} reported a typical growth in galaxy half-light radii by a factor of 5--6 since $z\sim2.3$.
They found that only 10\% of the massive spheroid-like galaxies at $z\sim2.3$ had galaxy sizes equal to that of elliptical galaxies of comparable mass at $z\sim0$.  They, therefore, argued against `monolithic collapse' as the formation channel because they found only 10\% are fully assembled by $z\sim2.3$.  
They noted an array of processes to possibly explain the size growth, including dry mergers, satellite accretion, expansion after mass loss from stellar winds, and the arrival of new, larger galaxies since $z\sim2.3$ onto the red sequence.

The notion of major mergers driving size evolution has been heavily challenged.
The observed number of major mergers (1:1 to 1:4 mass ratios) is too low to explain the change in number density of massive ($\log_{10}(M_*/\mathrm{M_{\odot}})\geq11.1$) galaxies \citep{Man2016}.
\citet{Taylor2010} found no galaxies in the SDSS DR7 with comparable sizes to the \cite{van_Dokkum2008} and \cite{Damjanov2009} red nuggets at higher redshifts.  
They also argued against {\it stochastic} major mergers as the critical driving force for the size evolution in early-type galaxies because it will result in some galaxies (those which experienced many mergers) having masses larger than observed in the local Universe. 
\citet{Carrasco2010}, who studied eight massive galaxies at $1 < z < 2$ from the POWIR survey \citep{Conselice2007}, reported extreme compactness within $R<2$~kpc (higher than the stellar densities of local galaxies).
With similar reasoning as \citet{Taylor2010}, they advocated that minor mergers and accretion are more likely the cause of the size growth.

While some kind of E-to-E scenario is widely accepted in the community \citep[e.g.,][]{Buitrago2008,Bezanson2009, Hopkins2009,Szomoru2011,Newman2012}, 
the mechanism which supports a dramatic E-to-E size growth remains a subject of debate.
\citet{Fan2008} had proposed that through mass loss (stellar winds, AGN jets), galaxies experience adiabatic expansion as the new gravitational equilibrium relaxes.
However, adiabatic expansion has been called into question by numerical simulations, as \citet{Ragone-Figueroa2011} found galaxies `puffing-up' due to stellar winds occurs
when the stellar population is younger than the formation of the red nuggets (by 0.5\,Gyr).
\citet{Hopkins2009} and \citet{Bezanson2009} advocated that, through minor mergers and accretion, red nuggets may build up a low-density three-dimensional envelope which hides the compact galaxies as the core of a larger {\it elliptical} galaxy \citep[see also][]{Hopkins2010, Szomoru2011,Lopez-Sanjuan2012,Oogi2013}.

\vspace{2mm}
{\it (ii) Compact massive quiescent galaxies experience only a mild, or no, evolution and now reside in local galaxy clusters: an E-to-E/S0 scenario.} 

(Galaxy cluster)-focused studies at intermediate-$z$ have challenged the universality of the significant size-growth scenario. 
\citet{Valentinuzzi2010} and \cite{Saracco2010} report on the abundance of `compact' galaxies in cluster environments and postulate that red nuggets live in dense environments nowadays.
Similarly, \cite{Poggianti2013B} found that the fraction of local compact massive galaxies are three times higher in clusters than in the field, 
arguing for the scenario of mild evolution (by a factor of $\sim$1.6 in size) for high-$z$ red nuggets.  
\citet{Damjanov2015B} echo the same point as they subsequently found compact massive quiescent galaxies at $0.1  < z < 0.4$ in the COSMOS field \citep[][]{Scoville2007} resided in denser environments than other quiescent galaxies.
\cite{Carollo2013} argued that the addition of more large newly quenched early-type galaxies, as time progresses, is responsible for the apparent size evolution \citep[a.k.a., the `progenitor bias',][]{van_Dokkum1996,Saglia2010}. 
In this picture, the size growth of quiescent galaxies since $z\sim2$ is of secondary importance.

\vspace{2mm}
{\it (iii) Compact massive quiescent galaxies experience size evolution via disc growth: a disc-cloaking (E-to-S0/S) scenario} 

\citet{2013pss6.book...91G} and 
\cite{Graham2015} expanded the possible evolutionary channels, where instead of only considering elliptical (E) galaxies as the end product of red nugget evolution, lenticular (S0) and early-type spiral (Sa) galaxies may arise by cloaking the compact spheroid with a large-scale disc built through minor mergers and gas accretion.
Upon entering a cluster of galaxies with a hot X-ray gas halo, any future disc growth via gas accretion and star formation would naturally be curtailed. 
The hypothesis of disc-cloaking was supported by a reported lower-limit to the number density of $n\sim6.9\times 10^{-6}\,\rm Mpc^{-3}$, based on 21 local compact massive spheroids within 90\,Mpc \citep{Graham2015}.
It was a lower-limit because the volume was not sampled completely; the 21 systems simply represented spheroids that had been reported in the literature as having small sizes and high masses and were known to the authors. 
Their finding was further supported by \cite{de_la_Rosa2016} using the automatic Bulge+Disc decompositions of local galaxies imaged by SDSS \citep{Mendel2014}.
From a sample of SDSS DR7 galaxies and a survey area of $8,032\,\rm deg^2$ at $0.025< z <0.15$, they found a number density of local compact bulges ($n\sim 0.28$--$3.12 \times10^{-4}\,\rm Mpc^{-3}$) comparable with the number density of the distant red nuggets across multiple size-mass sample-selection criteria \citep[][]{Barro2013,van_der_Wel2014,van_Dokkum2015}.
Because galaxies can have (truncated) non-exponential discs and more than two components, such as bars and rings, etc., automated Bulge+Disc decompositions often fail to provide a reliable measurement of the bulge size and luminosity, which is why we undertook our careful investigation.

\subsubsection{Differing selection criteria and assumptions}
\label{sec:diff_assumption}

The criteria of selecting for the parent sample, measurement of the galaxies' properties, and the definition of compactness, vary from one study to another.
It will be a monumental task to homogenise all the data presented in the literature.
Instead, we provide a brief description of various approaches and assumptions commonly made (see Table~\ref{tab:assumptions}), and we highlight a few potential causes of discrepancy.

\begin{table*}
   \caption{The underlying characteristics of different surveys spanning a range of redshifts.}   
   \begin{threeparttable}[b]
   \label{tab:assumptions}
   \begin{tabular}{llllll}
   \hline
   Data source & Unique selection & Galaxy model & Photometry & Adopted IMF &  Lower-mass \\
    &  conditions &  &   &  &  limit ($M_*/\rm M_{\odot}$) \\ 
   (1) & (2) & (3) & (4)  & (5) & (6) \\
   \hline
    $z\sim0$ \\
    \hline
   \citet{Bezanson2009} & `E'~$^\mathrm{[a]}$	&  $R^{1/n}~(n\leq4)$$^\mathrm{[a2]}$ & $V$-band & \citet{Kroupa2001} &  $10^{11}$\\  
     \citet{Trujillo2009} & `Q' \& `SF'~$^\mathrm{[b]}$	&  $R^{1/n}$ & $r$-band & \citet{Kroupa2001} & $8\times10^{10}$ \\
    \citet{Taylor2010} &  `Q' \& `SF'~$^\mathrm{[c]}$ &  $R^{1/4}$  &  $z'$-band & \citet{Chabrier2003} & $ 5\times 10^{10}$ \\
    \citet{Poggianti2013B} &   `Q' \& `SF'~$^\mathrm{[d]}$	&  $R^{1/n}$ & $H$-band & \citet{Kroupa2001} & $ 10^{10}$  \\
   \hline
   $0.2 < z < 1.0$ \\
   \hline
    \citet{Valentinuzzi2010} &  `Q' \& `SF'~$^\mathrm{[e]}$	&	 $R^{1/n}$  & $V$-band & \citet{Salpeter1955,Kroupa2001} & $5\times 10^{10}$ \\
    \citet{Carollo2013} & `E/S0' \& `Sa-cd'~$^\mathrm{[f]}$	&  $R_{1/2}$$^\mathrm{[b2]}$ & $I$-band & \citet{Salpeter1955} &  $ 3.1 \times 10^{10}$ 
    \\     
    \citet{Damjanov2015A} & ---	&	 $R^{1/n}$  & $I$-band  & \citet{Kroupa2001} & $ 1$--$5\times10^{10}$  \\
    \hline
   $1.0 < z < 3.0$ \\
   \hline
   \citet{Daddi2005}  & ---	&  $R^{1/n}$ & $i$-,$z$-band& \citet{Salpeter1955} & $10^{11}$ \\
    \citet{Trujillo2007} & $n=2.5$ divide~$^\mathrm{[g]}$	&  $R^{1/n}$ & $I$-band & \citet{Chabrier2003} & $10^{11}$ \\
   \citet{van_Dokkum2008} & ---	&  $R^{1/n}$ & $H$-band  & \citet{Kroupa2001} & $10^{11}$ \\  
   \citet{Damjanov2009} &  ---	&  $R^{1/4}$ or $R^{1/n}$ & $H$-band & \citet{Baldry2003}~$^\mathrm{[a3]}$ &  $2.5\times10^{10}$ \\  
     & ---	&  $R^{1/4}$ & $H$-band & \citet{Salpeter1955}~$^\mathrm{[b3]}$&  --- \\   
    \citet{Saracco2010} &  $n > 2$~$^\mathrm{[h]}$	&  $R^{1/n}$ & $z'$-band & \citet{Chabrier2003} &  $3\times10^{10}$ \\ 
    \citet{Cassata2011} & `Spheroidal`~$^\mathrm{[i]}$	&	 $R^{1/n}$  & $z$-,~$H$-band   & 	\citet{Salpeter1955} &  $10^{10}$ \\    
    \citet{Barro2013} &  ---	&	 $R^{1/n}$  & $H$-band  & 	\citet{Chabrier2003} &  $10^{10}$ \\
    \citet{van_der_Wel2014} &  ---	&	 $R^{1/n}$   & $H$-, $J$-band	 & \citet{Chabrier2003} &  $5\times 10^{10}$ \\
     \citet{van_Dokkum2015} & ---	&	 $R^{1/n}$   & $H_{160}$-band & \citet{Chabrier2003} &  $4\times 10^{10}$ \\
   \hline
   \end{tabular}
      \underline{Columns}: 
      (1) Data source; 
      (2) special conditions in the parent sample selection other than (i) low star-formation rate, (ii) stellar mass, and (iii) redshift; see the respective footnotes for more detailed descriptions.
      (3) size measurement model: $R^{1/4}$ is the \citet{1948AnAp...11..247D} profile and $R^{1/n}$ is the \citet{Sersic1968} profile; 
      (4) photometric passband in which the size measurement is taken; 
      (5) assumed initial mass function (IMF); and 
      (6) the lower-mass limit for the compact criteria. If no clear size-mass selection criteria were used, we show the lower limit in the parent sample selection instead. 

   \begin{tablenotes}
      \small
      \item
      \underline{Footnotes}:
      [a] \citet{Bezanson2009} selected the parent sample at $z\sim0$ based on morphologically `E' type galaxies from \citet{Tal2009} (see their section~2.1.2);
      [b], [c], [d], and [e]: In their parent sample selection, \citet{Trujillo2009}, \citet{Taylor2010}, \citet{Poggianti2013B} and \citet{Valentinuzzi2010} did not discriminate between quiescent (`Q') and star-forming (`SF') galaxies;
      [f]: \citet{Carollo2013} use the non-parametric ZEST+ morphological classification algorithms to select for `E/S0' and `Sa-cd' galaxies to construct the parent sample;
      [g]: \citet{Trujillo2007} divided their sample into two groups: the `spheroid-like' (S\'{e}rsic index $n>2.5$) and `disc-like' ($n<2.5$);
      [h]: \citet{Saracco2010} selects for galaxies from \citet[][]{Giavalisco2004} and excluded those with S\'{e}rsic index $n<2$;
      [i]: \citet{Cassata2011} select only the galaxies with a spheroidal morphology under visual inspection, which are the `galaxies with no signs of asymmetry and centrally concentrated' (see their section~2);
      [a2]: \citet{Bezanson2009} limited the fitting range of the S\'{e}rsic index $n \leq 4$ for their local galaxy sample (see their section~2.2);
      [b2]: \citet{Carollo2013} measured the half-light radius of each galaxy numerically (see their section~3.1).
      Here, we denote such by `$R_{1/2}$';
      [a3], [b3]: The stellar mass of \citet{Damjanov2009}'s GDDS and MUNICS red nuggets are calculated assuming the \citet{Baldry2003} and \citet{Salpeter1955} IMFs, respectively.
      \begin{flushleft}
      \end{flushleft}
    \end{tablenotes}
    \end{threeparttable}
\end{table*}

\vspace{2mm}
\noindent{\it Parent sample}

The majority of studies select their parent galaxy set based on 
the galaxies' stellar masses, redshifts, and star formation (or its proxies, e.g., colour-colour selection). 
Sometimes the sample is further restricted by morphology or brightness concentration (S\'ersic index). 
Column~(2) in Table~\ref{tab:assumptions} summaries such unique conditions for several studies. 
\citet{Trujillo2007} divided their sample of luminous galaxies into `spheroidal' (S{\'e}rsic index $n > 2.5$) and `disc-like' (S{\'e}rsic index $n < 2.5$) to illustrate the stronger size evolution in spheroids, and the lack thereof in disc galaxies.
The \cite{Bezanson2009} local sample is morphology-dependent, using the apparent `E' galaxies from \cite{Tal2009}.
Some studies did not construct the parent sample based on star formation \citep{Trujillo2007,Trujillo2009,Taylor2010}, where both quiescent and star-forming galaxies are considered.
The sample from \citet{Carollo2013} is based on the ZEST$+$ morphology classification \citep[an upgraded version from][]{Scarlata2007}, from which they select both massive early-type (`E/S0') and late-type (`Sa--cd') galaxies.
Studies that have a parent sample inclusive of all morphologies would not be biased by an E-to-E scenario presumption. 
Although, depending on the lower-mass threshold in their `compactness' criteria, individual works might be probing a different population of red nuggets. Indeed, for instance, applying a higher mass selection from \citet[][]{Trujillo2009} ($M_*/\rm M_{\odot} \sim 8\times10^{10}$) would return a lower number density compared to that of \citet{Poggianti2013B} ($M_*/\rm M_{\odot} \sim 10^{10}$).

\vspace{2mm}
\noindent{\it Size measurements}

The method of size measurement varies slightly between studies.
The most commonly used method is fitting a single S\'{e}rsic $R^\mathrm{1/n}$ model to the surface brightness profile of the galaxies \citep[e.g.,][]{Daddi2005, Trujillo2007, Trujillo2009,Bezanson2009,Valentinuzzi2010, Cassata2011, Mancini2010, Saracco2010, Poggianti2013A,Barro2013,van_der_Wel2014,van_Dokkum2015}.
From this, the effective radius $R_\mathrm{e}$ (essentially the half-light radius $R_{1/2}$) is generally used to represent the size of a galaxy in discussions. 
Some studies, however, fit the \citet{1948AnAp...11..247D} $R^{1/4}$ model instead.
For example, \citet{Taylor2010} rely on the parametric measurements from  \citet{Strauss2002}, who used an $R^{1/4}$ model to describe the galaxy surface brightness profiles (see also section~3.2 in \citealt{Taylor2010}, for a relevant discussion).
For red nuggets, \citet{Damjanov2009} fit the $R^{1/4}$ model to their MUNICS \citep[][]{Drory2001} data and both $R^{1/4}$ and $R^{1/n}$ models to their GDDS \citep[][]{Abraham2004} data.
If one assumes an $R^{1/4}$ model, the effective radius could be overestimated if the galaxy has a S\'{e}rsic index $n < 4$.
In Column~(3) of Table~\ref{tab:assumptions}, we list the measurement method.

The colour of an image also affects its size measurement \citep{2012MNRAS.421.1007K, 2014MNRAS.441.1340V, 2016MNRAS.460.3458K}. 
In Column~(4) of Table~\ref{tab:assumptions}, we show the band in which the study was conducted.
Importantly, the size of early-type galaxies is smaller in the red band as compared to the blue band.  As seen in \citet{La_Barbera2010}, there is a size decrease by 30\% from $g$- through $K$-band, and a 25\% decrease from the $H_\mathrm{160}$ to the $r_\mathrm{625}$ filter \citep{Marian2018}.

\vspace{2mm}
\noindent{\it Stellar mass}

The galaxy stellar mass can be estimated via broadband SED fitting \citep[e.g.,][]{Daddi2005,Longhetti2007,Kriek2008,Barro2013}\footnote{Among the works listed in Table~\ref{tab:assumptions}, only \citet{Poggianti2013B} did not perform SED fitting. Their data source from the PM2GC survey \citep{Calvi2011} estimates the galaxy stellar mass via the \citet{Bell2001} $M/L$ relation.}.
The underlying assumptions for such an operation differ between studies and,  as we have shown, the stellar mass calculation affects the reported number density.
One such factor is the assumed stellar population model.
Most works in Table~\ref{tab:assumptions} use the \citet{Bruzual2003} stellar population model, except for few studies:
\citet{Valentinuzzi2010} uses the \citet{Maraston2005} model, where the influence of thermally-pulsating asymptotic giant branch (TP-AGB) stars are considered; 
\citet{Saracco2009} and \citet{Cassata2011} use the updated Charlot \& Bruzual models from 2008 (CB08) and 2009 (CB09), respectively.
According to \citet{Wuyts2007}, assuming a \citet{Maraston2005} model over the \citet{Bruzual2003} model will result in a factor of 1.6 decrease in stellar mass.
Meanwhile, \citet{Salimbeni2009} found a 0.2\,dex decrease in estimated stellar mass if one adopts CB09 rather than the \citet{Bruzual2003} model.

Another factor is the choice of the initial mass function (IMF).
Here, we only show, in Column~(5) of Table~\ref{tab:assumptions}, the IMF used in the respective studies for reference purposes.  
The \citet{Chabrier2003} IMF is the most widely-used and also the most bottom-light compared to the other contemporaries. 
The CANDELS-based \citep{Grogin2011,Koekemoer2011}  size evolution studies \citep[][]{Barro2013,van_der_Wel2014,van_Dokkum2015} all assume the  \citet{Chabrier2003} IMF.
However, some low-to-intermediate redshift studies \citep[][]{Bezanson2009,Valentinuzzi2010,Poggianti2013B,Carollo2013} instead assume the more bottom-heavy \citet{Salpeter1955} IMF, or the pseudo-Kroupa IMF \citep{Kroupa2001} rather than the \citet{2002Sci...295...82K} IMF.
Using a \citet{Chabrier2003} IMF will yield 0.24 dex less mass than the  \citet{Salpeter1955} IMF \citep{Salimbeni2009}.
A bottom-heavy IMF could result in higher stellar mass, and therefore, more spheroids will be included by the `compact massive' selection criteria. 

\vspace{2mm}
\noindent{\it Compactness criteria}

As discussed in Section~\ref{sec:size-mass}, the compactness criteria are usually somewhat arbitrary or it borders the distribution of the general galaxy population's size-mass relation.
The lower-mass limit, in particular, can alter the number density of `compact' systems in one's sample. 
We have listed the lower-mass limit for each study in Column~(6) of Table~\ref{tab:assumptions}.
If the study did not apply a clear size-mass selection boundary, the lowest mass limit (corresponding to the magnitude limit) of the parent sample is provided. 
These lower-mass limits vary from $M_*/\rm M_{\odot} = 10^{10}$ to $10^{11}$. 
If a study assumes a bottom-heavy IMF and a lower-mass limit, one can expect a higher reported number density.

\subsection{The possibility of E-to-E evolution}
\label{sec:E-to-E}

Our work is fundamentally different from the previously mentioned works. 
By not treating galaxies as single-component systems, our analysis provides additional information on the galaxies' structural features. 
Based on the RC3 morphology, the frequency of alleged `E' galaxies in our local sample is abundant.
However, after performing analysis and decomposition of the galaxy light, the number of elliptical galaxies is revised downward, already weakening the argument that the high-$z$ red nuggets transformed into today's elliptical galaxies, simply because the actual number of elliptical galaxies is sparse. 
Only 13/28 originally labelled `E' galaxies are true ellipticals (11 E $+$ 2 ES, see Section~\ref{sec:morph_reclass}).  This reflects revelations from a quarter of a century ago, whereby the true number of pressure-supported elliptical galaxies was found to be notably lower than previously thought \citep[e.g.,][]{D'Onofrio_1995, Graham1998}. 

\subsubsection{The morphology of host galaxies with an embedded compact massive spheroid}
\label{sec:transistion}

The literature abounds with examples of what were formerly elliptical galaxies that are reclassified as disc galaxies.  Early examples are NGC~4111 \citep[E7 $\rightarrow$ Sa:][their p.31]{1937MWOAR...9....1A} and 
NGC~3115 \citep[E7 $\rightarrow$ S0:][]{1961ApJ...133..393B}.

\begin{figure}
\centering
	\includegraphics[clip=true, trim= 3mm 1mm 5mm 6mm, width=\columnwidth]{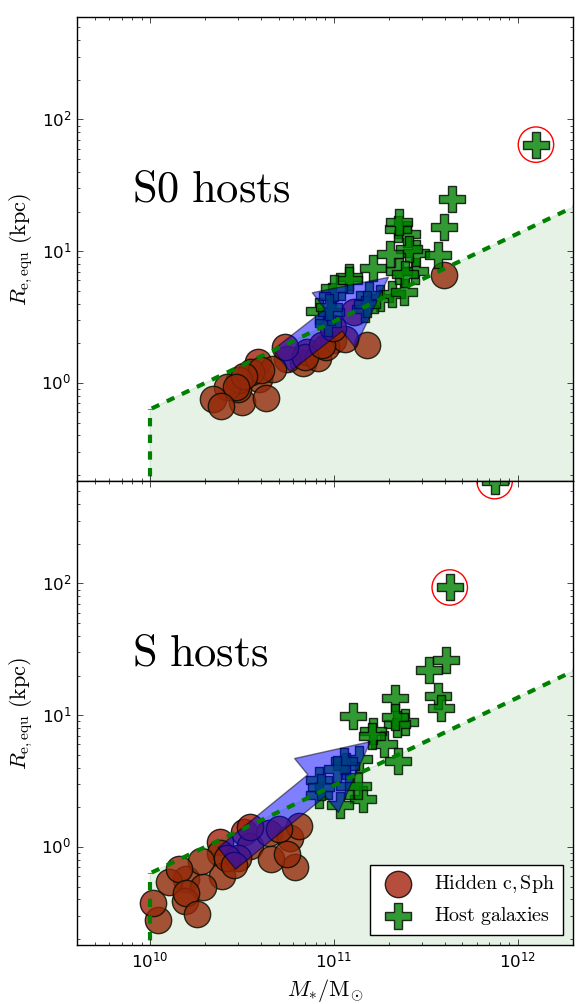}
    \caption{The size-mass (in RC15 mass) distribution of 55 compact massive spheroids according to the size and mass selection criteria shown in Equation~\ref{eq:Barro_cut} (green dashed line and the shaded area). 
    The host galaxies' morphologies are based on their reclassifications in Section~\ref{sec:morph_reclass}.
    The sample is separated into `S0 hosts' (upper panel) and `S hosts' (lower panel).
    The red solid circles are the compact massive spheroids hidden within the host galaxies (labelled `Hidden c,Sph') and the green plus signs are the host galaxies (labelled `Host galaxies').
    The size (circularised radius $R_\mathrm{e,eq}$) of the host galaxy is obtained via a single S\'{e}rsic fit to the surface brightness profile.
    The blue arrows indicate the direction of growth if the high-$z$ compact massive system were to assemble mass via the inside-out, disc cloaking process.
    We highlighted the three prominent outliers with red hollow circles.}
    \label{fig:morph_trans}
\end{figure}

Our results cast doubt on the evolutionary scenario of red nuggets transforming into elliptical galaxies.
In Figure~\ref{fig:morph_trans}, we plot the geometric-mean, i.e., circularised, half-light radius and total mass of the host galaxies, shown by the green plus signs (`$+$'), and their respective embedded spheroids (shown by the solid red circles). 
The three prominent outliers are marked with red hollow circles\footnote{For the sake of mimicking how most studies measure galaxy size, we use a single S\'{e}rsic function to measure the host galaxy's effective radius, although we know it is not the best measurement for some galaxies (see also Appendix~\ref{sec:Gal_can}). Case in point, the three outliers are NGC~3158 (S0), IC~983 (S), and NGC~3646 (S).
All have a disc component with a rather shallow profile and large scale length $h>20\,\rm kpc$.
Combined with a prominent bulge, this creates a curvature in the surface brightness profile that requires a single S\'{e}rsic function with an exceptionally high index ($n>5$).
This highlights a problematic issue with the conventional method of measuring galaxy size and shall be pursued further in future work.}.
The sample shown here are the 55 spheroids that satisfy the compact-massive selection criteria of \citet{Barro2013}, the most inclusive criteria.
We divided the sample into two categories, they are:
\begin{enumerate}
    \item
    spheroids embedded in lenticular galaxies\footnote{The morphologies here refers to the reclassification we introduced in Section~\ref{sec:morph_reclass}.} (`S0 hosts', see the upper panel in Figure~\ref{fig:morph_trans}); and
    \item
     spheroids embedded in spiral galaxies (`S hosts', see the lower panel in Figure~\ref{fig:morph_trans}).
\end{enumerate}

None of the compact massive spheroids, as identified using the selection criteria of \citet{Barro2013}, are pure elliptical galaxies.
A significant number of compact massive spheroids with stellar masses above (4--5)$\times 10^{10}$\,M$_\odot$ come from S0 host galaxies, while those with stellar masses lower than (4--5)$\times10^{10}$\,M$_\odot$ are slightly more likely to come from a spiral host galaxy. 
Additionally, none of the 11 E $+$ 2 ES galaxies (when using the  T11, Z09, or RC15 mass prescriptions) satisfy the `compact-massive' selection criteria of any paper\footnote{Only when applying the highest stellar mass estimation (IP13), did we find 3, 3, and 4 `compact massive' galaxies when using the \citet{Barro2013}, \citet{Damjanov2014}, and \citet{van_Dokkum2015} selection criteria, respectively. However, because IP13 assumes a different IMF compared to these studies, it is not exactly a fair comparison.}. 
What this means is that all of our compact massive spheroids are embedded inside either S0 or early-type S galaxies; none are compact massive elliptical \textit{galaxies}, although a few such galaxies do exist locally. 

\subsubsection{3-D envelopes}
\label{sec:3Denvelope}

\citet{Hopkins2009} proposed an evolutionary pathway for red nuggets, where, through dry minor mergers, the high-$z$ compact galaxies develop a pressure-supported 3-D envelope to become the cores of today's elliptical galaxies. 
This is a different evolutionary path to acquiring a rotating 2-D disc to become the bulges of today's lenticular galaxies.  
The answer to differentiating between these two evolutionary paths depends on understanding the surface density profiles of massive local galaxies as a function of radius. 
One can also resort to kinematic information to ascertain the presence of a rotating disc. 
\citet{Hopkins2009} investigated 180 early-type galaxies from \citet{Lauer2007} and \citet{Kormendy2009ApJS}, and found similarities between the inner part, $\sim$(1--$5) \rm \, k p c$, of the local early-type galaxy light profiles and the distant compact massive quiescent galaxies in the sample of \citet{van_Dokkum2008}.  
Earlier in the same year, \citet{Bezanson2009} had reached a similar conclusion, stating that the inner ($R<1$~kpc) stellar densities of local elliptical galaxies are similar to the high-$z$ red nuggets. 
Possibly due to the difficulty distinguishing E from S0 among ETGs, papers advocating for an E-to-E evolution kept emerging \citep[e.g.][]{Oldham2017,Tortora2020,Zhu2021}, despite the fact that the similarity of the profile's inner density did not actually prove the proposed E-to-E evolution over an E-to-S0 evolution. 
\citet{Arnold2011} suggested minor mergers were building a pressure-supported halo around NGC~3115. 
\citet{De2014PASA} reported structural similarity in the inner part at radius $R < 1\,\rm k p c$ of local early-type galaxies with the red nuggets at $2 < z < 7$, and virtually no correlation at $R > 10 \, \rm k p c$.
However, information on the $z\sim0$ number density of S0 galaxies with compact massive cores is needed. 
This required identifying which early-type galaxies are S0 or E, or ES.

While the studies mentioned above provide a legitimate argument for the inside-out growth scenario, our result provides additional insight into the nature of this evolution.
Those studies generally did not attempt to separate the early-type galaxies into a disc (lenticular) and no-disc (true elliptical) systems. By and large, they assumed evolution into larger pressure-supported systems rather than entertaining the notion of substantial disc growth to create a different type of galaxy with significant ordered motion.  
\citet{Barbosa2021} conducted a detailed kinematic and population analysis on a bona fide elliptical, NGC~3311, and found a compact core, a high-$z$ `relic' galaxy, is hidden inside $R<2\rm~kpc$. Similar studies would be much welcomed to confirm an E-to-E transition.
Our bulge-to-disc and disc-to-total ratios reveal that the discs, and their bars and ansae, are nowadays the dominant mass component in many local massive galaxies. This has considerable implications on many fronts, such as the inferred dark matter halos of massive early-type galaxies implied from their central velocity dispersion, $\sigma$, via the expression $M\sim\sigma^2 R_{\rm e,gal}$ \citep[e.g.,][]{1911lhcp.book.....P, 1937ApJ....86..217Z, 1958BOTT....2q...3P, 1961ApJ...134..910P}. Obviously, the larger $R_{\rm e,gal}$ radii, which are dominated by the size of the disc, is not applicable for use in this expression for a virialised, pressure-supported galaxy.  

\citet{Hopkins2009} used data from \citet{Kormendy2009ApJS}, a selected set of 27 elliptical (E) galaxies from the Virgo Cluster and with very few massive galaxies ($M_* > 3 \times 10^{11}\,{\rm M_\odot}$). 
They modelled all the E galaxies with a single S{\'e}rsic fit.
However, upon inspection, some of these galaxies exhibit features which may be a signature of a rotational component. For instance, there is a rising ellipticity $\epsilon$ profile in IC~3381, IC~3490, IC~3509, NGC~4387, NGC~4482 NGC~4486 and NGC~4636, a typical sign for an extended disc; high ellipticity section ($\epsilon>0.4$) in the $\epsilon$ profile: NGC~4464, NGC~4467, NGC~4478, NGC~4515, and NGC~4621 which could indicate the existence of intermediate scale disc. 
\citet{Hopkins2009} also used data from \citet{Lauer2007} based on 219 early-type galaxies taken from different sources  \citep{Lauer1995,Faber1997,Quillen2000,Ravindranath2001,Rest2001,Laine2002,Lauer2005}.
In terms of morphologies, roughly half of the sample was thought to be `E', a quarter of them `S0', and the other quarter `BCG' according to the RC3 classifications.
Most BCGs are ellipticals, yet without a multi-component analysis of kinematic data, the nature of the alleged `E' galaxies is not confirmed.

The local galaxy sample used by \citet{Bezanson2009} came from \citet{Tal2009}, with well-defined distances ($ 15\, \rm M p c< Dist. < 50\, \rm M p c$) and absolute magnitude ($\mathfrak{M}_B < - 20$\,mag).
They pointed out half of their sample exhibits morphological disturbance features, such as shells and tidal tails, implying heavy interactions with their environments.  

\citet{De2014PASA} supports the 3-D envelope scenario based on the three-S\'{e}rsic-component decompositions by \citet{Huang2013} on the Carnegie-Irvine Galaxy Survey data \citep{Ho2011}.
The 94 galaxies they investigated used to model listed as `E' in the RC3 classifications.
\citet{Huang2013} used three S\'{e}rsic functions to describe the inner, middle, and outer region of the galaxy.
They found a high correlation between the inner region of the local galaxies and the high-$z$ red nuggets but not the outer region, hence concluding the red nuggets went through a multi-stage buildup in an inside-out manner. 
This, however, does not rule out disc growth.
Indeed, as \citet{Huang2013} pointed out in their Table 1, the following galaxies have an edge-on disc: ESO 221-G026, NGC 3585, and NGC 7029, and these: IC~4797, NGC 584, NGC~3904, NGC 4033, NGC 4697, NGC~6673, NGC~7145, NGC~7192, NGC~7507 are possibly S0 galaxies.
Instead of a multi-layered elliptical galaxy, there could be many S0 galaxies in the mix, and the inner region of these galaxies is, in fact, the bulge. 

As we have established in Section~\ref{sec:morph_reclass}, without exploring the galactic substructure, the morphology of early-type galaxies is often mistaken.
Instead of comparing the high-$z$ quiescent galaxies to true elliptical galaxies, the comparisons are essentially being made using many local lenticular galaxies.
It is likely that many of the alleged low-density 3D envelopes are actually 2D rotating discs seen somewhat face-on. 
Rather than evidence of 3D envelope growth, many of the galaxies used in the above studies instead support the 2-D disc cloaking process.

\subsection{Disc growth} 
\label{sec:disk_growth} 

\subsubsection{Evidence of disc accretion}
\label{sec:disk_support}

Lyman-$\alpha$ gas clouds are observed within 500\,kpc of high-$z$ galaxies of all types \citep{Wakker2009, Prochaska2011, Thom2011, Stocke2013, Tumlinson2013}.
For instance, \citet{Bouche2013} used a background quasar to trace the nearby gas (damped Lyman absorber $26\, \rm kpc$ away) of a star-forming galaxy at $z=2.3$ with a typical rotational disc, according to its kinematics \citep{Forster_Schreiber2009}.
The kinematics of the gas shows a signature of cold accretion, where it appears to be low-metallicity, co-planar, and co-rotational compared to its disc.
In the local Universe, \citet{Coccato2013} examined the two counter-rotating discs in both NGC~3593 (S0/a) and NGC~4550 (S0), where the secondary discs differ from the primary discs in metallicity and stellar populations.
They favour a scenario of gas accretion forming the second disc $\sim$2--7$\, \rm G yr$ ago.

High-$z$ red nuggets are reported to have reached their peak abundance at $z \sim 1.2$--$1.6$, and dropped in number ever since \citep[][]{Barro2013,van_der_Wel2014,van_Dokkum2015}. 
The disappearance of these galaxies through the growth of discs may involve the creation of star-forming galaxies (should much of the disc stars be built from accreted gas) or not (when the bulk of the disc is built from accreted galaxies with little gas). 

There is an abundance of evidence from different approaches that confirm the prominence of discs in local early-type galaxies.
\citet{Kaviraj2007} investigated the UV and optical photometry of $\sim$2100 SDSS early-type galaxies at $ 0<z<0.11$ and found at least $\sim$30\% exhibit recent ($\sim$1\,Gyr ago) star formation activity.
\citet{Fabricius2014} performed a detailed analysis of NGC~7217, a spheroid-dominated galaxy (classified as an early-type spiral in \citealt{Buta1995}) with two distinct rotational components.
The active star formation of the rotational structures suggested an ongoing (re)growth of the stellar disc.
While this process is truncated in cluster environments, it can still occur in the field \citep[e.g.][]{Graham2017}.

While `cosmic noon' has passed, and there are less substantial cold gas clouds nearby for galaxies to accrete, it is an ongoing, downsizing phenomenon. 
The kinematics of the gas shows a signature of cold accretion, where it appears to be low-metallicity, coplanar, and corotational compared to its disc \citep[see also,][]{2013ApJ...769...74S, 2010ApJ...711..533K, 2019NatAs...3..822M, 2019MNRAS.485.1961Z, 2015ApJ...815...22K, 2017ApJ...834..148N}. 
\citet{Davis2011} examined the gaseous content of local early-type galaxies with CO and H\,\textsc{i} interferometric observations.  
The kinematic misalignment between the gas and stellar components implied that $42\%$ of the gas in field galaxies (fast rotators) is \textit{ex-situ} in origin, namely from cold accretion and minor mergers\footnote{Note that only 22\% of early-type galaxies in their survey has detectable molecular gas \citep{Young2011}. The gas content in local early-type galaxies is evident but not at all prominent.}.
\citet{Alatalo2013} investigated the morphology of molecular gas through CO imaging by the $\rm CARMA~ATLAS^{3D}$ survey, finding half of the 40 CO-rich early-type galaxies exhibit a CO disc.
They estimated (with a correction factor from \citealt{Kaviraj2012}) the lower-mass limit to the total accreted gas to be  $2.48 \times 10^{10}\, \rm M_{\odot}$ across 15 galaxies, a value consistent with the mass brought in by minor mergers \citep{Lotz2008} and thus the accretion of not just gas but also stars from smaller galaxies. 

As mentioned before, the extensive kinematic observations from the $\rm ATLAS^{3D}$ survey revealed that the majority of the local early-type galaxies are `fast rotators' \citep{Emsellem2011}. 
\citet{Krajnovic2011} shows that most early-type galaxies ($\sim$82\%) are `regular rotators' (defined in \textit{Kinemetry}, \citealt{Krajnovic2006, Krajnovic2008}), in the sense that the velocity maps are dominated by ordered rotation.
\citet{Krajnovic2013} performed Bulge+Disc decompositions on a set of $180$ of the non-barred galaxies.
They found that galaxies with high angular momentum ($\lambda_\mathrm{e}$) also have a large disc-to-total flux ratio.
The combination of kinematics and photometric results revealed that $83\%$ of their early-type galaxies have a rotational disc that accounts for around $\sim$40\% the stellar mass of a galaxy.

Our result highlighted the abundance of discs in massive local galaxies, providing strong support for the disc growth (cloaking) scenario.
The cyclical nature of galaxy metamorphosis has been recognised widely in simulations \citep{White1978, White1991, Navarro1991, Steinmetz2002, Bournaud2002}.
If disc cloaking were to occur to our spheroids, and, subsequently, build up an extended disc and other substructures in an inside-out manner, the systems will experience growth in both size and mass.
They would migrate outward (see the blue arrows, in Figure~\ref{fig:morph_trans}) from the compact massive region (the green shaded area) in the size-mass diagram and become the massive ($M_*/\rm M_{\odot} > 10^{11}$) local S0 and S galaxies.
Depending on the nature of the minor mergers (wet or dry) and the amount of cold stream gas accretion, the galaxies may also migrate outside of the `quiescent' region in the colour-colour diagram to enter the `star-forming' blue region because of the newly built rotational discs.
With time, star formation in these discs turns off, and a quenched, lenticular galaxy now moves back into the `quiescent' region with a larger overall size.

\subsection{Important caveats}
\label{sec:caveat}

\subsubsection{The role of IMF in stellar mass estimation}
\label{sec:IMF}

It is questionable whether there is a universal IMF among galaxies \citep{Cenarro2003, 2010ARA&A..48..339B, van_Dokkum2010IMF, Cappellari2012IMF, Ferreras2013, La_Barbera2013, 2014MNRAS.443L..69S, Spiniello2014}.
The slope of the low-mass end of the IMF is thought to be directly correlated with the velocity dispersion of galaxies \citep{Ferreras2013, Dominguez-Sanchez2019} and, consequently, the colour of galaxies \citep{Dutton2012, Pforr2012, Ricciardelli2012, Vazdekis2012}.
For example, \citet{2015MNRAS.451.1081M} analysed the `relic' galaxy NGC~1277 \citep{2016ApJ...819...43G} that is compact (1--2\,kpc) and has an exceptionally high velocity dispersion ($\sigma > 400\, \rm k m \,s^{-1}$, see also \citet{Ferre-Mateu2017} for its detailed features).
However, the IMF correlation with velocity dispersion has been called into doubt.
Importantly, from one BCG, NGC~3311, \citet{Barbosa2021IMF} presented an important case showing the IMF-$\sigma$ correlation to be invalid, a mere coincidence that old stars are found in the area of a galaxy with high dispersion. 
They found the tightest relations to be between stellar age-to-IMF and radius-to-IMF.
They found the IMF to be bottom-heavy and requiring a high $\Upsilon_*$. 
\citet{Ferre-Mateu2013} has shown the non-universality, whereby adjusting the slope of the IMFs according to the central velocity dispersion yields a more comparable star formation history among the early-type galaxies.
While \citet{Smith2014} pointed out the discrepancy between methods for probing the IMF of galaxies, based on the data from $\rm ATLAS^{3D}$ and \citet{Conroy2012b}, there was a broad agreement that elliptical galaxies have a more bottom-heavy IMF.

If a bottom-heavy IMF is indeed better suited for high dispersion galaxies, then among the MLCRs in Section~\ref{sec:stellar_mass}, IP13 perhaps portrays a more accurate stellar mass since it assumes the \citet{Kroupa1998IMF} IMF --- yielding $\log(M_*/L)$ ratios 0.225 dex lower than those obtained with the \citet{Salpeter1955} IMF, see \citet[][their Figure~12]{2006MNRAS.372.1149F} --- 
instead of the more bottom-light \citet{2002Sci...295...82K} IMF or the  
\citet{Chabrier2003} IMF 
\citep[see also][]{Ferreras2013, 2013pss5.book..115K}.
Although, there is plenty of evidence \citep{Vazdekis2003, Cenarro2003, Falcon-Barroso2003, van_Dokkum2010IMF, Conroy2012a, Spiniello2012, Conroy2012b,van_Dokkum2012} indicating there are more low-mass stars in elliptical galaxies than even the most bottom-heavy IMF \citep{Salpeter1955} predicts.  
This implies that even our highest mass estimation (the IP13 masses) is an underestimation of the real stellar mass.
Throughout this work, however, the \emph{relative} stellar mass compared to the high-$z$ measurements is more important to us.
Because the studies of high-$z$ red nuggets assume a bottom-light \citet{Chabrier2003} IMF, for comparison purposes, we need to apply the same assumption as we estimate the spheroids' stellar mass (T11 and RC15 MLCRs), even if the real stellar mass is larger than what was presented in Figure~\ref{fig:sizemass}.

\subsubsection{The similarity between local spheroids and distant red nuggets}
\label{sec:structure_similarity}
In order to fully prove that high-$z$ red nuggets and local spheroids are drawn from the same population, we need matching sizes, stellar masses (and thus stellar densities), 
internal dynamic features and stellar populations. 
Such a full comparison for our sample is beyond the scope of this paper but is perhaps not required given the extensive evidence in the literature. 

The bulges of lenticular galaxies and high-$z$ quiescent galaxies both have high velocity dispersions.
For example, \citet{Martinez-Manso2011} reported on a range of dispersions, $\sigma \sim 156$--$236 \rm \,k m\, s^{-1}$, for four red nuggets at $z\sim1$ taken from \citet{Trujillo2007}. 
\citet{van_de_Sande2013} measured  five galaxies at $z\sim2$ to have $\sigma \sim 290$--$450 \, \rm k m\, s^{-1}$.
These highest values appear as outliers after 
\citet{Belli2014} analysed 56 quiescent galaxies at $0.9 < z < 1.6$ and obtained each galaxy's mean velocity dispersion within $R_{\rm e,gal}$.  They found an average $\langle\sigma_{\mathrm{e}}\rangle$ value of 
219\,km\,s$^{-1}$. 
For comparison, in the local Universe, the MASSIVE survey \citep{Ma2014} reports on the central velocity dispersions $\sigma_\mathrm{c}$ for 90 early-type galaxies within $104 \rm \, M p c$. They are in the range of $\sigma_\mathrm{c} \sim 200$--$350 \rm \, k m \, s^{-1}$ \citep{Veale2017}. 
In the redshift range $0.15\lesssim z \lesssim 0.5$, \citet{Scognamiglio2020} detected 19 UCMGs ($M_*/\rm M_{\odot}>8\times10^{10}$ and $R_\mathrm{e}< 1.5 \rm kpc$) spanning a range of $200 \lesssim \sigma/\rm km~s^{-1} \lesssim 400$ \citep[see also,][]{Tortora2016,Tortora2018,Tortora2020,Barbosa2021}.

Local spheroids are also known to contain old stellar populations that existed at high-$z$.
\citet{Saracco2009} showed 32 early-type galaxies at $1 < z < 2$ which exhibit two distinct stellar populations: the old ($\sim$3.5\,Gyr) and young ($\sim$1\,Gyr) stars, implying the early-type galaxies exist at least since $z>3$--$3.5$.
\citet{Peletier2007} also reached the same conclusion from the perspective of stellar velocity dispersion dips.
\citet[][see their table~4]{MacArthur2009} found that in local spiral galaxy bulges, old stellar populations ($>$10\,Gyr) make up the majority of the mass budget.
Regardless of whether the bulges of local galaxies formed via outside-in (i.e., secular evolution) or inside-out (i.e., mergers and/or accretion) scenarios, the structural similarity to high-$z$ red nuggets is now well established.
Although, evidence from stellar population studies \citep{Proctor2002, Moorthy2006, Thomas2006, Jablonka2007, MacArthur2008, Saracco2009} favour the inside-out process in early-type galaxies.

The stellar densities are also comparable between local bulges and high-$z$ red nuggets \citep[][his figure~1]{2013pss6.book...91G}, which have half-light densities higher than local early-type galaxies of the same mass.  This is because the discs of local lenticular galaxies --- which tend to dominate these galaxies' stellar mass budget --- increase the galaxies' half-light radii but not in an economical space-filling fashion. 
The Bulge+Disc decompositions from \citet{de_la_Rosa2016} also show the size-mass relation overlaps (see their figure~3); the red nuggets at $0.18 < z < 1$ share almost the same relation and galaxies at $1 < z <1.8$ have a similar slope, but are slightly more compact. 
\citet{2013pss6.book...91G} further showed that local dwarf `compact elliptical' (cE) galaxies have similar sizes, stellar masses, and densities as local low-mass bulges, suggesting that they are the (largely) disc-free counterparts of lower-mass bulges, just as red nuggets are the (largely) disc-free counterparts of high-mass bulges.

\citet{Graham2015} also illustrated that the overlap in the (S{\'e}rsic index, $n$)--(effective radius, $R_{\rm e}$) diagram (their figure~3, primarily from Bulge+Disc decompositions of local galaxies) between bulges and high-$z$ red nuggets from \citet{Damjanov2011}.
Given the structural similarity, they, therefore, concluded that not all compact massive \textit{spheroids} have gone through significant size evolution since $z<2.5$.

\subsubsection{Contributions from the Virgo Cluster}
\label{sec:VCC}
Bin~3 is of particular interest. 
Bin~3 captured galaxies within $ \rm Dist.< 45\,\rm Mpc$, a space containing the Virgo Cluster. 
Not only is it complete down to the lowest galaxy stellar mass ($M_*$ (IP13) = $10^{11}$\,M$_\odot$), it also yields the highest number density for compact massive spheroids.

\begin{figure}
	\includegraphics[clip=true, trim= 5mm 6mm 4mm 4mm, width=1.0\columnwidth]{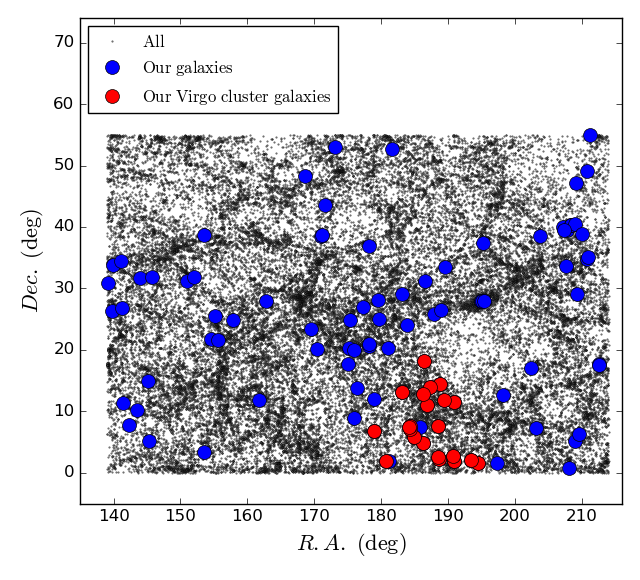}
    \caption{The spatial distribution of the galaxies.
    The parent sample selected by the angular boundaries from Equation~\ref{eq:RADEC} is marked in \textbf{black} dots.
    The 103 galaxies selected by the three bins (Equation~\ref{eq:Bin}) are marked as the \textcolor{blue}{blue} circles.
    Among these 103 galaxies, the ones that are the members of the Virgo Cluster are marked as \textcolor{red}{red} circles.
    }
    \label{fig:RADEC_VCC}
\end{figure}

In Figure~\ref{fig:RADEC_VCC}, we show the location of our host galaxies in the sky. 
Small black dots are all the galaxies in the SDSS field bounded by our angular selection (Equation~\ref{eq:RADEC}).
One can clearly see the large-scale filament structures in the local Universe manifesting. 
The blue points are the 103 host galaxies in our sample. 
Since we only select for massive galaxies, it is natural that most of them live in dense environments.
One can see the blue points largely coincide with the over-density regions.
The red points are the Virgo Cluster host galaxies.
There are 22 out of 52 galaxies in Bin~3 that are in the Extended Virgo Cluster Catalog \citep[EVCC,][]{Kim2014EVCC}.

Using the less restrictive \citet{Barro2013} and \citet{Damjanov2014} sample selection criteria, Bin~3 is more abundant with compact massive spheroids than the peak of the red nugget population ($n_\mathrm{c,Sph,Bin3} > \rm max(n_\mathrm{RN})$) across all stellar masses.
Note that the $n_\mathrm{c,Sph,Bin3}$ here corresponds to the respective `compact massive'  selection criteria as the high-$z$ nuggets' peak abundance ($\rm max(n_\mathrm{RN})$).
When using the more restrictive selection criteria from \citet{van_der_Wel2014} and \citet{van_Dokkum2015}, our result varies depending on the stellar mass estimates (Figure~\ref{fig:all_mass}).
With T11 and Z09 stellar mass, no compact massive spheroid was found in Bin~3 with the restrictive \citet{van_der_Wel2014}  and \citet{van_Dokkum2015} criteria.
With RC15 stellar mass, Bin~3 has a lower number of compact massive spheroids to the red nuggets, $n_\mathrm{c,Sph,Bin3} < \rm max(n_\mathrm{RN})$ (around 0.66--0.78 difference).
Finally, in IP13 stellar mass, $n_\mathrm{c,Sph,Bin3}$ is slightly higher than $\rm max(n_\mathrm{RN})$, by 38\% in \citet{van_der_Wel2014} criteria and 52\% in \citet{van_Dokkum2015} criteria.

The high number density in Bin~3 reinforces the idea that the cloaking process is more prevalent among the low-mass spheroids ($1\times 10^{10}< M_\mathrm{*,Sph} (\rm RC15)/\rm M_{\odot} < 4\times10^{10}$). 
Because Bin~3 has a lower stellar mass selection limit as compared to Bins~1 and 2, it samples some extra low-mass galaxies ($ 6.7\times10^{10} < M_\mathrm{*,gal}/\rm M_{\odot} (RC15)< 1.3\times10^{11}$).
The low-mass spheroids are likely to experience drastic disc growth since $z\sim1.5$, creating galaxies with stellar mass range $6.7\times10^{10} < M_\mathrm{*,gal}/\rm M_{\odot} (RC15) < 1.3\times10^{11}$, thereby increasing the number density in Bin~3.

Having roughly half of the galaxies in Bin~3 being Virgo Cluster members, we examine if the cluster affects the estimation of the number density of the compact massive spheroids.
We do so by removing the Virgo Cluster galaxies from our sample and recalculating the number density of compact massive spheroids in Bin~3. 
Table~\ref{tab:VCC_remove_n} shows the resulting number densities using the four criteria for defining compact massive spheroids \citep[][]{Barro2013,van_der_Wel2014,Damjanov2014,van_Dokkum2015} and using the four stellar mass estimates (T11, Z09, IP13, and RC15).

With the restrictive selection criteria \citep[][]{van_der_Wel2014,van_Dokkum2015}, we found a similar number of compact massive spheroids as before. 
With the less restrictive selection criteria \citep[][]{Barro2013,Damjanov2014}, the numbers of compact massive spheroids are roughly halved. 
Because the reduction  ($\sim$58\%--61\%) is comparable to the number of removed Virgo members ($\sim$42\%), we, therefore, conclude the Virgo Cluster over-density does not increase Bin~3's number density in any significant way.
Our wide-sky coverage prevented a potential bias as the Virgo Cluster takes up less than one-eighth of the area in the R.A. and Dec selection window (see Figure~\ref{fig:RADEC_VCC}).
Moreover, with the reduced number densities ($n_\mathrm{c,Sph,Bin3}\approx (0.3$--$5.0) \times10^{-4}\,\rm Mpc^{-3}$), Bin~3 still contains a comparable number of compact massive spheroids to the peak red nugget density 
($\rm max(n_\mathrm{RN}) \approx (1.8$--$2.4)\times10^{-4}\,\rm Mpc^{-3}$).       
Thus, even without the influence of the Virgo Cluster, there remains a sizable amount of compact massive spheroids in Bin~3, sufficient to account for the missing red nuggets.

\begin{table*}
   \caption{The number densities of Bin~3 after removing Virgo Cluster galaxies.}
   \label{tab:VCC_remove_n}
   \begin{tabular}{lllll}
   \hline
    MLCR     & T11 & Z09 & RC15 & IP13\\
    \hline
    Criteria     & $n_\mathrm{c,Sph,Bin3}/\rm Mpc^{-3}$ (\#) & $n_\mathrm{c,Sph,Bin3}/\rm Mpc^{-3}$ (\#) & $n_\mathrm{c,Sph,Bin3}/\rm Mpc^{-3}$ (\#)& $n_\mathrm{c,Sph,Bin3}/\rm Mpc^{-3}$ (\#)\\
   \hline
   \hline
    \citet{Barro2013}  & $3.1\times10^{-4}~(10)$ & $3.1\times10^{-4}~(12)$ & $4.3\times10^{-4}~(14)$ &   $6.4\times10^{-4}~(21)$ \\  
    \citet{van_der_Wel2014} & $0$ & $0$ & $0$ &  $9.2\times10^{-5}~(3)$\\
    \citet{van_Dokkum2015} & $0$ & $0$ & $3.1\times10^{-5}~(1)$ &  $1.2\times10^{-4}~(4)$\\
    \citet{Damjanov2014} & $3.7\times10^{-4}~(12)$ & $4.3\times10^{-4}~(14)$ & $5.0\times10^{-4}~(16)$ &  $7.1\times10^{-4}~(23)$\\
   \hline
   \end{tabular}
\end{table*}

\subsubsection{Some distant red nuggets may also have discs}
\label{sec:OGred}

It is unclear whether the original red nuggets \citep[from][]{Daddi2005,van_Dokkum2008,Damjanov2009} are devoid of any disc growth.
From visual inspection, some galaxies exhibit discy, high-ellipticity isophotes.
For example, galaxy numbers: `4950' ($z = 1.55 $) from \citet{Daddi2005};  `1030-1813' ($z = 2.56$), `1030-2559' ($z = 2.39$), `1256-142' ($z = 2.37$), and `HDFS1-1849' ($z = 2.31$) from \citet{van_Dokkum2008}; and `22-0189' ($z = 1.49$), `15-4367' ($z = 1.725$), '$\rm S7F5$\_$254$' ($z = 1.22$), and '$\rm S7F5$\_$045$' ($z = 1.45$) from \citet{Damjanov2009}.

 The existence of (nascent) discs in some high-$z$ galaxies is supported by \citet{van_der_Wel2011}.
They found a significant portion of the compact massive quiescent galaxies ($M_*/\mathrm{M}_\odot > 10^{10.8}$, and $R_{\rm e} < 2 \, \rm k p c$) are flattened in projection, thus concluding that  40\%--65\% of red nuggets at $z \sim 2 $ are disc-dominated.
\citet{Ferre-Mateu2012} also found elongated morphologies in red nuggets at $z > 1$ and classified them as fast rotators.
One may therefore question whether those red nuggets are genuinely single-component systems.
While spheroids can rotate, it is certainly the case that the low-resolution of images at high-$z$ will blend the bulges and discs together to create the appearance of a single-component system.  Indeed, even at low $z$, astronomers have struggled for a century to separate S0 galaxies from E galaxies.  
As such, the disc inclination angle and dust content will further complicate the situation.
The shrouding dust heated by star formation will redden the galaxies' light \citep[][]{Perez-Gonzalez2008}. 
Interestingly, a few notable local `relic' galaxies (NGC~1277, PCG~032873, and Mrk~1216), that are believed to be untouched by ex-situ processes, have strong stellar rotation. NGC~1277 has a rotational velocity $v_{r} \sim 300~\rm km~s^{-1}$ in the outer region and a central velocity dispersion $\sigma_0 > 300~\rm km~s^{-1}$ \citep[][]{2014ApJ...780L..20T}. In \citep[][see their Figure~2]{Ferre-Mateu2017}, both  PCG~032873 and Mrk~1216 exhibit a rotational velocity of $v_\mathrm{r} \sim 200~\rm km~s^{-1}$ and central dispersion of $\sigma_0 \sim 320~\rm km~s^{-1}$, implying, indeed, they are not single-component systems. These systems may have accreted and grown their discs. Until they transition through the ES to S0 phase, they may remain compact massive \textit{galaxies}. This may even conceivably explain some of the youthfulness seen in some ultra-compact massive galaxies \citep{2021A&A...654A.136S}, although we note this is just speculation on our part.

It will be interesting to perform structural decompositions on intermediate- and high-$z$ galaxies when high-quality images become available from facilities such as the European Southern Observatory's Very Large Telescope interferometer, equipped with PIONIER (1.6\,$\micron$, \citealt{Le_Bouquin2011PIONIER}), GRAVITY (2.0--2.4\,$\micron$ imager; 4\,mas seeing using the four 8-m Unit Telescopes, \citealt{Gillessen2010GRAVITY,Eisenhauer2011GRAVITY}), and MAVIS (optical, under development, \citealt{e31b3f4abec045d38febe6f157a1e2e6,Rigaut2020MAVISconcept}). 
Subsequently, cosmological simulations involving disc cloaking can be tested if the observed high-mass end of the spheroid mass function, and the spheroid size-mass relation can be reproduced, reported here for the local Universe. 
Careful analysis will enable one to measure the growth of discs in early-type galaxies, shifting the focus away from spheroids --- which appear to be relatively inert.

\subsection{Summary and Conclusions}\label{sec:conclusion}

We defined volume- and mass-limited samples of local galaxies from the SDSS that are nondiscriminatory against galaxy morphology  (Section~\ref{sec:data}). 
The galaxies have masses greater than $10^{11}$\,M$_\odot$ and reside within 110\,Mpc.  
We have discussed how the different mass-to-light ratios (Section~\ref{sec:stellar_mass}) and distance measurements (Section~\ref{sec:distance}) would affect the estimation of the stellar masses and sizes of the galaxies and their spheroids.
A detailed, case-by-case study of 103 galaxies was undertaken. 
We performed multi-component decompositions (Section~\ref{sec:decomposition}) on the surface brightness profiles of each galaxy.  Rather than using automated routines which blindly fit several S\'ersic components, we carefully examined each image and identified the morphological features that are present.  This was further supported by recourse to the literature, in particular, studies including kinematic data or high-resolution HST data.
This enabled a considerable improvement upon the conventional Bulge+Disc decomposition.  Our analysis encompasses the less prominent yet important substructures (i.e., bars, \textit{ansae}, spiral arms, nuclear discs, intermediate-scale discs, etc.), which can skew the bulge parameters if not taken into account.  
From our investigation, we have made the following observations. 

\begin{itemize}
    \item 
    New morphological classifications were required, and made, for the galaxies.
    Previously overlooked substructures, such as large- and intermediate-scale discs, are now incorporated into the description of the morphology (Section~\ref{sec:morph_reclass}).  This is important because the morphology reflects the physical processes which shaped the galaxies' evolution. 
\item 
    Implications from the low number of `true elliptical' galaxies in the local Universe rule out the dominance of the popular E-to-E evolutionary path whereby quasi-spherical 3-D envelopes accumulate around red nuggets for relatively low-mass red nuggets ($M_*/\rm M_{\odot} < 4 \times 10^{10}$).
   The possibility of such an E-to-E scenario is discussed in Section~\ref{sec:E-to-E}. 
    \item
    Morphology-dependent $B/T$ flux ratios are presented for the massive galaxies in Figure~\ref{fig:BT_ratio}  (Section~\ref{sec:B/T_ratio}). 
    \item 
    We provided the mass function of the local spheroids down to stellar mass $M_*/\rm M_{\odot} \sim (0.24$--$1.8) \times10^{10}$
    (Section~\ref{sec:mass-function}).
    Most massive spheroids ($M_*/\rm M_{\odot}(\rm RC15) > 3\times10^{11}$) are not simultaneously small ($R_\mathrm{e} < 2 \rm\,k p c$). 
    \item
    The size-mass distribution of the local spheroids is presented (Section~\ref{sec:size-mass}).
    The number density of `compact' and `massive' spheroids is obtained based on five different selection criteria.
     \item
     We compared the number density of local compact massive spheroids with intermediate- and high-redshift galaxies.
     The evolutionary trend for the number density of compact massive quiescent systems is presented in Figure~\ref{fig:nd_z}, based on four different sample selection criteria (Sections~\ref{sec:z2} and \ref{sec:z1}).
 \end{itemize}
 
Two important insights can be drawn from our results:

\begin{enumerate}
    \item
    The local elliptical galaxies are not as abundant as previously believed.
    Among the 28 supposed elliptical galaxies (E), 20 are actually lenticular galaxies (S0).
    There are only $\sim$11\% (11/103) true elliptical galaxies, plus 2 ES galaxies, in our sample.
    
    \item
    We found compact massive spheroids hidden in the local lenticular and early-type spiral galaxies.
    Working with our host galaxies' selection limit ($M_\mathrm{*,gal}/\rm M_{\odot} (RC15) > 6.7\times10^{10}$), we present a range of lower limits of the number density to the compact massive spheroid of $n_\mathrm{c,Sph} \sim (0.17$--$1.20)\times10^{-4}\,\rm Mpc^{-3}$, as defined by different definitions of `compact massive' systems.
    If relatively low-mass ($M_*/\rm M_{\odot}\sim (1$--$4)\times10^{10}$) spheroids are included, then there are sufficient numbers of compact massive spheroids ($n_\mathrm{c,Sph}\sim1.20\times10^{-4}\,\rm Mpc^{-3}$) to match the peak in the number density of red nuggets when defined using the same mass and size criteria.
  %  than true elliptical and ellicular galaxies ($n_\mathrm{E+ES} \sim 2.73\times10^{-5}~\rm Mpc^{-3}$, Section \ref{sec:comparsion}). 
    If we select for only high-mass spheroids ($M_*/\rm M_{\odot}>4 \times10^{10}$), the number densities for compact massive spheroids and true ellipticals are comparable ($n_\mathrm{c,Sph} \simeq n_\mathrm{E+ES}$).
    However, in the high-mass range, the uncertainty in stellar mass estimation makes it difficult to conclude which evolution dominates by number density alone (Section~\ref{sec:z3}).
    Interestingly, for $M_\mathrm{*,gal}/\rm M_{\odot} (RC15) > 6.7\times10^{10}$ and $\rm Dist. <45\,Mpc$ (Bin~3), we found more compact massive spheroids than the red nuggets' peak abundance: $n_\mathrm{c,Sph,Bin3} > \rm max(n_\mathrm{RN})$, inspite of the influence from Virgo Cluster galaxies.
\end{enumerate}

Collectively, this calls the frequency of the E-to-E scenario into doubt, while the disc-cloaking scenario is now more salient than ever.
An E-to-E evolution and disc-cloaking create distinctly different end products.  
If we are to truly understand galaxies, and the cosmological evolution of galaxies, this distinction is of key importance.

\section*{Acknowledgements}
\addcontentsline{toc}{section}{Acknowledgements}
This research was supported under the Australian Research Council's funding scheme DP17012923 
and by Tamkeen under the New York University Abu Dhabi Research Institute grant CAP$^3$.
This research has made use of the NASA/IPAC Extragalactic Database (NED), which is funded by the National Aeronautics and Space Administration and operated by the California Institute of Technology.

The project is made possible by using the following software packages:
\texttt{AstroPy} \citep{Astropy2013, Astropy2018},
\texttt{CMasher} \citep{van_der_Velden2020},
\texttt{IRAF} \citep{Tody1986IRAF, Tody1993IRAF},
\texttt{ISOFIT} \citep{Ciambur2015},
\texttt{Matplotlib} \citep{Hunter:2007},
\texttt{NumPy} \citep{Harris2020array_NumPy},
\texttt{Profiler} \citep{Ciambur2016},
\texttt{SAOImageDS9} \citep{Joye2003DS9},
\texttt{SciPy} \citep{2020SciPy},
\texttt{SExtractor} \citep{Bertin1996}, and
\texttt{TOPCAT} \citep{Taylor2005Topcat}.

All the scripts used in the analysis are available on GitHub (\url{https://github.com/dex-hon-sci/GalSpheroids}).

\section*{Data availability}
\addcontentsline{toc}{section}{Data availability}
The data underlying this article will be shared upon request to the corresponding author.

%%%%%%%%%%%%%%%%%%%%%%%%%%%%%%%%%%%%%%%%%%%%%%%%%%



%\bibliographystyle{mnras}

\bibliographystyle{mnras}
\bibliography{paperI_lib.bib}

\clearpage
\newpage

%%%%%%%%%%%%%%%%%%%%%%%%%%%%%%%%%%%%%%%%%%%%%%%%%%

%%%%%%%%%%%%%%%%% APPENDICES %%%%%%%%%%%%%%%%%%%%%

\appendix
\section{Distance correction}
\label{sec:dist_corr}

\begin{figure*}
\centering
	\includegraphics[clip=true, trim= 0mm 0mm 0mm 2mm, width=0.7\textwidth]{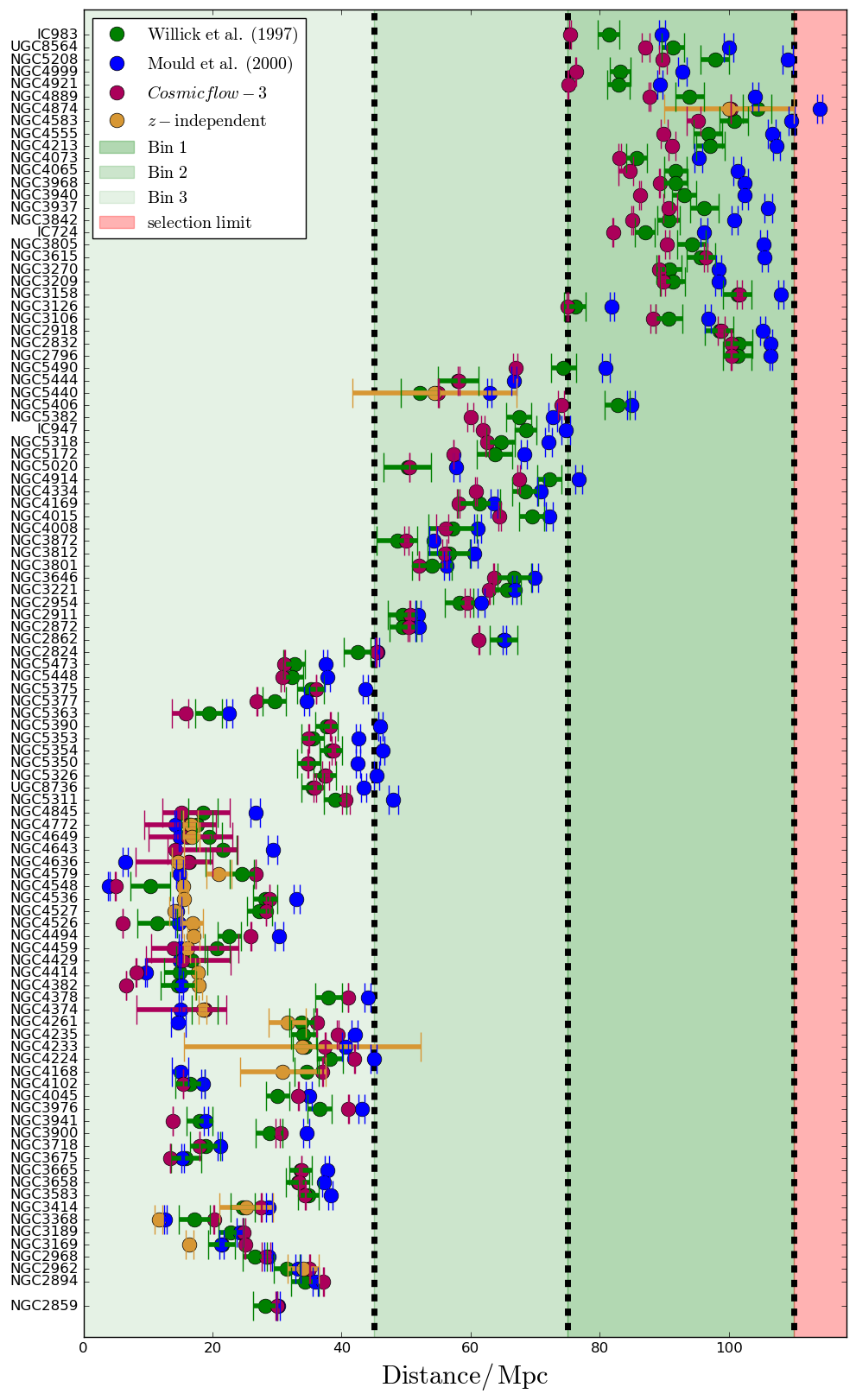}
    \caption{
    Distances using different models. The pink zone is the area outside of the selection volume.  The purple points are the \textit{Cosmicflow-3} \citep{Kourkchi2020} distances, the blue points are the M2000 distances, and the green points are the W1997 distances. The orange points are a blanket label for the redshift-independent measurements from various sources; see Table~\ref{tab:z-indi}. Galaxies are listed in descending (westward) order, according to right ascension. Note that several galaxies (NGC~5490, NGC~5406, NGC~4914, NGC~2824, NGC~5390, NGC~5354, NGC~5326, and NGC~5311) have their distances lie in different Bins. For those cases, we present the result in the Bins according to their \textit{Cosmicflow-3} distance as in both Table~\ref{tab:sample1}-\ref{tab:sample3} and Table~\ref{tab:result1}-\ref{tab:result3}.}
     \label{fig:dist_compare_bin1}
\end{figure*}

The variation among the four different distance measurements that we explored turns out not to be significant in regard to the number density of compact massive spheroids. 
We show these distance measurements in the forest plot of Figure~\ref{fig:dist_compare_bin1}.
The W1997, M2000, \textit{Cosmicflow-3} \citep{Kourkchi2020}, and $z$-independent distances are labelled in green, blue, purple, and orange points, respectively.
The mean differences between each set of distances span 3.16 to 7.54 Mpc, while the standard deviations vary from 2.38 to 4.95 Mpc. 
The values and sources of the various $z$-independent distances are listed in Table~\ref{tab:z-indi}.

\begin{table}
   \caption{Redshift-independent measurements.}
   \begin{threeparttable}[b]
   \label{tab:z-indi}
   \begin{tabular}{llll}
   \hline
   Name & Distance (Mpc) & Method & Sources  \\
   \hline
    \multicolumn{4}{c}{Bin~1} \\
    \hline
    NGC~4874 & $100.0\pm 10.0$	& SBF &	\citet{Liu2001}  \\
   \hline
   \multicolumn{4}{c}{Bin~2} \\
   \hline
    NGC~5440 & $54.425\pm 12.718$ & SNIa & \citet{Silverman2012}\\
   \hline
   \multicolumn{4}{c}{Bin~3} \\
   \hline
    NGC~4649 & $16.827\pm1.200$ & SBF & \citet{Tonry2001} \\
    NGC~4636 & $14.655\pm0.904$ &	SBF &	\citet{Tonry2001}\\
    NGC~4374 & $18.510\pm0.606$ &	SBF &	\citet{Blakeslee2009}\\
    NGC~4261 & $31.623\pm2.892$ &	SBF &	\citet{Tonry2001}\\
    NGC~4772 & $16.6\pm1.1 $	&	SNIa & 	\citet{Huang2017}\\
    NGC~4579 & $21.0\pm 2.0$	& SNIa	& \citet{Ruiz-Lapuente1996}\\
    NGC~4548 & $15.424\pm0.564$ & Cepheids & \citet{Kanbur2003}\\
    NGC~4536 & $15.567\pm0.749$ & Cepheids & \citet{Kanbur2003}\\
    NGC~4527 & $14.158\pm0.867$ & Cepheids & \citet{Kanbur2003}\\
    NGC~4526 & $16.904\pm1.631$ & SBF & \citet{Tonry2001}\\
    NGC~4494 & $17.061\pm0.887$ & SBF	& \citet{Tonry2001}\\
    NGC~4459 & $16.069\pm0.450$ & SBF & \citet{Mei2007}\\
    NGC~4414 & $17.693\pm0.362$ & Cepheids & \citet{Kanbur2003}\\ 
    NGC~4382 & $17.881\pm0.560$ & SBF & \citet{Blakeslee2009}\\
    NGC~4233 & $33.884\pm18.355$ & SBF	& \citet{Tonry2001} \\
    NGC~4168 & $30.903\pm6.594$ & SBF &\citet{Tonry2001}\\
    NGC~3414 & $25.235\pm4.142$ & SBF & \citet{Tonry2001} \\
    NGC~3368 & $11.695\pm0.609$ & Cepheids & \citet{Saha2006}\\
    NGC~3169 & $16.444\pm0.617$ & SNIa	& \citet{Mandel2011}\\ 
    NGC~2962 & $34.041\pm2.435$ & SBF &  \citet{Ajhar2001}\\
   \hline
   \end{tabular}
    SNIa = Supernovae Type~Ia \\
    SBF = Surface Brightness Fluctuations \\
   \begin{tablenotes}
      \small
      \item
      \begin{flushleft}
      \end{flushleft}
    \end{tablenotes}
    \end{threeparttable}
\end{table}

\textit{Cosmicflow-3} has the most sophisticated model to date, and it was built from the largest observational data set among the three model-based choices.
The main issue of \textit{Cosmicflow-3} is the triple-value region, where the gravity of the Virgo Cluster is significant enough to bend the velocity-distance relation into a cubic equation.
Galaxies residing within 7--$21 \, \rm M p c$, therefore, have two or three solutions (possible distances) for the same recessional velocity. 
In Figure \ref{fig:dist_compare_bin1}, we highlighted such objects with extended error bars (NGC~4374, NGC~4429, NGC~4459, NGC~4636, NGC~4643, NGC~4845, NGC~4649, and NGC~4772), in which the upper and lower limits represent the high and low estimation of the distance, respectively.

We used the redshift-independent distance as the primary distance to calculate the luminosity of the spheroids and galaxies in our sample. 
The only exceptions are NGC~4874 (an SBF distance), NGC~5440 (an SNIa distance), and NGC~4233 (an SBF distance), where all of them exhibit a large error ($>$10\,Mpc).
In these cases, we use the \textit{Cosmicflow-3} value instead.
In total, 19 galaxies' total and spheroidal stellar mass are obtained via $z$-independent distances.
Other than these 19 galaxies, we primarily used the \textit{Cosmicflow-3} distance, in total, for 81 galaxies.
There are an additional three galaxies that are within the triple-value region but which do not have $z$-independent measurements: NGC~4845, NGC~4643, and NGC~4429.
In all three cases, the W1997 distances lie within the triple-value region, with a reasonable agreement with the median value of the \textit{Cosmicflow-3} distance. 
For these galaxies, we elect to use the W1997 distances, derived from empirical TF-relation, instead.
The adopted distances for all 103 galaxies are provided in Table~\ref{tab:sample1}.

\section{The mass-to-light colour relations}
\label{sec:MLCR_suppl}
Here, we provide a supplementary description of possible error sources and the underlying assumptions that went into building the mass-to-light colour relations (MLCRs) presented in this work.  This speaks directly to the stellar masses that we have derived.

Several key factors contribute to the error budget of the estimated stellar mass.
\begin{itemize}
    \item 
The innate stellar population model degeneracy, between age and metallicity, 
is reported to constitute a small error of 0.1--0.2\,dex from idealised mock galaxy studies \citep{Gallazzi2009}.
    \item
A more prominent error comes from the systematic treatments of the `priors'.
The IMF provides the normalisation factor for the MLCRs, and its uncertainty affects the $M_*/L$ ratio. 
For example, there can be a change of up to 0.3\,dex in mass if one assumes a \citet{Chabrier2003} over a \citet{Salpeter1955} IMF (RC15).
    \item
Another obvious factor is the stellar population model.
The most widely-used prescription is the BC03 Stellar Population  Synthesis (SPS) model. 
It provides a comprehensive library of the stellar population that was validated by agreement with early data from the SDSS.
Over the years, the community has been continuously updating the model by including previously unconsidered stellar isochrones.
The most notable issue plaguing stellar population models is the influence of the thermally-pulsating asymptotic giant branch (TP-AGB) stars.
Multiple studies \citep[e.g.,][]{Maraston1998,Maraston2005} have pointed out that the excess light from the young ($\sim$1\,Gyr) TP-AGB star's emission can lead to an underestimation of the $M_*/L$ ratio.  
This effect is particularly relevant for near-infrared (NIR) studies, as the inclusion of TP-AGB stars \citep{2007IAUS..241..125B,2007ASPC..374..303B} in the renewed version of the BC03 SPS model in 2007 (commonly known as the CB07 SPS in the literature, see the \textit{MAGPHYS} \citealt{da_Cunha2008} library) results in $M_*/L$ ratio changes of 0.1\,dex and 0.4\,dex in the optical and NIR, respectively (Z09).
    \item
Treatment of the interstellar medium (ISM) and dust attenuation also changes the perceived light.
IP13 show the effect of omitting dust attenuation in their Figure~13.
In a dusty scenario, the mass-to-light relation can change by up to 0.5\,dex in the case of an inclined disc \citep[see also][]{2007MNRAS.379.1022D, 2008ApJ...678L.101D}. 
\end{itemize}

Given the need to appreciate how the stellar masses of the high-$z$ galaxies were obtained by different authors,  we identify and briefly describe each MLCR they used. 
\begin{itemize}
    \item 
 Z09 assumes the \citet{Chabrier2003} IMF, the CB07 SPS, and the dust attenuation from \citet{Charlot2000}.
 The $M_*/L$ ratio is produced using the fiducial mass reconstruction method by Monte Carlo sampling from the CB07 library, with parameters advised by \citet{Kauffmann2003A}.
 The $M_*/L$ ratio is then mapped onto the colour space of ($g-i$) and ($i-H$) to build their MLCR shown in Equation~\ref{eq:MLCRs_a}.
 They verified this by comparing it to nine local galaxies within 30\,Mpc.
    \item 
Both T11 and RC15 used the \citet{Chabrier2003} IMF and the BC03 SPS, which, as we have seen, has been the most commonly used assumption in the literature when calculating the stellar masses of high-$z$ galaxies.
Given the assumed dust attenuation law from \citet{Calzetti2000}, T11 calculated the stellar mass by fitting the SED from early Galaxy And Mass Assembly (GAMA DR1) broadband photometry data via Bayesian probability minimization.
Among the above-mentioned MLCRs, T11 is the only one that is built empirically from the ground up.
The GAMA data covers the intermediate-redshift range ($z <0.65$; median $z = 0.2$) and magnitudes $r_\mathrm{petro} < 19.4$--19.8\,mag.
However, claim has been made that T11's MLCR underestimated the $M/L$ ratio for early-type spirals and ellipticals \citep{Schombert2022}.
    \item 
RC15 constructed their MLCR via mock galaxies using the \textit{MAGPHYS} library \citep{da_Cunha2008}. 
The mock galaxy SEDs undergo Bayesian fitting on $girzH$ bands to recover the stellar masses.
The MLCR yields excellent agreement with the observational SHIVir survey \citep['Spectroscopy and $H$-band imaging of Virgo Cluster galaxies';][]{McDonald2011} data, having mean residuals of $-0.01$\,dex and $0.05$\,dex (see their Figure~6).

With the same IMF and SPS as T11, RC15 is different in several ways.
First, RC15 assume a different dust treatment \citep{Charlot2000} with two-component models accounting for the dust in molecular clouds and the ambient ISM.
Second, RC15 suggests that T11 did not include bursts in their model SFHs, while \textit{MAGPHYS} contains random burst models superimposed in the SFH.
Both works are validated by their respective observational data: one of which focuses on the nearby Virgo Cluster galaxies and the other on galaxies at intermediate redshifts.
The considerable discrepancy between them (Figure~\ref{fig:ML}) might arise from the use of different data sets.
    \item 
Finally, IP13 constructed their MLCR uniquely by choosing a \citet{Kroupa1998IMF} IMF and the \citet{Marigo2008} isochrone model (the same used in CB07). 
Once again, the choice of \citet{Marigo2008} isochrones stems from the concern over the TP-AGB influence.
Their equation is constructed assuming a dust-free scenario. While the effect of dust is significant in some colour, they concluded that $(g-i)$ remains a good tracer for the galaxies' stellar masses.
The consequence of using such a combination is the slower decline of stellar mass as a function of age (see their Figure~1) in comparison to the more popular BC03 SPS.
\end{itemize}

\section{Galaxy Data}
\label{sec:Gal_can}

Here, we present the basic information for the galaxy sample identified in Section~\ref{sec:distance}.
Tables~\ref{tab:sample1}--\ref{tab:sample3} also provide the new morphological classifications (Column~9) based on our multi-component decomposition, as described in~Section~\ref{sec:decomposition}.

In Tables~\ref{tab:sample1}--\ref{tab:sample3}, Column~(8) lists the past morphology assignment while Column~(9) provides our new classifications.
The labels for the new classes are inspired by the `morphology grid' from \citet{Graham2019class} for high surface brightness galaxies.
It represents the progression from the traditional Hubble-Jeans sequence \citep{Jeans1919,Lundmark1925,Hubble1926,1926CMWCI.324....1H,1927Obs....50..276H,1936rene.book.....H}, with an emphasis on the commonly-overlooked features in early-type galaxies, such as bars and the continua of disc sizes (nuclear, intermediate, and large-scale).
The grid includes the ellicular (ES) class of galaxies \citep{Liller1966}, denoting the spheroids with an embedded intermediate-scale disc, and it removes the redundant E4--E7 classes because they are essentially all lenticular galaxies \citep{Liller1966,Gorbachev1970SvA,Michard1984,Capaccioli1990,van_den_Bergh1990}.
For simplicity's sake, we omit the subdivision of spiral arm shapes (a, b, c, d, and m) which are not as relevant to our investigation.
We label the galaxies as ellipticals without a disc (E0--E3), elliculars (EAS or EBS), lenticulars (SA0, SAB0, or SB0), and spirals (SA, SAB, or SB), in which A and B are the distinctions between barless and barred galaxies, and AB is designated for weak bars, defined here as those with lower central surface brightness ($\mu_0$) than the disc.

\begin{figure}
\centering
	\includegraphics[clip=true, trim= 1mm 2mm 1mm 2mm, width=0.85\columnwidth]{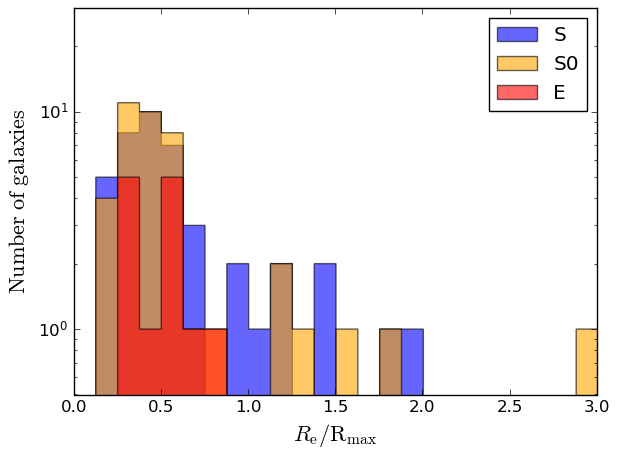}
    \caption{
    The distribution of galaxy circularized effective radius ($R_\mathrm{e}$)--maximum radius ($R_\mathrm{max}$) ratio.
    We obtained $R_\mathrm{e}$ via a single S\'ersic fit and the $R_\mathrm{max}$ is our cutoff radius for the fit.
    The sample is separated by their morphological type and colour coded with red, orange, and blue for E, S0, and S galaxies, respectively.
    Here, we only show the distribution between $ 0 < R_\mathrm{e}/R_\mathrm{max} < 3.0$, where 96 galaxies are present.
    }
    \label{fig:ReRmax}
\end{figure}

In accordance with the common practice in the literature, we performed a single S\'{e}rsic function fit to determine the effective radius ($R_\mathrm{e}$) for each galaxy. 
In Figure~\ref{fig:ReRmax}, we show the distribution of $R_\mathrm{e}/R_\mathrm{max}$ for our galaxies, separated by their morphology.
We choose the cutoff radius, $R_\mathrm{max}$, to fitting the light profile at the radius where the $S/N$ ratio is low and no longer useful for decomposition.
The majority (83/103) of our galaxies have $R_\mathrm{e}/R_\mathrm{max}<1$. 
There are 20 galaxies which have effective radii exceeding their cutoff radii ($R_\mathrm{e}/R_\mathrm{max}>1$).  
They are exclusively S0 and S galaxies.
The outliers illustrate the problem of using a single S\'{e}rsic fit to measure the size of a galaxy. 
A shallow disc profile combined with a prominent bulge can result in a long-tailed surface brightness profile. 
The parameterised effective size of such a galaxy, from a single S\'{e}rsic fit, will be an overestimate of the actual half-light radius. 
The effect of inflated galaxy radii in S0 and S galaxies is not widely appreciated and shall be addressed in a separate work.

\begin{table*}
   \caption{Galaxy sample and basic data.}
   \begin{threeparttable}[b]
   \label{tab:sample1}
   \begin{tabular}{lrrrrrccccrr}
   \hline \hline
   Name & RA & Dec & Dist.  & $\mathfrak{m}_g$ & $\mathfrak{m}_i$   & Seeing & Morph. & Morph. & $\mathfrak{M}_{i}$ & $M_{*,\rm gal}$ & $R_\mathrm{max}$ \\
           &  deg   &    deg     &  Mpc    &  mag & mag &  \arcsec  & (old) & (new)  & mag & $10^{10}$\,M$_{\odot}$ & \arcsec \\
   (1) & (2) & (3) & (4) & (5) & (6) & (7) & (8) & (9) & (10) & (11) & (12) \\
   \hline
   Bin 1 &  &  &  &  &  &   &  &  & & \\
    \hline
	IC~983&212.5&17.7&89.6&12.3&11.1&0.8&SBbc&SB&-23.7&62.7&72.0\\
	UGC~8564&203.7&38.5&100.0&13.4&11.9&0.8&S0&SA0&-23.1&59.9&50.0\\
	NGC~5208&203.1&7.3&109.2&13.1&11.9&0.9&S0&SA0&-23.3&48.6&36.0\\
	NGC~4999&197.4&1.7&92.8&11.9&10.8&0.9&Sb&SB&-24.1&81.3&50.0\\
	NGC~4921&195.4&27.9&89.3&12.1&10.9&0.8&Sab&SAB&-23.9&81.4&68.0\\
	NGC~4889&195.0&28.0&104.1&12.1&10.9&1.0&E&E3&-24.2&109.8&47.0\\
	NGC~4874&194.9&28.0&99.5&12.7&11.5&1.1&E&SA0&-23.5&52.3&90.0\\
	NGC~4583&189.5&33.5&109.7&13.6&12.2&1.4&S0&SA0&-23.0&60.1&34.0\\
	NGC~4555&188.9&26.5&106.6&12.2&11.0&1.3&E&E2&-24.1&89.1&57.0\\
	NGC~4213&183.9&24.0&107.3&12.6&11.4&0.9&E&SA0&-23.8&64.3&42.0\\
	NGC~4073&181.1&1.9&95.3&12.5&11.3&1.0&E&E3&-23.6&60.7&68.0\\
	NGC~4065&181.0&20.2&101.3&12.6&11.4&0.8&E&SA0&-23.6&54.3&27.0\\
	NGC~3968&178.9&12.0&102.4&12.1&10.9&1.1&Sbc&SAB&-24.1&84.8&57.0\\
	NGC~3940&178.2&21.0&102.5&12.9&11.6&1.5&E&SA0&-23.4&54.8&50.0\\
	NGC~3937&178.2&20.6&106.1&12.2&11.0&1.0&S0&SA0&-24.1&80.1&69.0\\
	NGC~3842&176.0&19.9&100.8&12.5&11.3&0.9&E&SA0&-23.7&63.1&62.0\\
	IC~724&175.9&8.9&96.1&12.6&11.4&1.2&Sa&SA&-23.5&58.1&43.0\\
	NGC~3805&175.2&20.3&105.3&12.9&11.6&0.9&S0&EAS&-23.5&63.3&64.0\\
	NGC~3615&169.5&23.4&105.6&12.8&11.6&1.0&E&E3&-23.5&57.7&60.0\\
	NGC~3270&157.9&24.9&98.4&12.8&11.5&1.3&SABb&SAB&-23.4&58.5&33.0\\
	NGC~3209&155.2&25.5&98.4&12.4&11.2&1.0&E&SA0&-23.7&66.2&52.0\\
	NGC~3158&153.5&38.8&108.0&12.7&11.5&1.3&E&SA0&-23.7&68.8&105.0\\
	NGC~3126&152.1&31.9&81.9&12.7&11.4&1.1&SABb&SB0&-23.2&58.9&23.0\\
	NGC~3106&151.0&31.2&96.7&12.4&11.2&1.4&S0&SA0&-23.7&60.9&36.0\\
	NGC~2918&143.9&31.7&105.3&12.8&11.5&1.1&E&SB0&-23.6&65.3&33.0\\
	NGC~2832&139.9&33.7&106.4&12.6&11.4&1.3&E&E3&-23.7&59.9&95.0\\
	NGC~2796&139.2&30.9&106.5&13.2&11.8&1.0&Sa&SA0&-23.3&71.1&38.0\\
   \hline
   \end{tabular}
   \begin{tablenotes}
      \small
      \item 
      \begin{flushleft}
      In descending (westward) order, according to right ascension.
      \underline{Columns}: 
      (1) the galaxy name; 
      (2) the right ascension angle in J2000 coordinates (in degrees); 
      (3) the declination angle in J2000 coordinates (in degrees); 
      (4) the adopted (luminosity) distance described in part (iii) of Section~\ref{sec:distance};
      (5) the  $g$-band apparent magnitude and 
      (6) the $i$-band apparent magnitude from the NASA-Sloan ATLAS catalogue (\url{http://www.nsatlas.org/}) after correcting both for Galactic dust extinction and the K(z)-correction; 
      (7) the seeing (in arcsec) measured by \texttt{imexam}; 
      (8) the RC3 Hubble morphology classification; 
      (9) our new classification described in Section~\ref{sec:morph_reclass}; 
      (10) the absolute $i$-band magnitude calculated using columns (4) and (6), with Galactic dust and K-correction from the NASA-Sloan ATLAS catalogue applied; and
      (11) the total galaxy stellar mass using the IP13 ($g-i$)-dependent $M_*/L$ ratio applied to the SDSS-derived $i$-band magnitude. 
      (12) the cutoff radius (in arcsec) in circularised equivalent-axis.
      \textcolor{red}{}
      \end{flushleft}
    \end{tablenotes}
    \end{threeparttable}
\end{table*}

\begin{table*}
   \caption{Galaxy sample \emph{continued}.}
   \begin{threeparttable}[b]
   \label{tab:sample2}
   \begin{tabular}{lrrrrrccccrr}
   \hline \hline
   Name & RA & Dec & Dist.  & $\mathfrak{m}_g$ & $\mathfrak{m}_i$  & Seeing & Morph. & Morph. & $\mathfrak{M}_{i}$ & $M_{*,\rm gal}$ & $R_\mathrm{max}$\\
           &  deg   &    deg     &  Mpc    &  mag & mag &   \arcsec  & (old) & (new)  & mag & $10^{10}$\,M$_{\odot}$ &\arcsec \\
   (1) & (2) & (3) & (4) & (5) & (6) & (7) & (8) & (9) & (10) & (11) & (12) \\
   \hline
   Bin 2 &  &  &  &  &  &   &  &   \\
   \hline
	NGC~5490&212.5&17.5&80.9&12.4&11.2&0.9&E&SA0&-23.4&50.9&63.0\\
	NGC~5444&210.9&35.1&66.7&12.2&11.0&1.0&E&E2&-23.2&38.9&85.0\\
	NGC~5440&210.8&34.8&54.4&12.3&11.1&0.9&Sa&SA&-22.6&28.1&83.0\\
	NGC~5406&210.1&38.9&84.8&12.3&11.1&0.8&Sbc&SAB&-23.5&54.1&53.0\\
	NGC~5382&209.6&6.3&72.7&13.1&11.9&0.8&S0&EAS&-22.4&19.4&42.0\\
	IC~947&208.1&0.8&74.7&12.5&11.3&1.3&S0&SA0&-23.1&36.5&12.7\\
	NGC~5318&207.7&33.7&72.0&13.0&11.7&1.3&S0&SB0&-22.6&33.4&50.0\\
	NGC~5172&202.3&17.1&68.3&12.3&11.2&0.7&SABc&SB&-22.9&24.4&29.0\\
	NGC~5020&198.2&12.6&57.8&11.5&10.6&1.2&Sbc&SB&-23.2&20.7&90.0\\
	NGC~4914&195.2&37.3&76.7&11.7&10.7&1.0&E&SAB0&-23.8&51.6&78.0\\
	NGC~4334&185.8&7.5&70.9&12.6&11.5&1.0&Sab&SB&-22.8&26.2&45.7\\
	NGC~4169&183.1&29.2&63.6&12.7&11.4&0.8&S0&SB0&-22.6&25.0&47.0\\
	NGC~4015&179.7&25.0&72.2&13.1&11.8&0.9&E&SA0&-22.5&26.5&37.0\\
	NGC~4008&179.6&28.2&61.0&12.4&11.2&0.9&E&SA0&-22.7&23.7&64.0\\
	NGC~3872&176.5&13.8&54.2&11.8&10.6&0.8&E&SA0&-23.0&34.4&75.0\\
	NGC~3812&175.3&24.8&60.6&12.3&11.1&0.8&E&E0&-22.8&28.5&40.0\\
	NGC~3801&175.1&17.7&56.2&12.4&10.9&1.1&S0&SA0&-22.8&56.1&46.0\\
	NGC~3646&170.4&20.2&70.0&12.0&10.8&1.1&Sc&SA&-23.4&50.5&29.0\\
	NGC~3221&155.6&21.6&66.8&12.8&11.5&0.7&Sc&SB&-22.6&28.4&54.0\\
	NGC~2954&145.1&14.9&61.6&12.6&11.4&1.1&E&SA0&-22.5&20.2&47.0\\
	NGC~2911&143.4&10.2&51.8&11.5&10.0&1.1&S0&SA0&-23.6&108.0&128.0\\
	NGC~2872&141.4&11.4&51.9&12.2&11.0&1.0&E&E2&-22.6&26.0&38.0\\
	NGC~2862&141.2&26.8&65.2&12.9&11.6&0.9&SBbc&SB&-22.4&22.3&34.0\\
	NGC~2824&139.8&26.3&45.5&11.1&9.9&0.9&S0&SA0&-23.4&44.8&23.3\\
   \hline
   \end{tabular}
   \begin{tablenotes}
      \small
      \item 
      \begin{flushleft}
      \underline{Columns}: See Table~\ref{tab:sample1}.
      \end{flushleft}
    \end{tablenotes}
    \end{threeparttable}
\end{table*}

\begin{table*}
   \caption{Galaxy sample \emph{continued}.}
   \begin{threeparttable}[b]
   \label{tab:sample3}
   \begin{tabular}{lrrrrrccccrr}
   \hline \hline
   Name & RA & Dec & Dist.  & $\mathfrak{m}_g$ & $\mathfrak{m}_{i}$   & Seeing & Morph. & Morph. & $\mathfrak{M}{i}$ & $M_{*,\rm gal}$ & $R_\mathrm{max}$\\
           &  deg   &    deg     &  Mpc    &  mag & mag  &  \arcsec  & (old) & (new) & mag & $10^{10}$\,M$_{\odot}$ &\arcsec\\
   (1) & (2) & (3) & (4) & (5) & (6) & (7) & (8) & (9) & (10) & (11) & (12) \\
   \hline
   Bin 3 &  &  &  &  &  &   &  &  \\
   \hline
	NGC~5473&211.2&54.9&37.5&12.1&10.9&0.9&S0&SB0&-22.0&15.0&77.0\\
	NGC~5448&210.7&49.2&37.8&12.0&10.8&0.8&Sa&SB&-22.1&15.8&52.0\\
	NGC~5375&209.2&29.2&43.6&12.0&10.9&0.9&SBab&SB&-22.3&15.6&85.0\\
	NGC~5377&209.1&47.2&34.6&11.4&10.3&1.0&Sa&SB&-22.4&17.6&113.0\\
	NGC~5363&209.0&5.3&22.6&10.4&9.2&1.2&S0-a&SA0&-22.5&22.4&100.0\\
	NGC~5390&208.9&40.5&46.0&11.2&10.0&0.8&Sbc&SB&-23.3&48.3&82.0\\
	NGC~5353&208.4&40.3&42.6&11.2&10.0&1.1&S0&SB0&-23.1&38.8&28.0\\
	NGC~5354&208.4&40.3&46.3&11.0&9.8&1.0&S0&SA0&-23.5&49.3&69.0\\
	NGC~5350&208.3&40.4&42.5&10.9&9.8&1.0&Sbc&SB&-23.3&32.4&53.0\\
	NGC~5326&207.7&39.6&45.4&12.2&11.0&1.0&Sa&SA&-22.3&20.1&37.0\\
	UGC~8736&207.3&39.5&43.4&12.2&10.4&0.9&Sc&SA&-22.8&54.3&17.0\\
	NGC~5311&207.2&40.0&48.0&12.3&10.9&0.9&S0-a&SA0&-22.5&29.4&78.0\\
	NGC~4845&194.5&1.6&26.7&12.3&10.9&1.1&SABa&SA&-21.3&13.2&70.0\\
	NGC~4772&193.4&2.2&26.0&11.7&10.5&1.0&SABa&SA&-21.6&10.1&112.0\\
	NGC~4649&190.9&11.6&16.8&9.7&8.4&1.1&E&SA0&-22.7&28.1&155.0\\
	NGC~4643&190.8&2.0&29.3&10.8&9.6&0.9&S0-a&SB0&-22.7&25.7&124.0\\
	NGC~4636&190.7&2.7&14.7&9.3&8.2&1.1&E&SA0&-22.7&23.8&210.0\\
	NGC~4579&189.4&11.8&21.0&10.4&9.2&1.0&Sb&SB&-22.4&19.0&75.0\\
	NGC~4548&188.9&14.5&15.4&10.2&9.0&0.8&Sb&SB&-22.0&13.6&156.0\\
	NGC~4536&188.6&2.2&15.6&11.2&9.7&1.1&SABb&SB&-21.3&13.2&89.0\\
	NGC~4527&188.5&2.7&14.2&11.2&9.5&1.2&SABb&SB&-21.2&13.0&125.0\\
	NGC~4526&188.5&7.7&16.9&10.3&9.1&1.0&S0&SB0&-22.0&13.7&72.0\\
	NGC~4494&187.9&25.8&17.1&10.4&9.3&0.8&E&E2&-21.9&10.2&146.0\\
	NGC~4459&187.3&14.0&16.1&10.6&9.4&0.9&S0&E2&-21.6&10.7&68.0\\
	NGC~4429&186.9&11.1&14.9&10.4&9.1&1.1&S0-a&SA0&-21.7&12.0&108.0\\
	NGC~4414&186.6&31.2&17.7&10.3&9.2&1.2&Sc&SA&-22.0&10.3&140.0\\
	NGC~4382&186.4&18.2&17.9&10.0&8.9&1.4&S0-a&SA0&-22.3&12.1&120.0\\
	NGC~4378&186.3&4.9&44.0&11.9&10.7&1.1&Sa&SA&-22.5&23.7&96.0\\
	NGC~4374&186.3&12.9&18.5&10.0&8.8&1.1&E&SA0&-22.6&23.7&228.0\\
	NGC~4261&184.8&5.8&31.6&11.0&9.8&1.1&E&E2&-22.7&27.2&120.0\\
	NGC~4235&184.3&7.2&42.0&12.0&10.7&0.8&Sa&SA&-22.4&23.7&53.0\\
	NGC~4233&184.3&7.6&33.9&12.5&11.2&0.8&S0&SA0&-21.4&10.2&66.0\\
	NGC~4224&184.1&7.5&45.0&12.4&11.0&1.0&SABa&SA0&-22.2&23.4&71.0\\
	NGC~4168&183.1&13.2&30.9&11.4&10.3&1.1&E&SA0&-22.1&12.0&100.0\\
	NGC~4102&181.6&52.7&18.6&11.8&10.3&1.0&SABb&SB&-21.0&10.8&87.0\\
	NGC~4045&180.7&2.0&34.9&12.1&10.8&1.0&Sa&SB&-21.9&15.0&63.0\\
	NGC~3976&179.0&6.7&43.1&12.2&10.9&0.8&Sb&SB&-22.2&17.6&26.0\\
	NGC~3941&178.2&37.0&19.0&10.6&9.5&1.4&S0&SB0&-21.9&10.0&96.0\\
	NGC~3900&177.3&27.0&34.5&11.7&10.6&1.1&S0-a&SA0&-22.1&11.3&81.0\\
	NGC~3718&173.1&53.1&21.2&11.6&10.0&1.2&Sa&SA&-21.6&19.0&105.0\\
	NGC~3675&171.5&43.6&15.4&10.6&9.3&1.1&Sb&SA&-21.6&11.9&135.0\\
	NGC~3665&171.2&38.8&37.7&11.0&9.8&1.0&S0&SA0&-23.1&38.0&120.0\\
	NGC~3658&171.0&38.6&37.3&12.0&10.7&0.9&S0&SB0&-22.1&16.2&62.0\\
	NGC~3583&168.5&48.3&38.3&11.9&10.7&1.0&Sb&SB&-22.2&13.8&86.0\\
	NGC~3414&162.8&28.0&25.2&11.3&10.1&0.9&S0&SA0&-21.9&12.4&107.0\\
	NGC~3368&161.7&11.8&11.7&9.8&8.5&1.1&Sab&SB&-21.8&13.2&188.0\\
	NGC~3189&154.5&21.8&24.2&11.1&9.8&0.9&Sa&SB&-22.2&19.0&76.0\\
	NGC~3169&153.6&3.5&16.4&11.4&9.9&1.2&Sa&SA&-21.2&12.5&158.0\\
	NGC~2968&145.8&31.9&28.6&12.6&10.9&1.1&Sa&SB&-21.4&15.4&77.0\\
	NGC~2962&145.2&5.2&34.0&12.0&10.7&1.3&S0-a&SB0&-22.0&17.4&71.0\\
	NGC~2894&142.4&7.7&35.9&13.3&11.7&0.9&Sa&SB&-21.1&11.4&48.0\\
	NGC~2859&141.1 &34.5&30.1& 11.6&10.4&0.9&S0-a&SB0&-22.0&11.6&70.0\\
   \hline
   \end{tabular}
    \begin{tablenotes}
      \small
      \item 
      \begin{flushleft}
      \underline{Columns}: See Tables~\ref{tab:sample1} and \ref{tab:sample2}.
      \end{flushleft}
    \end{tablenotes}
    \end{threeparttable}
\end{table*}

\section{Local spheroids' parameters}
\label{sec:sph_para}

We present the structural parameters and the masses of the bulges/spheroids in our host galaxies in Tables~\ref{tab:result1}--\ref{tab:result3}. 
Column~(2) is the distance we used to calculate the absolute magnitude and stellar masses (see Appendix~\ref{sec:dist_corr} and (iv) in Section~\ref{sec:distance}).
Column~(3) showcases the function we used to model the spheroids, either a S\'{e}rsic (S) or a core-S\'{e}rsic (cS) function (see \citet{Graham2003coreSersic}.
Columns~(4)--(6) provide the parameters of the S\'{e}rsic and core-S\'{e}rsic functions. 
In column (4), it represents the surface brightness (i.e. $\mu_\mathrm{e}$) at effective radius $R_\mathrm{e}$ (column~(5)) in equivalent axis.
Column~(6) shows the S\'{e}rsic index $n$ of the spheroids.
Columns~(7) and (8) give their apparent and absolute magnitudes, respectively.
Columns~(9)--(12) are the spheroids' stellar masses based on the four $M_{*}/L$ ratios discussed in Section~\ref{sec:stellar_mass}.
\begin{table*}
   \caption{Spheroid sample.  }
   \begin{threeparttable}[b]
   \label{tab:result1}
   \begin{tabular}{lrrrrrrrrrrr}
   \hline \hline
   Name & Dist. & Type &$\mu_\mathrm{e}$ & $R_\mathrm{e}$ & $n$  & $\mathfrak{m}_{\rm Sph}$ & $\mathfrak{M}_{\rm Sph}$ & $\log_{10}\left(\frac{M_*}{\rm M_\odot}\right)_\mathrm{\small T11}$  & $\log_{10}\left(\frac{M_*}{\rm M_\odot}\right)_\mathrm{\small Z09}$ & $\log_{10}\left(\frac{M_*}{\rm M_\odot}\right)_\mathrm{\small RC15}$ & $\log_{10}\left(\frac{M_*}{\rm M_\odot}\right)_\mathrm{\small IP13}$  \\
       & Mpc &    &    &    kpc     &      &  mag & mag &  &     &  &  \\
    (1) & (2) & (3) & (4) &  (5) & (6) & (7) & (8) & (9)  & (10) & (11) & (12)\\
   \hline
   Bin 1 &  &  &  &  &  &   &  &  & & \\
   \hline
	IC~983&75.38&S&19.20&1.31&1.70&13.42&$-$20.97&$10.33\pm0.15$&$10.43\pm0.13$&$10.50\pm0.12$&$10.67\pm0.09$\\
	UGC8564&87.05&S&19.86&2.13&2.70&13.12&$-$21.58&$10.76\pm0.11$&$10.94\pm0.08$&$11.00\pm0.08$&$11.17\pm0.06$\\
	NGC~5208&89.72&S&18.67&1.42&1.61&13.08&$-$21.69&$10.64\pm0.15$&$10.75\pm0.13$&$10.81\pm0.11$&$10.98\pm0.09$\\
	NGC~4999&76.28&S&19.15&1.08&2.16&13.65&$-$20.76&$10.20\pm0.17$&$10.28\pm0.15$&$10.36\pm0.13$&$10.52\pm0.10$\\
	NGC~4921&75.09&S&20.03&1.70&2.19&13.56&$-$20.82&$10.31\pm0.14$&$10.43\pm0.12$&$10.50\pm0.11$&$10.67\pm0.08$\\
	NGC~4889&87.72&cS&25.40&101.89&10.13&10.23&$-$24.48&$11.77\pm0.14$&$11.89\pm0.12$&$11.96\pm0.11$&$12.13\pm0.08$\\
	NGC~4874&100.26&cS&20.05&3.71&1.78&12.65&$-$22.36&$10.90\pm0.15$&$11.01\pm0.13$&$11.08\pm0.11$&$11.24\pm0.09$\\
	NGC~4583&95.22&S&19.06&1.08&1.23&14.36&$-$20.53&$10.34\pm0.11$&$10.53\pm0.08$&$10.59\pm0.08$&$10.76\pm0.06$\\
	NGC~4555&89.89&S&21.07&7.74&3.41&11.46&$-$23.31&$11.26\pm0.15$&$11.35\pm0.14$&$11.42\pm0.12$&$11.59\pm0.09$\\
	NGC~4213&91.24&cS&21.02&4.62&5.05&12.35&$-$22.45&$10.90\pm0.16$&$10.99\pm0.14$&$11.06\pm0.13$&$11.23\pm0.10$\\
	NGC~4073&83.08&cS&23.72&40.11&6.34&10.25&$-$24.35&$11.70\pm0.15$&$11.81\pm0.13$&$11.88\pm0.11$&$12.05\pm0.09$\\
	NGC~4065&84.68&S&19.12&1.43&1.62&13.39&$-$21.25&$10.41\pm0.16$&$10.49\pm0.15$&$10.56\pm0.13$&$10.73\pm0.10$\\
	NGC~3968&89.26&S&18.50&0.89&1.36&14.02&$-$20.73&$10.21\pm0.16$&$10.30\pm0.14$&$10.37\pm0.13$&$10.54\pm0.10$\\
	NGC~3940&86.28&S&20.78&3.23&4.05&12.82&$-$21.86&$10.68\pm0.15$&$10.78\pm0.13$&$10.86\pm0.12$&$11.02\pm0.09$\\
	NGC~3937&90.69&S&19.01&1.71&2.53&12.81&$-$21.98&$10.68\pm0.17$&$10.75\pm0.15$&$10.82\pm0.14$&$10.99\pm0.10$\\
	NGC~3842&85.05&S&18.89&1.51&1.49&13.12&$-$21.52&$10.56\pm0.15$&$10.66\pm0.13$&$10.73\pm0.12$&$10.90\pm0.09$\\
	IC~724&82.06&S&17.63&0.60&0.95&13.97&$-$20.60&$10.20\pm0.15$&$10.31\pm0.13$&$10.38\pm0.11$&$10.55\pm0.09$\\
	NGC~3805&90.42&S&23.68&26.4&4.65&11.24&$-$23.54&$11.41\pm0.14$&$11.54\pm0.12$&$11.61\pm0.10$&$11.78\pm0.08$\\
	NGC~3615&96.41&cS&21.83&9.98&7.99&11.49&$-$23.43&$11.34\pm0.15$&$11.45\pm0.13$&$11.52\pm0.11$&$11.69\pm0.09$\\
	NGC~3270&89.11&S&18.50&1.02&1.01&13.88&$-$20.86&$10.35\pm0.14$&$10.47\pm0.11$&$10.54\pm0.10$&$10.71\pm0.08$\\
	NGC~3209&89.78&S&18.67&1.53&2.15&12.80&$-$21.97&$10.73\pm0.15$&$10.83\pm0.13$&$10.90\pm0.12$&$11.07\pm0.09$\\
	NGC~3158&101.55&cS&20.49&6.62&3.65&11.48&$-$23.56&$11.41\pm0.14$&$11.53\pm0.12$&$11.60\pm0.11$&$11.77\pm0.08$\\
	NGC~3126&75.01&S&18.33&0.71&1.72&13.85&$-$20.52&$10.27\pm0.12$&$10.43\pm0.10$&$10.49\pm0.09$&$10.66\pm0.07$\\
	NGC~3106&88.19&cS&18.82&1.28&2.12&13.35&$-$21.38&$10.47\pm0.16$&$10.56\pm0.14$&$10.64\pm0.13$&$10.80\pm0.10$\\
	NGC~2918&98.82&S&19.04&1.88&1.67&13.06&$-$21.91&$10.76\pm0.14$&$10.88\pm0.12$&$10.95\pm0.10$&$11.12\pm0.08$\\
	NGC~2832&100.35&cS&21.72&12.69&5.18&11.15&$-$23.9&$11.50\pm0.15$&$11.60\pm0.14$&$11.67\pm0.12$&$11.84\pm0.09$\\
	NGC~2796&100.39&cS&19.13&1.95&2.78&12.88&$-$22.13&$10.94\pm0.12$&$11.12\pm0.09$&$11.18\pm0.08$&$11.85\pm0.07$\\
   \hline
   \end{tabular}
   
   \begin{tablenotes}
      \small
      \item 
      \begin{flushleft}
      \underline{Columns}: 
      (1) the name of the galaxy in descending (westward) order, according to Right Ascension; 
      (2) The selected distance as described in Section~\ref{sec:distance} updated with the newly available \textit{Cosmicflow-3} distances (\citet{Kourkchi2020}, see also Appendix~\ref{sec:dist_corr} and (iv) in Section~\ref{sec:distance});
      (3) Using an SDSS $i$-band image, the spheroid's geometric mean axis, equivalent to a circularised radius, was fit with either a \citet{Sersic1968} (S) or a core-S\'ersic (cS) model \citep[see][]{Graham2003coreSersic}; 
      (4) the surface brightness $\mu_\mathrm{e}$ at $R_\mathrm{e}$ (in mag\,arcsec$^{-2}$);
      (5) the effective radius (in kpc); 
      (6) the S\'ersic index;
      (7) the apparent magnitude of the spheroid in the $i$-band (AB mag) with the Galactic extinction and K-correction applied; 
      (8) the absolute magnitude of the spheroid in the $i$-band (AB mag); 
      the stellar mass of the spheroid using the 
      $g$-$i$ colour-dependent $M_*/L_{i}$ 
      ratios from (9) T11, 
      (10) Z09, 
      (11) RC15, and 
      (12) IP13 (see Equation~\ref{eq:MLCRs}). 
      \end{flushleft}
    \end{tablenotes}
    \end{threeparttable}
\end{table*}

\begin{table*}
   \caption{Spheroid sample \emph{continued}.}
   \begin{threeparttable}[b]
   \label{tab:result2}
   \begin{tabular}{lrrrrrrrrrrr}
   \hline \hline
   Name & Dist. &Type &$\mu_\mathrm{e}$ & $R_\mathrm{e}$ & $n$  & $\mathfrak{m}_{\rm Sph}$ & $\mathfrak{M}_{\rm Sph}$ & $\log_{10}\left(\frac{M_*}{\rm M_\odot}\right)_\mathrm{\small T11}$  & $\log_{10}\left(\frac{M_*}{\rm M_\odot}\right)_\mathrm{\small Z09}$ & $\log_{10}\left(\frac{M_*}{\rm M_\odot}\right)_\mathrm{\small RC15}$ & $\log_{10}\left(\frac{M_*}{\rm M_\odot}\right)_\mathrm{\small IP13}$  \\
        &  Mpc & &    &    kpc     &      &  mag & mag &  &     &    \\
    (1) & (2) & (3) & (4) &  (5) & (6) & (7) & (8) & (9)  & (10) & (11)& (12)\\
   \hline
   Bin 2 &  &  &  &  &  &   &  & & \\
   \hline
	NGC~5490&66.93&S&19.25&2.17&2.62&11.88&$-$22.24&$10.86\pm0.15$&$10.98\pm0.12$&$11.05\pm0.11$&$11.22\pm0.09$\\
	NGC~5444&58.11&cS&24.26&31.04&9.52&10.44&$-$23.38&$11.32\pm0.15$&$11.43\pm0.13$&$11.50\pm0.11$&$11.67\pm0.09$\\
	NGC~5440&54.94&S&18.64&1.18&1.98&12.34&$-$21.36&$10.55\pm0.14$&$10.69\pm0.11$&$10.75\pm0.10$&$10.92\pm0.08$\\
	NGC~5406&74.03&S&18.02&0.57&1.20&14.19&$-$20.15&$10.02\pm0.15$&$10.12\pm0.13$&$10.19\pm0.11$&$10.36\pm0.09$\\
	NGC~5382&60.04&cS&22.28&7.22&9.83&11.57&$-$22.33&$10.87\pm0.15$&$10.97\pm0.13$&$11.04\pm0.12$&$11.21\pm0.09$\\
	IC~947&61.88&S&17.69&0.50&1.63&13.57&$-$20.39&$10.09\pm0.15$&$10.19\pm0.13$&$10.26\pm0.12$&$10.43\pm0.09$\\
	NGC~5318&62.57&S&19.30&1.57&2.26&12.57&$-$21.41&$10.61\pm0.13$&$10.77\pm0.10$&$10.83\pm0.09$&$11.00\pm0.07$\\
	NGC~5172&57.35&S&19.82&0.92&1.80&14.20&$-$19.60&$9.71\pm0.17$&$9.77\pm0.16$&$9.85\pm0.14$&$10.02\pm0.11$\\
	NGC~5020&50.51&S&18.29&0.73&0.64&13.37&$-$20.15&$9.79\pm0.22$&$9.79\pm0.23$&$9.88\pm0.20$&$10.05\pm0.15$\\
	NGC~4914&67.45&S&19.86&3.45&3.26&11.42&$-$22.72&$10.96\pm0.17$&$11.03\pm0.16$&$11.10\pm0.14$&$11.27\pm0.11$\\
	NGC~4334&60.86&S&17.72&0.54&0.51&13.94&$-$19.98&$9.92\pm0.16$&$10.02\pm0.14$&$10.09\pm0.12$&$10.26\pm0.09$\\
	NGC~4169&58.12&S&19.24&1.94&2.17&11.92&$-$21.90&$10.74\pm0.14$&$10.86\pm0.12$&$10.92\pm0.11$&$11.09\pm0.08$\\
	NGC~4015&64.45&S&18.81&0.89&2.00&13.43&$-$20.61&$10.26\pm0.13$&$10.40\pm0.11$&$10.46\pm0.10$&$10.63\pm0.08$\\
	NGC~4008&56.19&S&19.14&1.57&2.34&12.21&$-$21.54&$10.58\pm0.16$&$10.70\pm0.14$&$10.77\pm0.12$&$10.94\pm0.09$\\
	NGC~3872&50.00&S&17.80&0.77&1.86&12.18&$-$21.32&$10.43\pm0.16$&$10.52\pm0.15$&$10.59\pm0.13$&$10.76\pm0.10$\\
	NGC~3812&55.94&cS&21.10&3.88&6.60&11.65&$-$22.09&$10.78\pm0.15$&$10.89\pm0.13$&$10.96\pm0.12$&$11.13\pm0.09$\\
	NGC~3801&52.01&S&20.44&1.71&2.41&13.07&$-$20.51&$10.34\pm0.11$&$10.54\pm0.08$&$10.60\pm0.08$&$10.77\pm0.06$\\
	NGC~3646&63.60&S&18.59&0.78&1.41&13.66&$-$20.36&$10.09\pm0.15$&$10.19\pm0.13$&$10.26\pm0.12$&$10.43\pm0.09$\\
	NGC~3221&62.85&S&19.09&0.68&1.02&14.57&$-$19.42&$9.77\pm0.14$&$9.91\pm0.11$&$9.97\pm0.10$&$10.14\pm0.08$\\
	NGC~2954&59.51&cS&18.69&1.27&1.75&12.44&$-$21.44&$10.48\pm0.16$&$10.57\pm0.14$&$10.64\pm0.13$&$10.81\pm0.10$\\
	NGC~2911&50.62&cS&21.16&4.33&4.43&11.41&$-$22.11&$10.97\pm0.11$&$11.16\pm0.08$&$11.22\pm0.08$&$11.39\pm0.06$\\
	NGC~2872&50.40&S&20.88&5.01&3.65&10.90&$-$22.61&$11.01\pm0.15$&$11.12\pm0.13$&$11.19\pm0.11$&$11.35\pm0.09$\\
	NGC~2862&61.24&S&18.62&0.68&1.15&13.98&$-$19.96&$9.96\pm0.14$&$10.07\pm0.12$&$10.14\pm0.11$&$10.31\pm0.08$\\
	NGC~2824&45.38&S&17.70&0.29&1.30&14.24&$-$19.04&$9.52\pm0.16$&$9.61\pm0.15$&$9.68\pm0.13$&$9.85\pm0.10$\\
   \hline
   \end{tabular}
    \begin{tablenotes}
      \small
      \item 
      \begin{flushleft}
      \underline{Columns}: See Table~\ref{tab:result1}.
      \end{flushleft}
    \end{tablenotes}
    \end{threeparttable}
\end{table*}

\begin{table*}
   \caption{Spheroid sample \emph{continued}.}
   \begin{threeparttable}[b]
   \label{tab:result3}
   \begin{tabular}{lrrrrrrrrrrr}
   \hline \hline
   Name & Dist. & Type & $\mu_\mathrm{e}$ & $R_\mathrm{e}$ & $n$  & $\mathfrak{m}_{\rm Sph}$ & $\mathfrak{M}_{\rm Sph}$ & $\log_{10}\left(\frac{M_*}{\rm M_\odot}\right)_\mathrm{\small T11}$  & $\log_{10}\left(\frac{M_*}{\rm M_\odot}\right)_\mathrm{\small Z09}$ & $\log_{10}\left(\frac{M_*}{\rm M_\odot}\right)_\mathrm{\small RC15}$ & $\log_{10}\left(\frac{M_*}{\rm M_\odot}\right)_\mathrm{\small IP13}$  \\
        & Mpc &  &     &    kpc     &      &  mag & mag &  &     &  &   \\
    (1) & (2) & (3) & (4) &  (5) & (6) & (7) & (8) & (9)  & (10) & (11) & (12)\\
   \hline
   Bin 3 &  &  &  &  &  &   &  &  &  &\\
   \hline
	NGC~5473&31.17&S&17.39&0.39&1.35&12.49&$-$19.98&$9.99\pm0.14$&$10.12\pm0.11$&$10.18\pm0.10$&$10.35\pm0.08$\\
	NGC~5448&30.82&S&19.12&0.53&1.97&13.33&$-$19.11&$9.63\pm0.14$&$9.75\pm0.12$&$9.81\pm0.11$&$9.98\pm0.08$\\
	NGC~5375&36.06&S&20.07&1.36&1.82&12.57&$-$20.22&$9.99\pm0.16$&$10.08\pm0.15$&$10.15\pm0.13$&$10.32\pm0.10$\\
	NGC~5377&26.89&S&19.01&0.61&2.47&12.47&$-$19.67&$9.78\pm0.16$&$9.87\pm0.14$&$9.94\pm0.13$&$10.11\pm0.10$\\
	NGC~5363&15.80&S&19.12&1.21&2.24&9.97&$-$21.02&$10.36\pm0.15$&$10.46\pm0.13$&$10.53\pm0.12$&$10.70\pm0.09$\\
	NGC~5390&38.20&S&19.46&0.75&2.58&13.22&$-$19.69&$9.86\pm0.14$&$9.98\pm0.12$&$10.05\pm0.11$&$10.21\pm0.08$\\
	NGC~5353&34.85&S&17.36&0.44&0.84&12.61&$-$20.10&$10.01\pm0.15$&$10.12\pm0.12$&$10.19\pm0.11$&$10.36\pm0.09$\\
	NGC~5354&38.76&S&19.57&1.87&2.55&11.39&$-$21.55&$10.56\pm0.15$&$10.66\pm0.13$&$10.73\pm0.12$&$10.90\pm0.09$\\
	NGC~5350&34.77&S&18.99&0.35&0.98&14.71&$-$17.99&$9.06\pm0.18$&$9.12\pm0.17$&$9.19\pm0.15$&$9.36\pm0.11$\\
	NGC~5326&37.55&S&18.59&0.82&2.47&12.14&$-$20.73&$10.28\pm0.14$&$10.40\pm0.12$&$10.47\pm0.11$&$10.64\pm0.08$\\
	UGC~8736&35.81&S&21.58&0.63&1.82&15.75&$-$17.02&$8.99\pm0.10$&$9.20\pm0.07$&$9.26\pm0.07$&$9.43\pm0.06$\\
	NGC~5311&40.64&S&19.49&1.26&2.59&12.26&$-$20.79&$10.38\pm0.12$&$10.53\pm0.10$&$10.60\pm0.09$&$10.77\pm0.07$\\
	NGC~4845&18.51&S&20.83&0.95&1.98&12.64&$-$18.70&$9.63\pm0.11$&$9.83\pm0.09$&$9.88\pm0.09$&$10.05\pm0.07$\\
	NGC~4772&16.60&S&19.51&0.74&1.72&11.67&$-$19.43&$9.73\pm0.15$&$9.85\pm0.13$&$9.91\pm0.11$&$10.08\pm0.09$\\
	NGC~4649&16.83&cS&24.85&5.37&0.87&9.71&$-$21.42&$10.54\pm0.15$&$10.66\pm0.12$&$10.73\pm0.11$&$10.90\pm0.09$\\
	NGC~4643&21.64&S&18.43&0.82&1.71&10.95&$-$20.73&$10.23\pm0.16$&$10.32\pm0.14$&$10.39\pm0.12$&$10.56\pm0.10$\\
	NGC~4636&13.11&cS&20.82&2.09&2.15&10.00&$-$20.59&$10.16\pm0.16$&$10.25\pm0.14$&$10.33\pm0.12$&$10.49\pm0.10$\\
	NGC~4579&21.00&S&18.92&1.13&2.82&10.40&$-$21.21&$10.40\pm0.16$&$10.49\pm0.14$&$10.57\pm0.13$&$10.74\pm0.10$\\
	NGC~4548&15.42&S&19.01&0.50&1.68&11.85&$-$19.09&$9.58\pm0.15$&$9.68\pm0.13$&$9.75\pm0.12$&$9.92\pm0.09$\\
	NGC~4536&15.57&S&17.65&0.20&1.48&12.60&$-$18.36&$9.49\pm0.11$&$9.69\pm0.08$&$9.75\pm0.08$&$9.92\pm0.06$\\
	NGC~4527&14.16&S&18.03&0.28&1.52&12.04&$-$18.71&$9.67\pm0.10$&$9.88\pm0.08$&$9.94\pm0.07$&$10.11\pm0.06$\\
	NGC~4526&16.90&S&18.59&0.85&0.43&11.11&$-$20.03&$9.95\pm0.15$&$10.06\pm0.14$&$10.13\pm0.12$&$10.30\pm0.10$\\
	NGC~4494&17.06&S&20.67&3.70&2.93&9.14&$-$22.02&$10.71\pm0.17$&$10.79\pm0.15$&$10.86\pm0.13$&$11.03\pm0.10$\\
	NGC~4459&16.07&S&20.87&3.21&4.00&9.30&$-$21.73&$10.64\pm0.15$&$10.75\pm0.13$&$10.82\pm0.12$&$10.99\pm0.09$\\
	NGC~4429&16.76&S&19.37&1.42&1.38&10.23&$-$20.89&$10.33\pm0.15$&$10.45\pm0.13$&$10.51\pm0.12$&$10.68\pm0.10$\\
	NGC~4414&17.69&S&17.16&0.29&1.26&11.65&$-$19.58&$9.69\pm0.18$&$9.75\pm0.17$&$9.83\pm0.15$&$10.00\pm0.11$\\
	NGC~4382&17.88&cS&18.57&1.04&2.61&9.94&$-$21.32&$10.34\pm0.19$&$10.38\pm0.19$&$10.46\pm0.16$&$10.63\pm0.12$\\
	NGC~4378&41.00&S&19.10&1.37&2.30&11.76&$-$21.31&$10.50\pm0.14$&$10.62\pm0.12$&$10.69\pm0.11$&$10.86\pm0.08$\\
	NGC~4374&18.51&S&19.57&2.61&1.75&9.20&$-$22.14&$10.79\pm0.15$&$10.89\pm0.14$&$10.96\pm0.12$&$11.13\pm0.09$\\
	NGC~4261&31.62&cS&23.29&20.79&8.88&9.03&$-$23.47&$11.36\pm0.15$&$11.47\pm0.13$&$11.54\pm0.11$&$11.71\pm0.09$\\
	NGC~4235&39.40&S&19.37&0.84&2.36&13.04&$-$22.10&$10.01\pm0.13$&$10.16\pm0.11$&$10.22\pm0.10$&$10.39\pm0.08$\\
	NGC~4233&37.45&S&18.59&0.94&0.69&12.44&$-$20.43&$10.20\pm0.13$&$10.34\pm0.11$&$10.40\pm0.10$&$10.57\pm0.08$\\
	NGC~4224&42.00&S&19.56&1.13&2.87&12.56&$-$20.56&$10.28\pm0.12$&$10.43\pm0.10$&$10.50\pm0.09$&$10.67\pm0.07$\\
	NGC~4168&37.02&S&19.98&2.46&1.99&11.18&$-$21.66&$10.52\pm0.18$&$10.58\pm0.17$&$10.65\pm0.15$&$10.82\pm0.11$\\
	NGC~4102&15.46&S&16.35&0.12&0.60&12.84&$-$18.11&$9.41\pm0.10$&$9.61\pm0.08$&$9.67\pm0.07$&$9.84\pm0.06$\\
	NGC~4045&33.27&S&17.58&0.26&0.89&13.85&$-$18.76&$9.50\pm0.14$&$9.63\pm0.11$&$9.70\pm0.10$&$9.87\pm0.08$\\
	NGC~3976&41.06&S&17.47&0.37&1.51&13.25&$-$22.06&$9.92\pm0.14$&$10.04\pm0.12$&$10.11\pm0.11$&$10.28\pm0.08$\\
	NGC~3941&13.85&S&17.20&0.27&1.52&11.21&$-$19.49&$9.68\pm0.17$&$9.75\pm0.16$&$9.83\pm0.14$&$9.99\pm0.10$\\
	NGC~3900&30.53&S&19.28&0.77&1.34&12.79&$-$19.63&$9.71\pm0.18$&$9.78\pm0.16$&$9.85\pm0.14$&$10.02\pm0.11$\\
	NGC~3718&17.97&S&18.86&0.28&2.10&13.22&$-$18.05&$9.40\pm0.10$&$9.62\pm0.07$&$9.67\pm0.07$&$9.84\pm0.06$\\
	NGC~3675&13.44&S&18.22&0.22&2.15&12.45&$-$18.19&$9.30\pm0.13$&$9.44\pm0.11$&$9.50\pm0.10$&$9.67\pm0.08$\\
	NGC~3665&33.84&S&19.68&2.51&1.31&10.90&$-$21.75&$10.67\pm0.15$&$10.78\pm0.13$&$10.85\pm0.11$&$11.02\pm0.09$\\
	NGC~3658&33.44&S&16.79&0.21&1.15&13.45&$-$19.18&$9.64\pm0.14$&$9.76\pm0.12$&$9.83\pm0.11$&$10.00\pm0.08$\\
	NGC~3583&34.37&S&17.93&0.31&1.12&13.81&$-$18.87&$9.46\pm0.16$&$9.54\pm0.14$&$9.62\pm0.13$&$9.78\pm0.10$\\
	NGC~3414&27.53&S&18.76&0.93&2.49&11.35&$-$20.85&$10.28\pm0.15$&$10.38\pm0.13$&$10.45\pm0.12$&$10.62\pm0.09$\\
	NGC~3368&11.70&S&17.83&0.37&1.45&10.80&$-$19.54&$9.80\pm0.14$&$9.93\pm0.12$&$9.99\pm0.11$&$10.16\pm0.08$\\
	NGC~3189&24.88&S&16.83&0.31&1.32&11.91&$-$20.07&$10.04\pm0.14$&$10.17\pm0.11$&$10.24\pm0.10$&$10.41\pm0.08$\\
	NGC~3169&16.44&S&17.87&0.46&1.58&11.06&$-$20.02&$10.15\pm0.11$&$10.35\pm0.08$&$10.40\pm0.08$&$10.57\pm0.06$\\
	NGC~2968&28.33&S&19.24&0.81&2.15&12.28&$-$19.98&$10.17\pm0.10$&$10.39\pm0.08$&$10.44\pm0.07$&$10.61\pm0.06$\\
	NGC~2962&34.04&S&18.84&0.75&2.01&12.38&$-$20.28&$10.11\pm0.14$&$10.23\pm0.12$&$10.30\pm0.11$&$10.47\pm0.08$\\
	NGC~2894&37.18&S&18.95&0.89&2.26&12.25&$-$20.60&$10.42\pm0.10$&$10.64\pm0.07$&$10.69\pm0.07$&$10.86\pm0.06$\\
	NGC~2859&30.00&S&17.88&0.67&1.50&11.63&$-$20.75&$10.21\pm0.16$&$10.30\pm0.15$&$10.37\pm0.13$&$10.54\pm0.10$\\
   \hline
   \end{tabular}
   \begin{tablenotes}
      \small
      \item 
      \begin{flushleft}
      \underline{Columns}: See Tables~\ref{tab:result1} and \ref{tab:result2}.  
      \end{flushleft}
    \end{tablenotes}
    \end{threeparttable}
\end{table*}

% Don't change these lines
\bsp	% typesetting comment
\label{lastpage}

\end{document}